\documentclass{CUP-JNL-DAP}%

\usepackage{graphicx}
\usepackage{multicol,multirow}
\usepackage{amsmath,amssymb,amsfonts}
\usepackage{mathrsfs}
\usepackage{amsthm}
\usepackage{apacite}
\usepackage{rotating}
\usepackage{appendix}
\usepackage[authoryear]{natbib}
\usepackage{caption}
\usepackage{subcaption}

\articletype{RESEARCH ARTICLE}
\jname{Data Centric Engineering}
\jyear{2020}
\begin{document}

\begin{Frontmatter}

\title[Sequential Calibration]{Continuous calibration of a digital twin: comparison of particle filter and Bayesian calibration approaches}

\author*[1,2]{Rebecca Ward}\email{rmw61@cam.ac.uk}
\author[1,2]{Ruchi Choudhary}
\author[1]{Alastair Gregory}
\author[2]{Melanie Jans-Singh}
\author[1,2]{Mark Girolami}

\authormark{Ward, R \textit{et al}}

\address*[1]{\orgdiv{Data-centric Engineering}, \orgname{The Alan Turing Institute}, \orgaddress{\street{The British Library, 96 Euston Road}, \postcode{NW1 2DB}, \state{London}, \country{UK}}}
\address[2]{\orgdiv{Engineering Department}, \orgname{University of Cambridge}, \orgaddress{\street{Trumpington Street}, \postcode{CB2 1PZ}, \state{Cambridge}, \country{UK}}}

\received{10 October 2011}


\abstract{Assimilation of continuously streamed monitored data is an essential component of a digital twin; the assimilated data are used to ensure the digital twin represents the monitored system as accurately as possible. One way this is achieved is by calibration of simulation models, whether data-derived or physics-based, or a combination of both. Traditional manual calibration is not possible in this context hence new methods are required for continuous calibration.  In this paper, a particle filter methodology for continuous calibration of the physics-based model element of a digital twin is presented and applied to an example of an underground farm. The methodology is applied to a synthetic problem with known calibration parameter values prior to being used in conjunction with monitored data.  The proposed methodology is compared against static and sequential Bayesian calibration approaches and compares favourably in terms of determination of the distribution of parameter values and analysis run-times, both essential requirements. The methodology is shown to be potentially useful as a means to ensure continuing model fidelity.  }

\begin{impact}
This study addresses the problem of ensuring that a simulation model is able to continuously assimilate data and identify variation in the underlying model parameters.  As the availability of monitored data rapidly increases across diverse industries, the requirement for tools that are able to make timely use of the data is also on the increase. While the technology for developing the simulation models already exists, in many fields there is a dearth of tools that can facilitate rapid interpretation of streamed data to ensure continuing model fidelity. The methodology presented here potentially offers one way to fill this gap by proposing a sequential calibration process using a particle filter approach. The proposed approach is illustrated by application to a simulation model of an underground farm.      

\end{impact}

\end{Frontmatter}

\section[Introduction]{Introduction}
The technological advancement and drop in price of monitoring equipment has led to a boom in availability of monitored data across industrial fields as diverse as aviation, manufacturing and the built environment.  This facilitates the development of digital twin technology. The specification of what constitutes a digital twin is still evolving, but the essence comprises a monitored system together with a computational model of the system which demonstrably replicates the system behaviour in the context of interest (\cite{Worden_2020}). The computational model may be either data-derived or physics-based; for engineering systems it is more common for a physics-based model to be used as this offers the advantage of representing the salient physical laws that govern the system. The intention is that the computational model can give information that may not be easily accessible from the system directly, and can be used to explore performance when it is impractical to run physical tests.  The greatest potential for digital twinning perhaps lies in systems which are continuously operational generating live streamed data that inform the model, in which case the computational model can be simulated in (close to) real-time to advise changes to operational parameters for improved efficiency (\cite{Madni_2019}). 


Recent advancements in machine learning techniques are leading to a new generation of computational models that combine the physics and the data.  On the one hand, physics enhanced machine learning can improve a data-derived model by constraining it with the governing scientific laws (\cite{Choudhary_2020}).  A more common practice is to start with a physics-based model and incorporate the data to calibrate the model and ensure it matches reality as closely as possible over the range of the available data. This then means that when exercising the model beyond the realm of the data, for example when exploring scenarios that cannot practically be tested with the real system, we can have greater confidence in the model predictions. Calibration can be performed manually but it is time consuming and is not always practical. Indeed, in the situation where model parameters are not static but dynamic, it may be impossible owing to the dimensions of the parameter space and the conflation between different reasons for the differences between model outputs and the data. In the case of a digital twin, an automated calibration process that forms an integral part of the system model is favourable, and thus a computationally expensive and complex calibration process is not well suited. 

Bayesian calibration (BC) offers a formal way to combine measured data with model predictions to improve the model.  In the field of building energy simulation considered here, the Bayesian framework proposed by Kennedy and O'Hagan (\cite{KOH_2001}) (KOH) has been explored in some depth (\cite{Higdon_2004, Heo_2012, Li_2016, Chong_2017}), specifically for the identification of static model parameters and the quantification of uncertainty.  Bayesian calibration has been shown to be the optimal approach as it incorporates prior knowledge into the calibration process which can significantly improve parameter identifiability.  In addition, the KOH methodology enables not only the inference of uncertain (and important) model parameters but also of the model deficiency (model bias) and errors in observations. As such, the KOH methodology is considered the 'gold standard' for calibration of computer models. However, the KOH methodology can be computationally expensive, increasingly so as the numbers of calibration parameters and data points increase, owing to the typical Markov chain Monte Carlo (MCMC) implementation which generates a random walk through the target distribution and is inherently inefficient (\cite{Chong_2017}). This has implications for use in continuous calibration over short timescales - the run-time of the calibration must be shorter than the time interval between acquisition of new data points, and yet there must be sufficient data points to characterise the parameter space.  Nonetheless there has been some exploration of its extension to dynamically varying parameters. For example, \citet{Chong_2019} used the KOH formulation in conjunction with data from building energy management systems to continuously calibrate a building energy model, updating the model every month, demonstrating that prediction of future performance is improved with continuous calibration. The study required selection of a reduced sample size and a more efficient MCMC algorithm to overcome the computational challenge presented by the large dataset.

An alternative approach that potentially offers a time-efficient solution to the problem of continuous calibration is Particle Filtering (PF). This is a sequential Bayesian technique in which a large sample of possible parameter estimates are filtered according to their likelihood given some data.  The use of particle filters for static parameter estimation is described in some detail by \citet{Andrieu_2004}, and the approach has been used for sequential data assimilation in other fields for which the memory effect has a significant impact on parameter values, e.g. hydrology (\cite{Moradkhani_2005}), or for estimation of dynamic state and parameter values in nonlinear state-space models e.g. \cite{Cheng_2017}. The novelty of the study presented here is its potential impact on digital twin technology.  The parameters are not memory-dependent but are time-varying and we require up to date estimates of the parameter values to ensure that forecasting using the physics-based model is as accurate as possible. This is particularly important if the physics-based model can only be a relatively simple representation of the real system - which is often the case. To this end, we explore whether the PF approach can offer a suitable calibration mechanism.  We use monitored data in conjunction with the particle filter to calibrate uncertain model parameters - we then use the estimates of the model parameters in conjunction with the model to predict model performance with a quantification of the uncertainty. The PF is also compared against a sequential BC model using the KOH formulation. While the KOH approach is typically used to get the best estimates of static parameter values, in this study we also explore whether it is feasible to use it with sequentially changing datasets. We demonstrate that while both methods are suitable, the PF method is quicker without losing accuracy. 

The digital twin considered here is a farm constructed in a previously disused underground tunnel, in which salad crops are grown hydroponically. A physics-based simulation model of the farm has been developed that calculates temperature and relative humidity as a function of external weather conditions and farm operational strategies. This is a relatively simple model based on a single thermal zone that simulates heat and mass transfer between the different system components. An extensive programme of monitoring has also been carried out, so data are available for calibration of the model. The aim of the digital twin (or the physics-based simulation model) is that at any time it can be used in conjunction with forecast weather data and operational scenario of the farm to predict environmental conditions within the tunnel with a view to alerting the farm operators to future potentially unsatisfactory conditions. Calibration is a critical component of the digital twin system in this instance, as it is the only mechanism by which the model can be adapted to represent the farm environment to an adequate degree of realism. 

The paper is laid out as follows; in the following section the calibration approaches are compared in more detail. We then describe the physics-based model of the underground farm. The feasibility of the PF and sequential KOH approaches are first tested with synthetic data and known parameters, and thereafter implemented with monitored data from the farm.  In the discussion we explore the extent to which the parameter values identified a) are the values that give the best fit of the model to the data, and b) are indicative of 'real' values, and compare the approaches used in terms of their applicability in the context of a digital twin.


\section[Calibration approach]{Calibration approach}
In this study we consider a static Bayesian calibration approach as formulated by \cite{KOH_2001} (KOH) to be the basis against which we compare the viability of the proposed particle filter approach.  We thus also use the static KOH approach in a sequential manner to explore the comparison as detailed below.  Both approaches make use of Bayes' formula, i.e. for a parameter $Y$ and an observed data point $y(x)$,

\begin{equation}
    P(Y|y(x)) = \frac{P(y(x)|Y).P(Y)}{P(y(x))}
\end{equation}

or, the posterior probability of the parameter value $Y$, given the observed data point $y(x)$ is equal to the \textit{likelihood} of the observed data point, $P(y(x)|Y)$, multiplied by the probability of $Y$ before making the observation - the prior probability, $P(Y)$ - all normalised by a normalisation factor $P(y(x))$.

\subsection[BC]{The KOH Calibration Framework}

In the Bayesian calibration formulation proposed by Kennedy and O'Hagan (\cite{KOH_2001}), a numerical simulator $\eta(x,\theta^*)$ can be related to field observations $y(x)$ by the following equation:

\begin{equation}
    y(x) = \eta(x,\theta^*) + \delta(x) + \epsilon + \epsilon_n
    \label{eqn:KOH}
\end{equation}

where $x$ are observable inputs into the model or \textit{scenarios} for example location of sensor or time of sensing, and $\theta^*$ represents the true but unknown values of the parameters $\theta$ which characterise the model.  This formulation inherently expects a discrepancy, or bias, between the model and reality, accounted for by $\delta(x)$.  $\epsilon$ represents the observation error and $\epsilon_n$ the numerical error.  Calibration of the model aims to identify the parameters $\theta^*$. Whereas traditional calibration approaches require multiple runs of the computer simulation with systematic variation of the input parameters and subsequent identification of the combination of parameters that gives the closest match to reality, the KOH framework offers a more efficient way to identify the uncertainty associated with the calibration parameters.  Rather than using an exhaustive iterative approach, it is common practice to use an emulator to map the model inputs to the outputs.  Gaussian process (GP) models are commonly used as the basis for the emulator, with separate models used for the simulator $\eta(x,\theta^*)$ and the discrepancy term $\delta(x)$ in the above equation. These GP models have their own hyperparameters, specifically precision hyperparameters $\lambda_\eta$ and $\lambda_b$ that relate to the magnitude of the emulator and model discrepancy respectively, and $\beta_\eta$ and $\beta_b$ that determine the correlation strength in each along the dimensions of $x$ and $\theta$ and determine the smoothness of the emulator and discrepancy function.  The random error terms for measurement and numerical error are included as unstructured error terms and are incorporated into the GP covariance matrix with precision hyperparameters $\lambda_e$ and $\lambda_{en}$. All of these parameters are uncertain and require a prior probability distribution to be specified in the KOH approach.  We have used prior distributions suggested in previous papers for the hyperparameters as detailed in Table \ref{tab:KOHHyperparameters} \citep{Menberg_2019}.

\begin{table}[]
    \centering
    \begin{tabular}{lcc}
    \TCH{Model hyperparameter} & \TCH{Prior distribution} \\
    \hline
    \TCH{$\lambda_\eta$}  & \TCH{Gamma(10,10)}  \\
    \TCH{$\lambda_b$}  & \TCH{Gamma(10,0.3)}  \\
    \TCH{$\lambda_e$}  & \TCH{Gamma(10,0.03)}  \\
    \TCH{$\lambda_{en}$}  & \TCH{Gamma(10,0.001)}  \\
    \TCH{$\beta_\eta$}  & \TCH{Beta(1,0.5)}  \\
    \TCH{$\beta_b$}  & \TCH{Beta(1,0.4)}  \\
    \hline
    \end{tabular}
    \caption{KOH approach hyperparameter prior distributions}
    \label{tab:KOHHyperparameters}
\end{table}

The practical application of the KOH approach to calibration of building energy simulation models is described in detail by \citet{Chong_2018}, and the implementation of the framework is as described by \citet{Guillas_2009}.  The procedure is to compare measured data (observations) against outputs from computer simulations generated using plausible ranges of uncertain input parameters under known scenarios.  By exploring the likelihood of the data given the simulation output, the posterior distribution of each calibration parameter is derived. The process, in sum, requires the following steps:

\begin{itemize}
    \item Sensitivity analysis to determine the model parameters that have the greatest impact on the simulation output.
    \item Perform calibration runs - run simulations over the plausible range of the parameters identified in the sensitivity analysis, varying one parameter at a time.
    \item Acquire field data - monitored observations corresponding to simulation output,
    \item Fit GP emulator to the field data and simulation output by calculating the mean and covariance function as described by \citet{Higdon_2004},
    \item Explore the parameter space using MCMC and extract the posterior distributions of the calibration parameters.
\end{itemize}

In this process, all the monitored data and simulation model outputs are included in the model at once, in the GP emulator, hence the run time of calibration is very dependent on the number of data points, increasing in proportion to the square of the size of the dataset. This means that it is impractical to perform Bayesian calibration with a large number of observations.  

The KOH framework also has at its heart an assumption of stationarity i.e. the calibration parameters, $\theta$, are assumed to be constant over the range of the data.  We are interested in parameters that are not necessarily constant over the timescales considered.  However, they may be essentially constant over shorter time periods, so it is possible to use the KOH framework to estimate the parameters over these shorter time periods. Given that assumption, it is feasible to implement the KOH calibration over successive time periods: at each timestep the oldest data point is removed and a new data point added. In that instance the posterior distribution for the previous timestep may be used as the prior distribution for the new dataset.  In essence this is a similar approach to the particle filter with the exception that data points are not considered singly but in groups over which the calibration parameter values are assumed to be constant. The potential benefit of this is that it facilitates exploration of the error terms in detail.

\subsection[PF]{Particle filtering}
A particle filter is a sequential Bayesian inference technique in which a large sample of possible parameter estimates are filtered according to their likelihood given some data. As an example, consider a virus spreading through a community of people. With no additional knowledge we can only guess which members of the population have the virus.  If then we get some information - say measured temperatures for each member, where an increased temperature is one of the symptoms of the virus - we can update the likelihood of each of our guesses according to their measured temperature.  We then re-sample from our population taking the increased likelihood into account, generating a new sample where each member again has an equal probability of having the virus and predict what the new temperatures will be based on our updated estimates of the location of the virus hosts.  When we get new temperature information we update the likelihood again and repeat the re-sampling and prediction.  





In the context of this study, the particles consist of possible values of the uncertain model inputs/parameters and the information is the monitored data. Based on the data, we update the likelihood of the particles i.e. we look to see how close the monitored data is to the outputs of the model with the parameter values assigned to each particle. We then re-sample the particles taking the likelihood into account and generate a new set of particles with equal weight. We repeat the process by updating the likelihoods of the new particle set using the next value of monitored data and the values predicted by the model using the parameter values from the new particles - and we then re-sample.  This repeating process results in a set of values for the model parameters that continuously update in line with the monitored data.  The re-sampling is an essential part of the filter to avoid degeneracy i.e. to avoid the situation where a few particles dominate the posterior distribution, as discussed in detail by \cite{Doucet_2008}.

Mathematically, the framework for the particle filtering approach is as follows: consider the model $\eta (\theta)$, where $\theta$ are parameters of the model.  We conjecture that observed data are given by $y \sim \rho \eta + \delta + \epsilon$, where $\rho$ is a scaling parameter, $\delta$ is a mean-zero Gaussian process representing the mismatch between the model and the data and $\epsilon$ is the measurement error with variance $\sigma^2$. We assume that the smoothness of the emulator and model mismatch are determined by a length scale, $l$. 

\begin{enumerate}
    \item Start by sampling $N$ different particles of the hyperparameters $l$, $\rho$ and model input parameters $\theta$ from the prior distributions $p(l)$, $p(\rho)$ and $p(\theta)$. Denote these particles $\{l_j\}_{j=1:N}$, $\{\rho_j\}_{j=1:N}$ and $\{\theta_j\}_{j=1:N}$ respectively.
    \item Obtain the model outputs $d_i$ taken at the coordinates $X_i^d$ and with the model input parameters $t_i$.  Also consider the monitored data $Y_i$ taken at the coordinates $X_i^Y$.
    \item Define the covariance matrices (for each of the particles),
    $K_j = \begin{bmatrix}
    k_\eta(X_i^d,X_i^d,t_i,t_i|l_j) & \rho_j k_\eta(X_i^d,X_i^Y,t_i,\theta_j|l_j) \\
    \rho_j k_\eta(X_i^Y,X_i^d,\theta_j,t_i|l_j) & \rho_j^2 k_\eta(X_i^Y,X_i^Y,\theta_j,\theta_j|l_j) + k_Y (X_i^Y,X_i^Y,\theta_j,\theta_j|l_j) + \sigma^2 \mathbf{I}
    \end{bmatrix}$
    
    using the covariance functions $k_\eta(x,x',t,t'|l)$ associated with the computer model emulator, $\eta$ and $k_Y(x,x',t,t'|l)$ associated with the model-data mismatch $\delta$. Note that these are conditioned on the hyperparameter, $l$.
    \item For each of the particles, compute the mean-zero Gaussian process marginal likelihoods $\xi_j = p([d_i,Y_i]^T|0,K_j)$. These will be used to update the posterior distribution of  $l$, $\rho$ and $\theta$ in the next step.
    \item Compute the normalized \textit{weights} $w_j = \xi_j/\left({\Sigma_{j=1}^N}\xi_j\right)$. Re-sample using these weights, such that the new sample of particles all have even weight.
    \item Increase $i$ by one, and repeat steps (2) to (5) with this new sample of particles.
\end{enumerate}

Throughout this process, the set of particles associated with the model input parameters provide an empirical approximation to their posterior distribution. The implementation of this approach has been written in Python and has used the Gaussian Process tools included in the Scikit-learn package \citep{scikit-learn}.  In this implementation, we assume a value for the scaling parameter, $\rho = 1$, to match the context in Equation \ref{eqn:KOH}. For the benefit of brevity in explaining the approach, we also set the variance of the measurement error, denoted by $\sigma^2$, to the mean of the prior distribution in Table \ref{tab:KOHHyperparameters}, although this isn't essential.

\section[Model]{Digital Twin of the Underground Farm}
Growing Underground is an unheated hydroponic farm developed in disused tunnels in London, UK. The farm grows salad crops for sale to restaurants and supermarkets across London using artificial lighting while relying on the relatively stable thermal ground conditions.  The tunnels were built in the 1940s to act as air raid shelters but with the intention of ultimately being incorporated into the London Underground rail network.  This never happened and the tunnels have been disused since acting as temporary housing for migrants in the 1960s.  They are 33m below ground, a location where the deep soil temperature is approximately constant, and enjoy some heat transferred from the nearby rail tunnels.  The heat source however is predominantly the LED growing lights which are typically switched on overnight and turned off during the day time so as to compensate for external daily temperatures tending to be colder at night and warmer in the day. The farm is ventilated to the outside air using the original ventilation system designed in the 1940s. There are two ventilation shafts which operate in tandem, one at Carpenter's Place (CP) at the main tunnel entrance and one at Clapham Common (CC) at the far end of the tunnel. These are manually operated and both affect the ventilation rate experienced in the farm.

It is important to maintain environmental conditions at optimum levels for plant growth and to this end the space is monitored for temperature, relative humidity and CO$_2$ levels (\cite{JansSingh_2019}). The purpose of the monitoring is both to facilitate the maintenance of adequate conditions for optimal crop growth and also to help optimise the energy use.  The tunnel environment is complex to understand, primarily because of the age and lack of information relating to the design of the tunnels and change in system properties over time. Traditional control systems developed for the horticultural industry cannot be used here as there is limited scope for automatic control of the tunnel environment.  The aim of the digital twin is to combine the monitored data with a physics-based model which simulates the heat and mass exchanges within the tunnel and with the external environment. By combining both monitoring and simulation we aim to develop a system which can be used to analyse performance, to predict future environmental conditions in conjunction with forecast weather data, and to help optimise the designs for expansion of the farm.

\subsection{The Simulation Model}
A simple physics-based numerical model has been developed representing a 1D slice through the central section of the farm.  The model calculates temperature and relative humidity of the tunnel air as a function of time by solving heat and mass exchange equations pertinent to the tunnel geometry, subject to the temperature, moisture content and CO$_2$ concentration of the incoming ventilation air and the deep soil temperature. The primary source of heat input is the LED lights which are typically operated overnight and switched off during daytime. For example, the time-dependent temperature variation can be calculated through a 1D slice of the tunnel using the heat balance equation:

\begin{equation}
    \frac{dT_j}{dt} = \frac{A_g}{m_jc_j} \sum_i q_{i,j}
\end{equation}

which gives the change in temperature, $T$  $(K/s)$ in each layer $j$ - such as the growing medium, vegetation, air and tunnel lining - as a function of the heat flows $q$  $(W/m^2)$ between the different layers.  Here, $i$ represents the different heat transfer processes such as radiation, convection, etc., $A_g$ is the surface area ($m^2$), $m$ is the mass of each layer ($kg$) and $c$ is the heat capacity ($J/kgK$).  

In a similar manner, the moisture content of the air can be calculated from the mass balance equation:

\begin{equation}
    \frac{dC_a}{dt} = \frac{A_g}{h_{fg}V} \sum_k q_{k}
\end{equation}

where $C_a$ is the change in moisture content of the air ($kg/m^3/s$), $h_{fg}$ is the latent heat of condensation of water ($J/kg$), $V$ is the volume of the unit ($m^3$) and $q_k$ are the heat flows ($W/m^2$) from the $k$ different latent heat transfer processes.  The moisture content and temperature are inextricably linked as changes to the moisture content of the air are associated with latent heat transfer. Hence the equations require solving in parallel to determine the temperatures and relative humidities within the tunnel. 

Optimal conditions for crop growth in the tunnel require maintenance of stable temperatures and relative humidity. The two routes for heat exchange from the tunnel are conduction into the surrounding ground and ventilation to the outside air. Both routes are dependent on the temperature differential.  The relative stability of the deep soil temperature, here assumed to be a constant $14^oC$, means that heat is typically lost to the surrounding ground, but this happens slowly owing to the ground thermal capacity and over time approaches an approximately steady-state temperature gradient. By comparison, heat exchange with the outside happens rapidly through bulk air transfer at rates of the order of $1 - 10 ACH$ (Air Changes per Hour).  
Within the model, heat transfer due to ventilation is simulated according to the equation:

\begin{equation}
    QV_{ie} = N/3600 \, V \rho_i c_i \Delta T
\end{equation}

where $QV_{ie}$ is the heat flux from inside to outside ($W$), $N$ is the ventilation rate in $ACH$, $rho_i$ is the density of the humid air ($kg/m^3$), $c_i$ is the thermal capacity of the humid air ($J/kgK$) and $\Delta T$ is the temperature difference between the inside of the tunnel and the outside ($K$).  As the tunnel environment may have a different relative humidity from outside, ventilation also gives rise to changes in the moisture content of the internal air. Within the farm, the heat transfer between the different farm layers by conduction and radiation is simulated according to the temperatures of the different layers. Heat transfer to the internal air, and hence then exchanged with the outside through the ventilation, is via convection from each of the farm layers.  In the model, convection is simulated using the following equation (here given for the vegetation-air interface): 

\begin{equation}
    QV_{iv} = A_v Nu \, \lambda \, \Delta T /d_v
\end{equation}

where $QV_{iv}$ is the heat transfer from the air to the vegetative surface ($W$), $A_v$ is the area of the surface ($m^2$), $\lambda$ is the thermal conductivity of air ($W/m/K$) and $\Delta T$ is the temperature difference across the interface ($K$). $Nu$ is the dimensionless Nusselt number which is a function of the speed of the air travelling over the convective surface.

The temperature and relative humidities within the farm are directly dependent on the operational conditions such as ventilation rate and internal air speed.  Only partial information is available concerning the design of the tunnel ventilation system, and these parameters are difficult to measure accurately. In addition, the simplicity of the model means that parameters may represent a mean effect of different physical components e.g. the combined effect of infiltration and controlled ventilation are represented in the model as a single bulk ventilation rate.  Calibration of the model and estimation of the parameters therefore becomes an important step in improving the model accuracy.  But the parameters are not constant and therefore it becomes necessary to re-calibrate the model to infer the change in parameter values.  The problem is that it is not always obvious when parameters are changing - some changes, such as manual  alterations to the ventilation settings are known, but the impact of the alteration in terms of the change to the ventilation rate is not a linear function of the dial settings.  

This need to re-calibrate, together with the availability of continuously monitored data, makes development of a sequential calibration process particularly desirable.  

\subsection{Sensitivity study}
For calibration of the model, the first step is to ascertain which model output is the most important with respect to farm environment.  From the viewpoint of the growers the air temperature and the relative humidity are the main quantities of interest.  Plants grow best in optimal temperature conditions and if the the relative humidity gets too high, growth of mould can occur affecting crop yield. So the calibration of the model focuses on these model outputs.

 A sensitivity study has been carried out using the Morris method (\cite{Menberg_2016}) to identify the relative significance of each input parameter with respect to the quantities of interest. The 8 input parameters of primary importance for the heat and mass balance and their plausible ranges are given in Table \ref{tab:UncertainParameters}.  The parameter ranges given in the table have been derived either from measurements, are anecdotal, or have been estimated based on an understanding of the farm processes.  For example, the fraction of planted area approaching harvest has been estimated based on the records of the number of newly planted trays and the number of trays being harvested at any time. Equally, the mean mat saturation is estimated based on the watering strategy employed in the farm in which the trays are flooded periodically and then drained over a period of time. The internal air speed has been measured (\cite{JansSingh_2019}).  The ventilation rate range is based on the tunnel ventilation system having been designed for an air change rate of $4 ACH$ but with the knowledge that additional uncontrolled ventilation can occur through the access shafts.  The fraction of the lighting power that is emitted as heat has been estimated according to values found in the literature (e.g. \cite{LEDMag_2005}), and the surface temperature of the lights is based on estimates provided by farm personnel.  The characteristic dimensions of the mat and leaf control whether the heat convection is primarily characterised by bulk heat transfer from the entire tray area - in which case the characteristic dimension is similar to that of the tray i.e. $1$, or whether it is more characterised by the dimension of the leaf or the gaps between plants.    
 
 The Morris method uses a factorial sampling strategy to discretise the parameter space in order to ensure that the entire space is covered efficiently.  In this way for 10 trajectories for each of the 8 parameters, 80 different combinations of the parameters have been run with each combination only one parameter value different from the previous one.  A statistical evaluation of the elementary effects (EEs), i.e. the impact of a change in one parameter by a pre-defined value on the output of the model, has been performed using the absolute median and standard deviation of the EEs \citep{Menberg_2019}.   

\begin{figure}[t]
\centering
\begin{subfigure}{.5\textwidth}
  \centering
  \includegraphics[width=.9\linewidth]{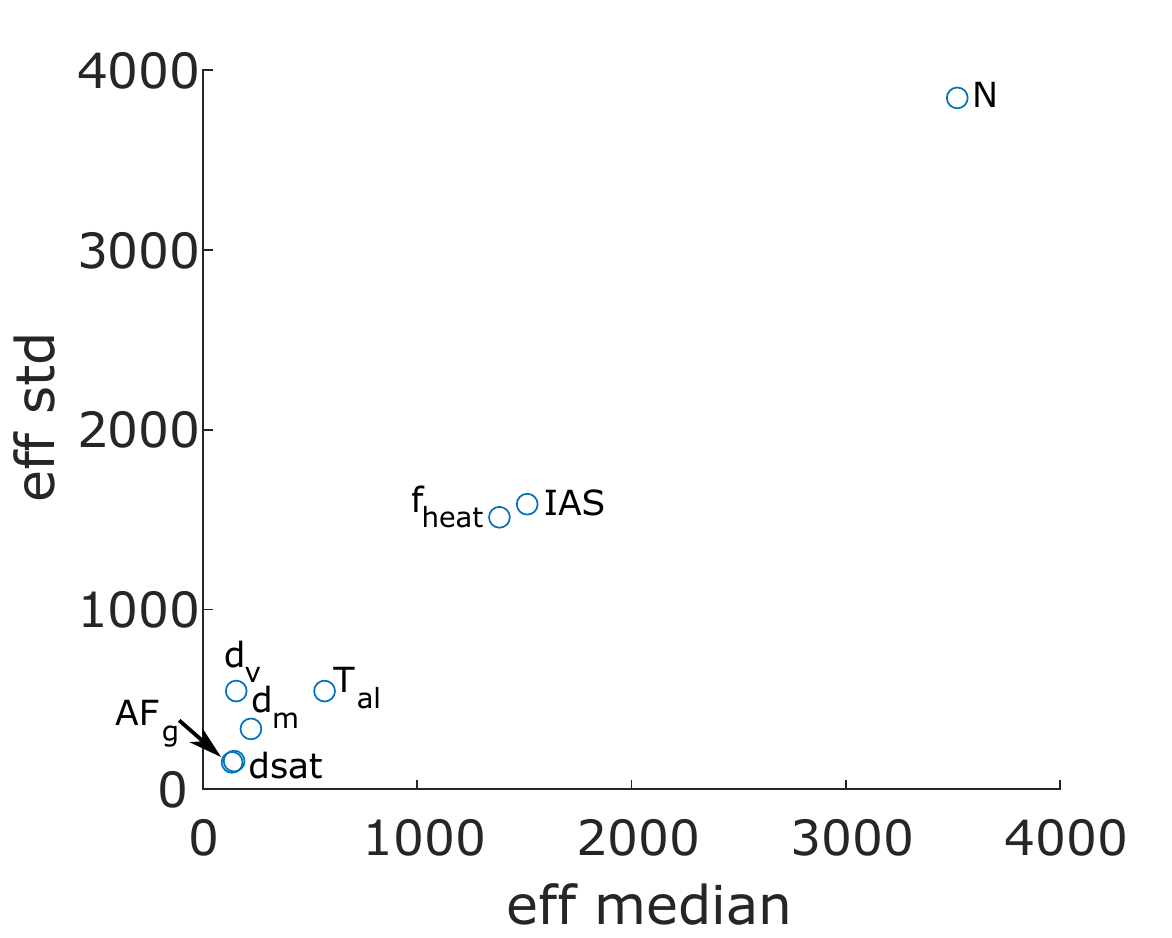}
  \caption{Sensitivity analysis results for temperature, T}
  \label{fig:sub1}
\end{subfigure}%
\begin{subfigure}{.5\textwidth}
  \centering
  \includegraphics[width=.9\linewidth]{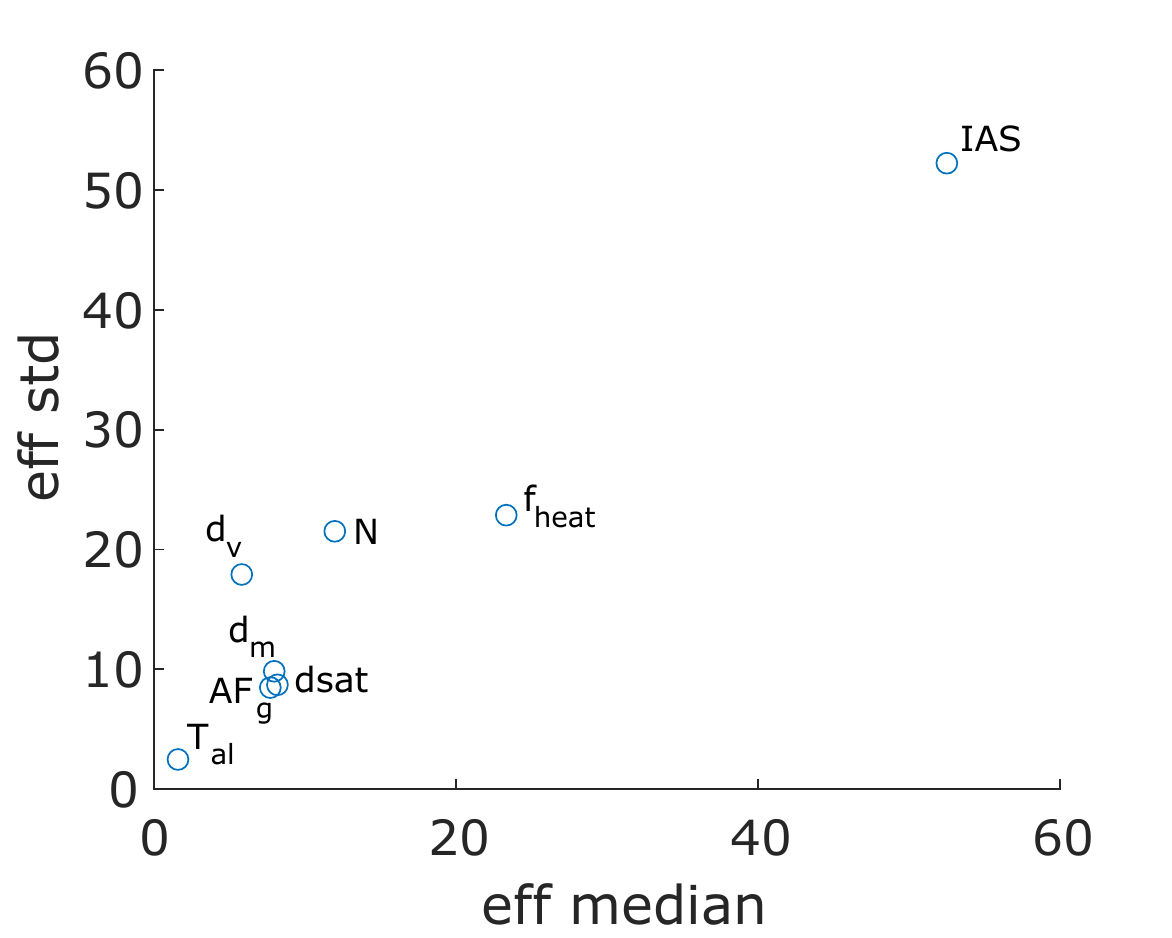}
  \caption{Sensitivity analysis results for relative humidity, RH}
  \label{fig:sub2}
\end{subfigure}
\caption{Sensitivity analysis results showing the impact of each parameter on the simulation output - higher values imply a more significant effect}
\label{fig:SensitivityAnalysis}
\end{figure}

The results of the sensitivity study are summarised in Figure \ref{fig:SensitivityAnalysis}. This shows the standard deviation as a function of the absolute median of the EEs when considering the summed absolute difference between the predicted and monitored a) temperature, and b) relative humidity, over a 30 day period.  A higher value for the standard deviation for a specific parameter implies that changes to that parameter have a more significant impact on the output of the model. Figure \ref{fig:sub1} indicates that the most influential parameters for temperature are the ventilation rate, $N$, and the internal air speed, $IAS$, closely followed by the fraction of energy emitted as heat by the lights, $f_{heat}$.  For relative humidity,  Figure \ref{fig:sub2} indicates that the internal air speed is very significant, followed again by $f_{heat}$. The ventilation rate, $N$ is the next most significant parameter. The heat fraction of the lighting has a significant impact on the simulation output and we could calibrate the model for this parameter to get a better estimate of the true value.  However, as we cannot influence this value in any way, we choose to focus on the two parameters over which we have some control i.e. the ventilation rate and internal air speed.  This approach is directed by the digital twin ethos in that we wish to be able to forecast the impact of changes to the operational conditions on the tunnel internal environment.

Relative humidity is itself a function of temperature, so rather than calibrating the model separately for both parameters, we have chosen to calibrate for relative humidity alone. 

\begin{table}[]
    \centering
    \begin{tabular}{lccc}
    \TCH{Model parameter} & \TCH{Symbol} & \TCH{Min} & \TCH{Max} \\
    \hline
    \TCH{Fraction of planted area approaching harvest} & \TCH{$AF_g$} & \TCH{0.25} & \TCH{0.375} \\
    \TCH{Internal air speed ($m/s$)} & \TCH{IAS} & \TCH{0.1} & \TCH{0.85} \\
    \TCH{Ventilation rate ($ACH$)} & \TCH{N} & \TCH{1} & \TCH{10} \\
    \TCH{Fraction of lighting power output as heat} & \TCH{$f_{heat}$} & \TCH{0.7} & \TCH{0.9} \\
    \TCH{Surface temperature of lights ($^oC$)} & \TCH{$T_{al}$} & \TCH{22.5} & \TCH{27.5} \\
    \TCH{Characteristic leaf dimension ($m$)} & \TCH{$d_v$} & \TCH{0.01} & \TCH{1} \\
    \TCH{Characteristic mat dimension ($m$)} & \TCH{$d_m$} & \TCH{0.1} & \TCH{1} \\
    \TCH{Mat saturation fraction} & \TCH{$dsat$} & \TCH{0.4} & \TCH{0.6} \\
    \hline
    \end{tabular}
    \caption{Uncertain parameters assessed in model sensitivity analysis}
    \label{tab:UncertainParameters}
\end{table}

\section{Calibration results} 

\subsection[Toy]{Toy problem}
\begin{figure}[t]%
\FIG{\includegraphics[width=0.6\textwidth]{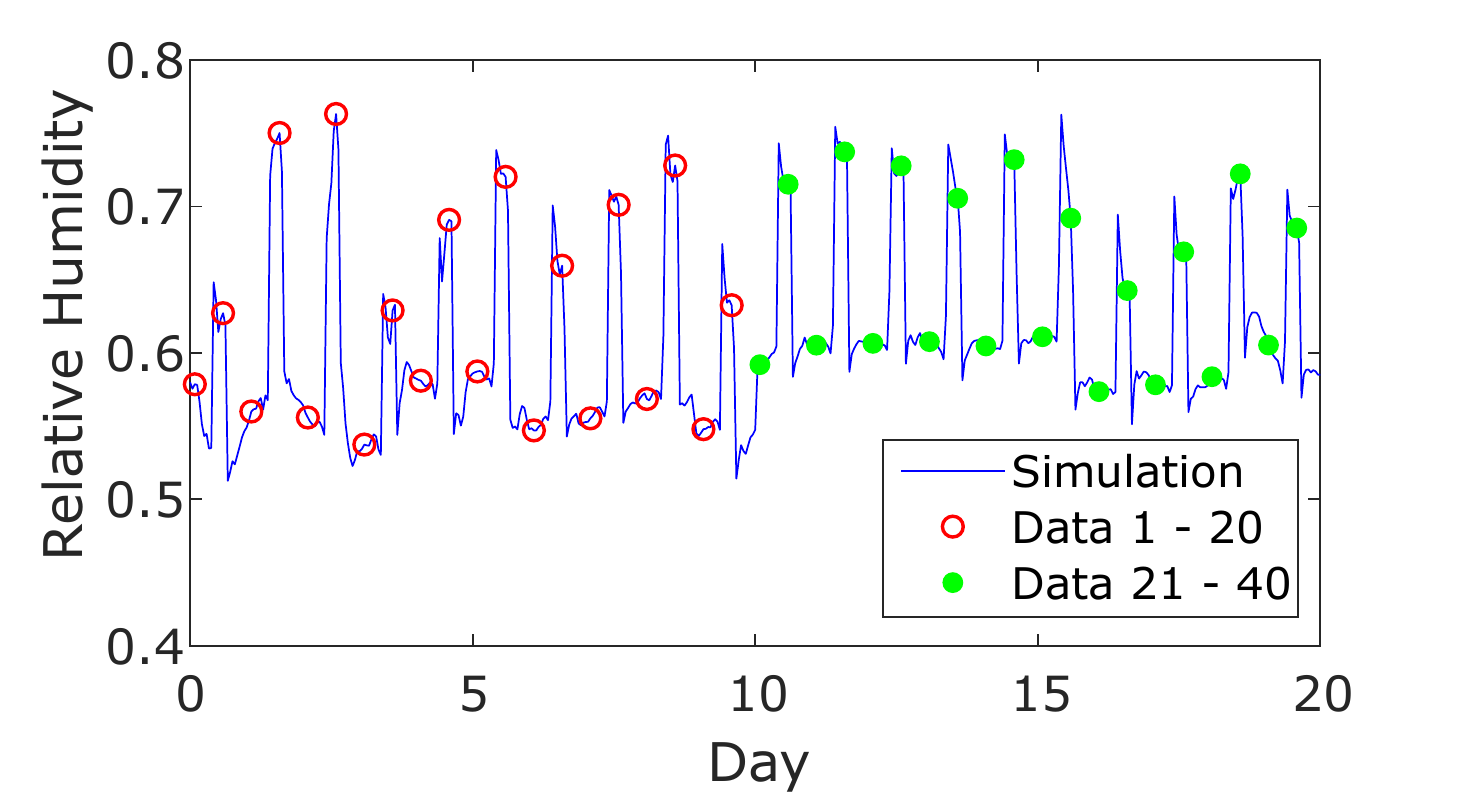}}
{\caption{Synthetic data: the solid line is the output of the simulation model with fixed parameters $N=4 ACH$, $IAS=0.3 m/s$. The data points used for the calibration are indicated as red circles (time period 1) and green dots (time period 2)}
\label{fig:ToyData_2}}
\end{figure}

As a test of the approach we start by fixing values for $N$ and $IAS$ and run the simulation for 20 days to generate synthetic test data. These data are illustrated in Figure \ref{fig:ToyData_2}. The simulation gives an hourly output, plotted in blue, but for calibration of the model we select values every 12 hours (shown as circles) so as to retain a manageable amount of data for the KOH approach. As a result, we have two data points per day: one when the LED lights are switched on and the second when the LED lights are switched off. The internal air speed is fixed at a constant value of $0.3 m/s$.  The ventilation rate as model input is initially fixed, then changed halfway through the simulation period, so that for the first 20 data points the ventilation rate is $4 ACH$, dropping to $2 ACH$ for the remaining period. The values chosen are for illustration but we believe they are typical of the magnitude of the ventilation rate in the tunnel.  The purpose of changing the ventilation rate halfway through the simulation is to investigate whether the sequential KOH and the PF approach are capable of inferring the change in parameter value.

\begin{figure}[h]
     \centering
     \begin{subfigure}[b]{0.48\textwidth}
        \centering
         \includegraphics[width=\textwidth]{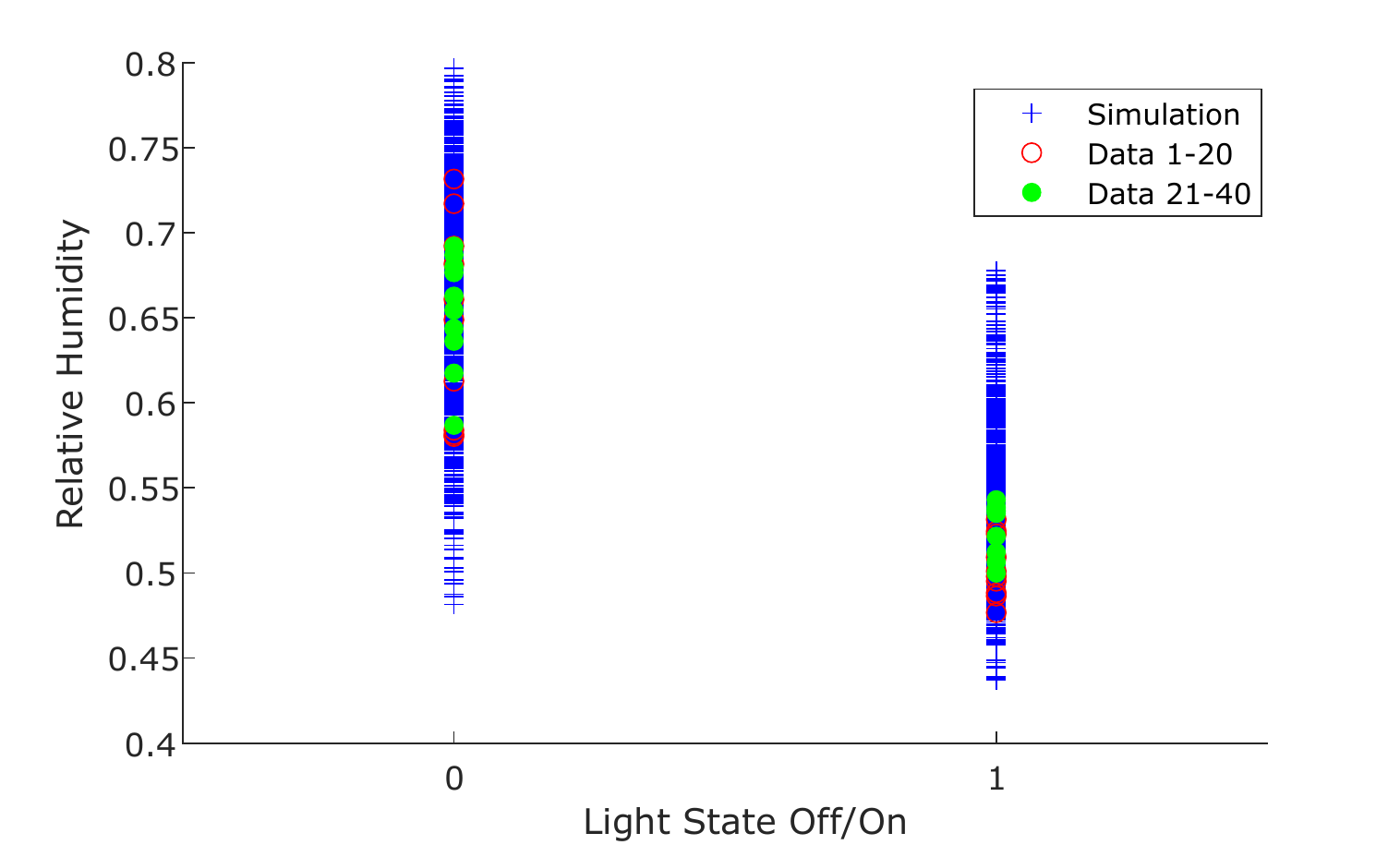}
         \caption{Light State Off=0, On=1}
         \label{fig:ToyLightState}
     \end{subfigure}
     \hfill
     \begin{subfigure}[b]{0.48\textwidth}
         \centering
         \includegraphics[width=\textwidth]{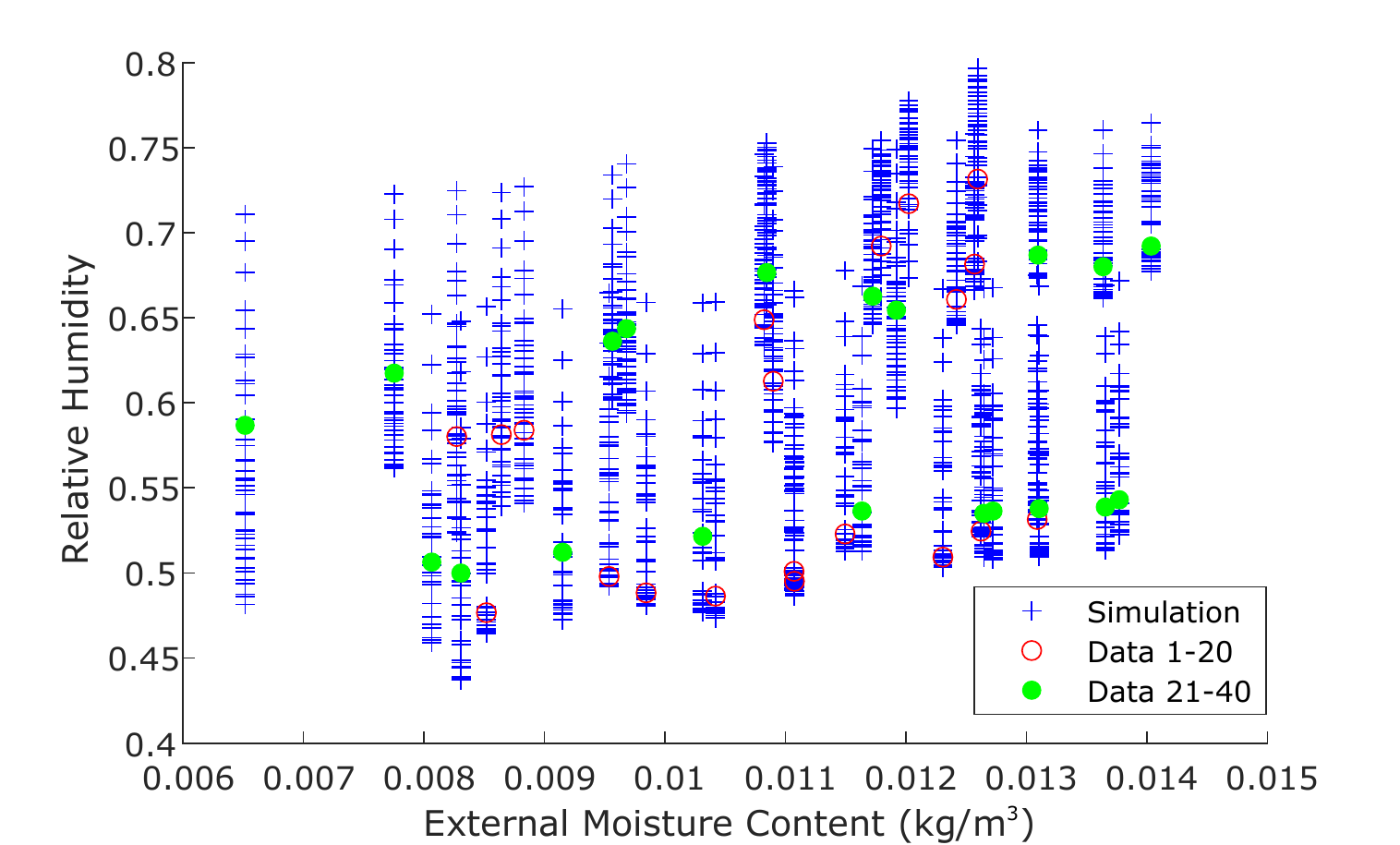}
         \caption{External Moisture Content, $C_w (kg/m^3)$}
         \label{fig:ToyExtCw}
     \end{subfigure}
        \caption{Toy Problem: Synthetic data and model relative humidity outputs plotted as a function of a) Light state and b) External moisture content, i.e. scenarios $x$ for the KOH approach}
        \label{fig:ToyDataKOH}
\end{figure}

As aforementioned, the synthetic test data have been generated from the simulation with fixed parameters. We also require outputs from the simulation over a range of parameter values for input into the KOH and PF calibration frameworks.  For this we run the simulation with a range of values of the two parameters of interest: for ventilation rate we use a range from $1$ to $10 ACH$ in steps of $1.5 ACH$, and for internal air speed we use a range from $0.1$ to $0.85 m/s$ in steps of $0.15 m/s$: taking all combinations gives a total of 42 calibration runs. Again, this toy problem is an illustration, but the range of values selected are in line with the values we expect to be realistic in the tunnel situation.

We use the light state - on/off - in conjunction with the external air moisture content as scenarios under which the model is executed.  Within the KOH formulation these are defined as $x$ and we seek posterior estimates of the parameters $\theta$, in this case $N$ and $IAS$.  The rationale for selecting light state as one of the $x$ values is that the lights are the main providers of heat in the system and as we expect to observe a diurnal variation in temperature and hence relative humidity according to whether the lights are on or off, it is essential to know the light state.  The external moisture content of the air makes a less significant contribution, but it does directly impact on the relative humidity depending on the ventilation rate. 

The test data are plotted as a function of these two parameters in Figure \ref{fig:ToyDataKOH}, together with the output of the simulations (shown in blue), showing that the synthetic test data fall within the range of outputs from the simulation runs, as expected.

\begin{figure}[t]
     \centering
     \begin{subfigure}[b]{0.48\textwidth}
        \centering
         \includegraphics[width=\textwidth]{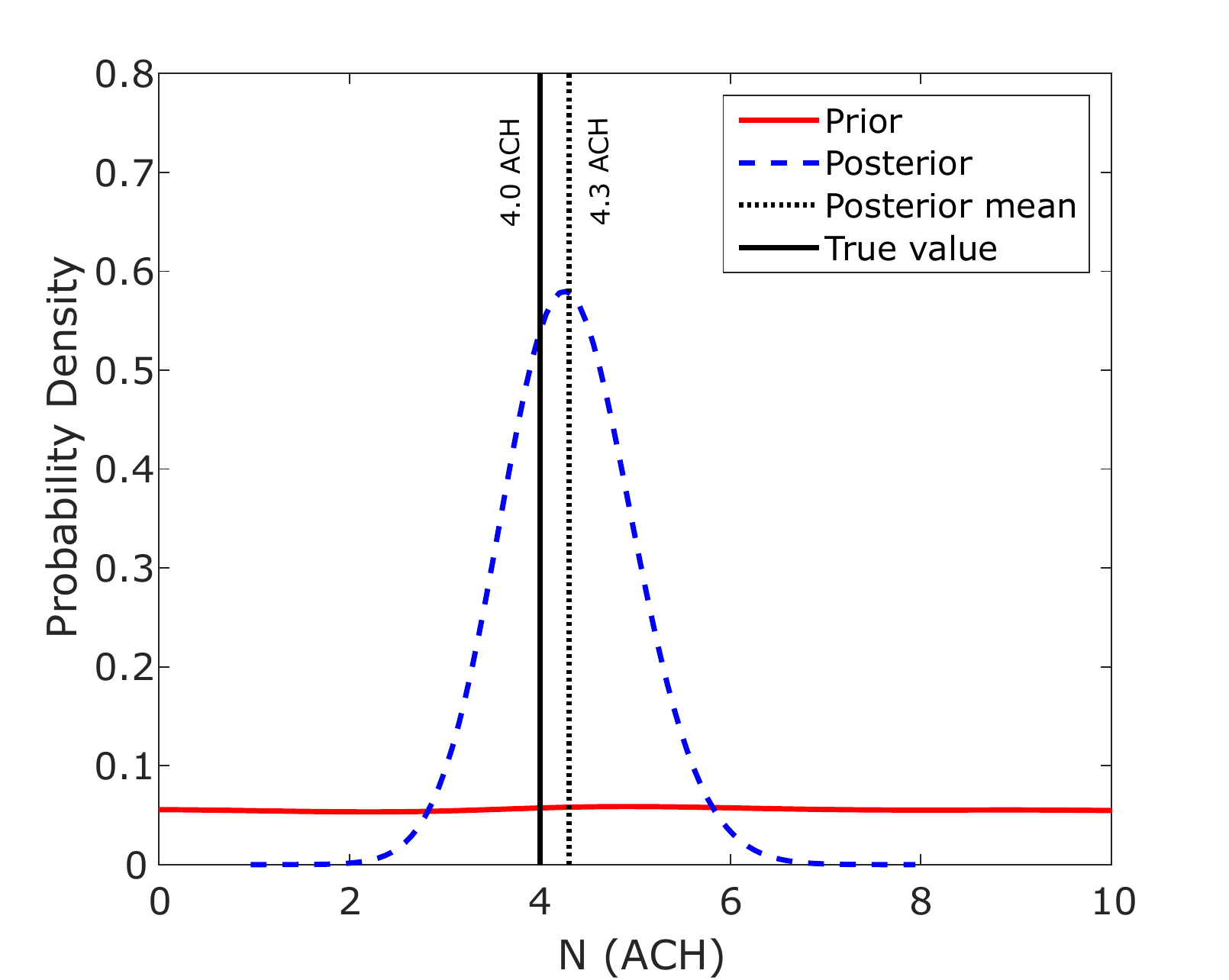}
         \caption{Ventilation rate, N $(ACH)$}
         \label{fig:ToyKOHRH1_N}
     \end{subfigure}
     \hfill
     \begin{subfigure}[b]{0.48\textwidth}
         \centering
         \includegraphics[width=\textwidth]{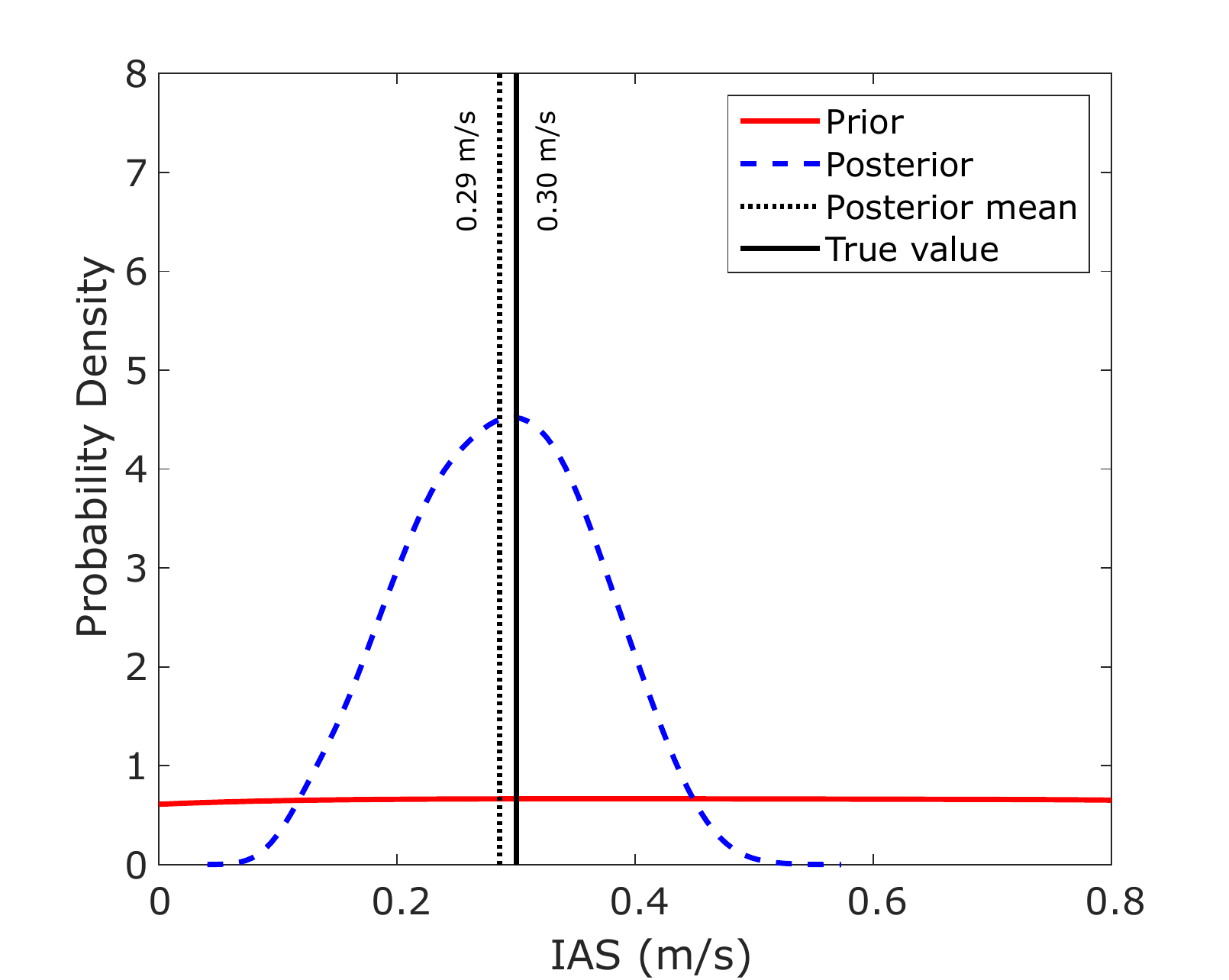}
         \caption{Internal air speed, IAS $(m/s)$}
         \label{fig:ToyKOHRH1_IAS}
     \end{subfigure}
        \caption{Toy problem: prior and posterior parameter probability distributions for KOH approach, period 1 only}
        \label{fig:ToyKOHRH1}
\end{figure}

\begin{figure}[]
     \centering
     \begin{subfigure}[b]{0.48\textwidth}
        \centering
         \includegraphics[width=\textwidth]{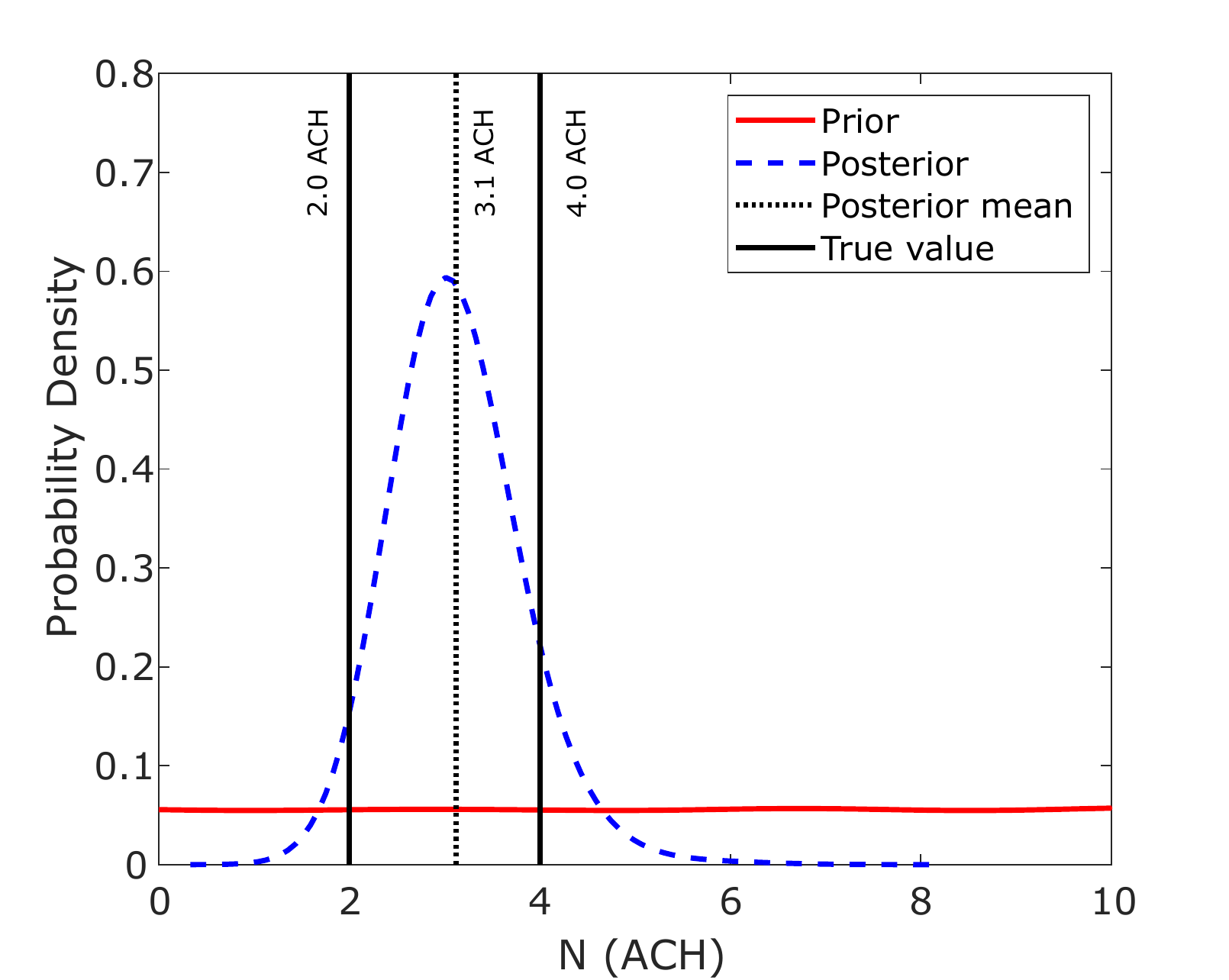}
         \caption{Ventilation rate, N $(ACH)$}
         \label{fig:ToyKOHRH_N}
     \end{subfigure}
     \hfill
     \begin{subfigure}[b]{0.48\textwidth}
         \centering
         \includegraphics[width=\textwidth]{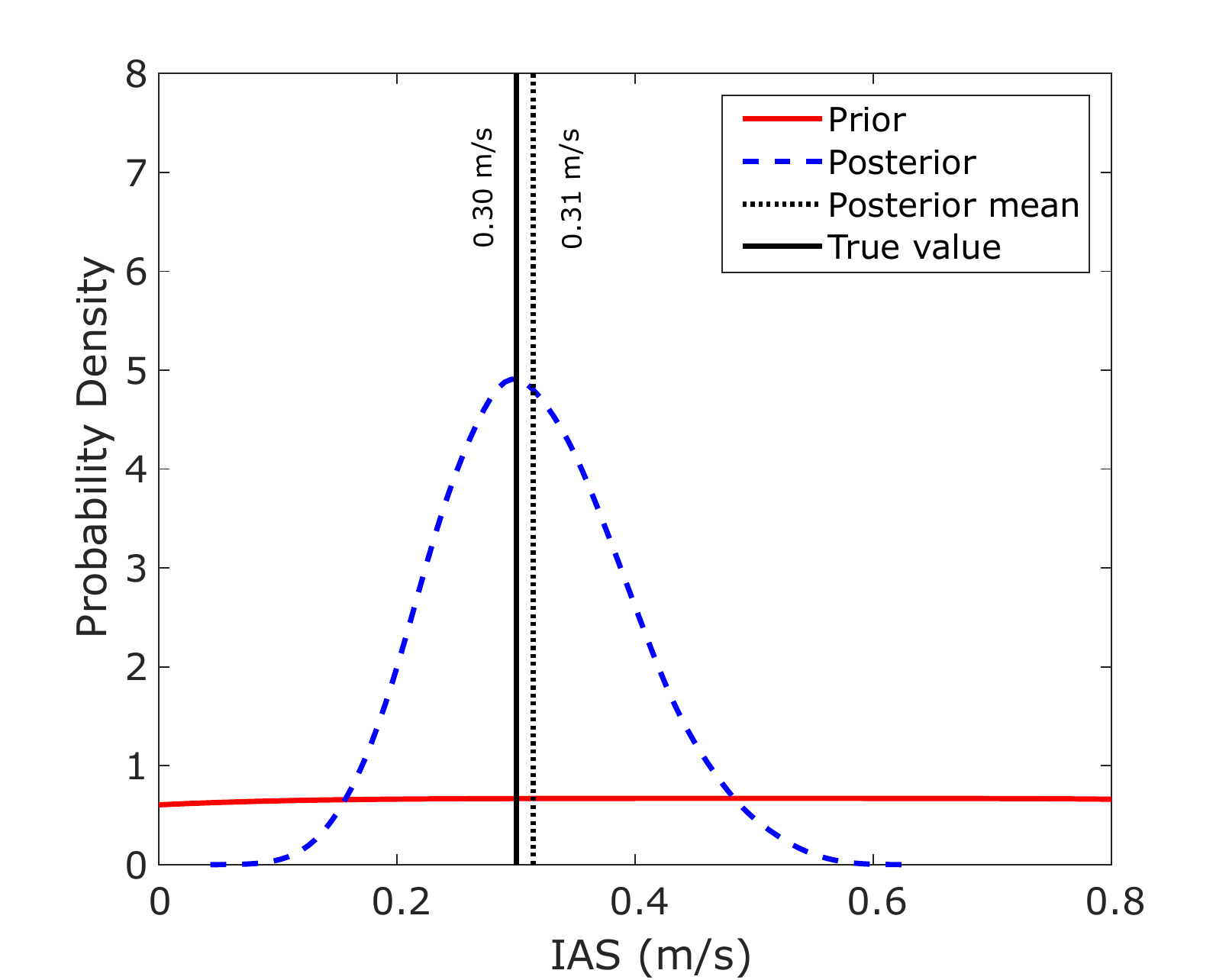}
         \caption{Internal air speed, IAS $(m/s)$}
         \label{fig:ToyKOHRH_IAS}
     \end{subfigure}
        \caption{Toy problem: prior and posterior parameter probability distributions for KOH approach, periods 1-2}
        \label{fig:ToyKOHRH}
\end{figure}

The KOH approach uses a MCMC technique for exploration of the parameter space: for all of the simulations performed here we have used 3 chains of 5000 iterations as this number has typically given good levels of convergence.
Considering the first 20 data points which form the first half of the test data (shown as red open circles on Figure \ref{fig:ToyData_2}), using the KOH approach with a uniform prior distribution for each of the two calibration parameters gives the posterior distributions for the calibration parameters shown in Figure \ref{fig:ToyKOHRH1}.  These plots show the uniform prior distributions as a solid red line, with the posterior distributions shown as dashed blue lines.  Also indicated on the figure are the true values, $N = 4 ACH$ and $IAS = 0.3 m/s$. 
As expected, the KOH calibration results in estimates for the calibration parameters which are centred about a mean value close to the true value.  Extending the KOH approach to all 40 data points gives the posterior distributions illustrated in Figure \ref{fig:ToyKOHRH}.  This shows nicely the issue with the KOH approach for time-varying parameters.  The simulation results in a value for $N$ midway between the two true values - $N = 4 ACH$ for the first half of the simulation and $N = 2 ACH$ for the second half - whereas the mean value of $IAS$ is close to the true value which remains static throughout the simulation.

\begin{figure}
        \centering
        \begin{subfigure}[b]{0.48\textwidth}
            \centering
            \includegraphics[width=0.7\textwidth]{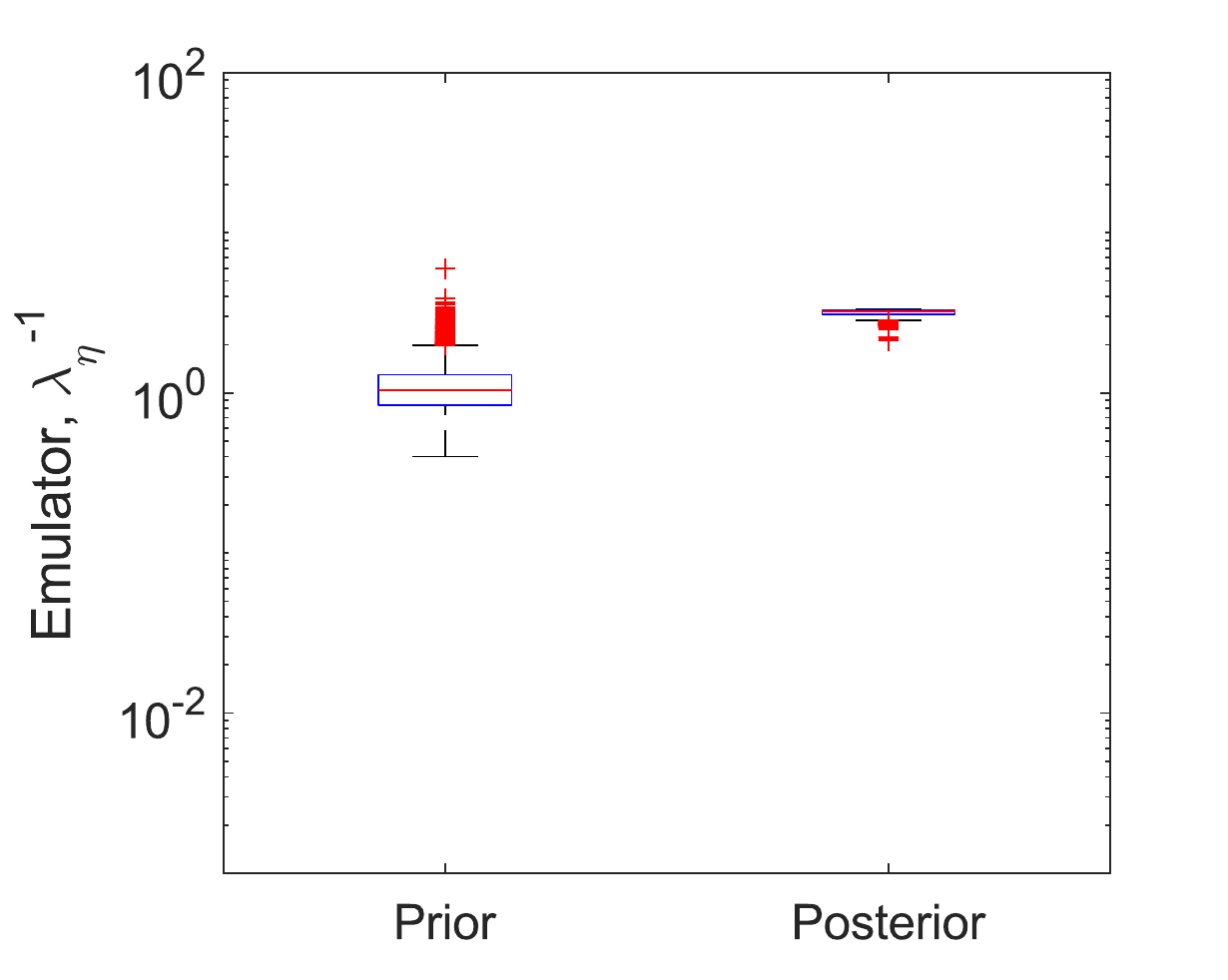}
            \caption{Emulator, $1/\lambda_{\eta}$}%
            \label{fig:HPs_emulator}
        \end{subfigure}
        \hfill
        \begin{subfigure}[b]{0.48\textwidth}  
            \centering 
            \includegraphics[width=0.7\textwidth]{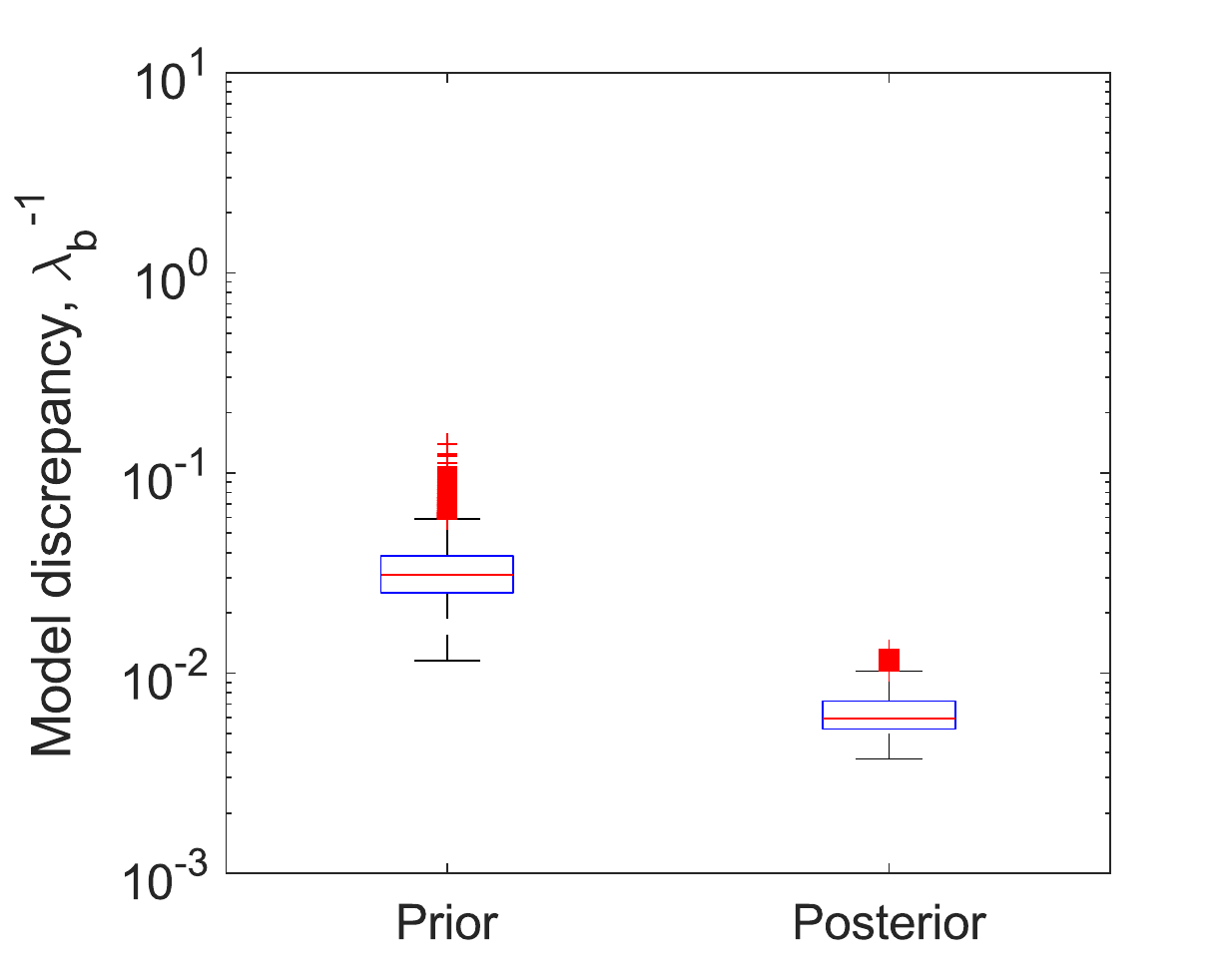}
            \caption{Model discrepancy, $1/\lambda_b$}
            \label{fig:HPs_Model discrepancy}
        \end{subfigure}
        \vskip\baselineskip
        \begin{subfigure}[b]{0.48\textwidth}   
            \centering 
            \includegraphics[width=0.7\textwidth]{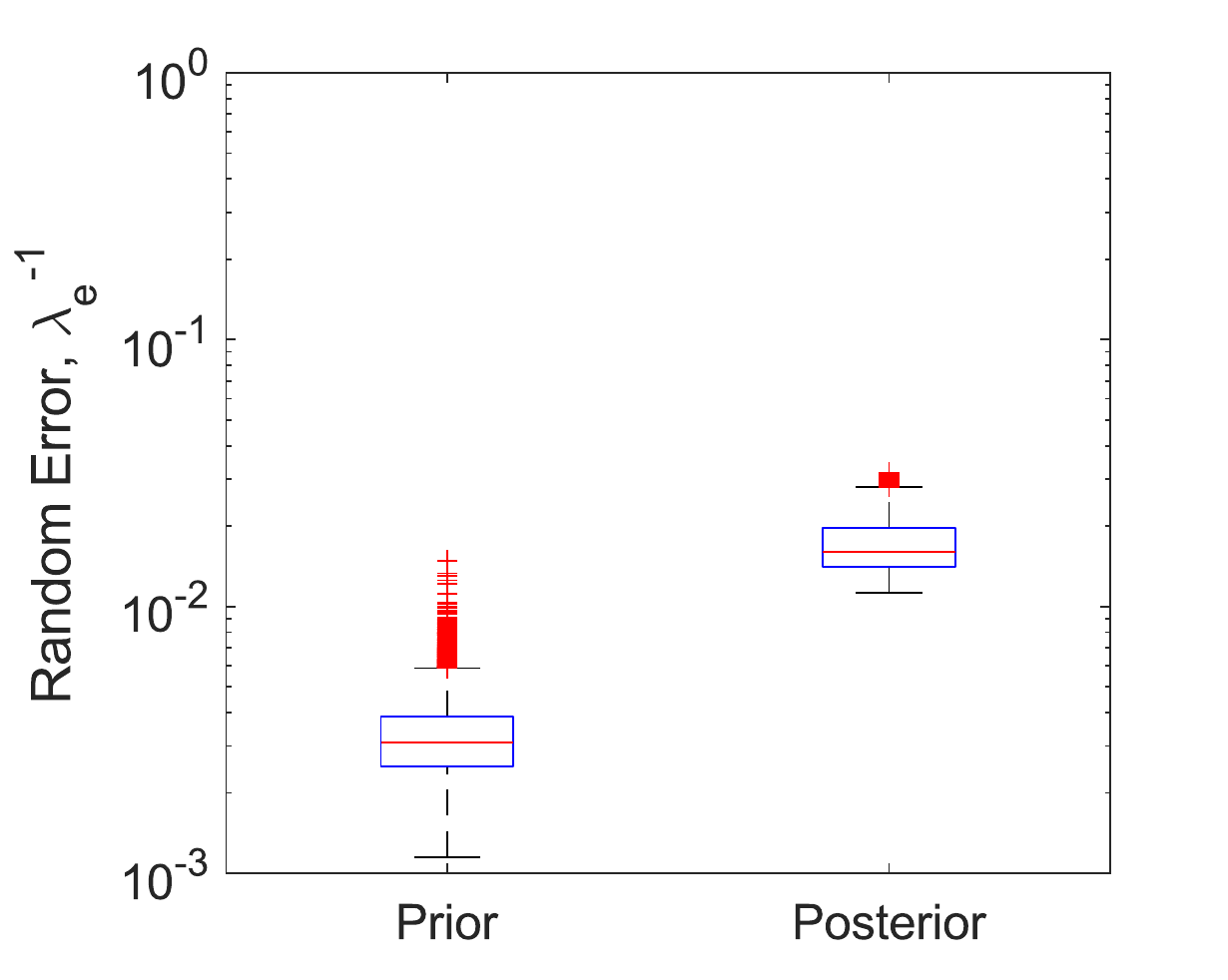}
            \caption{Random Error, $1/\lambda_e$}
            \label{fig:HPs_Random error}
        \end{subfigure}
        \hfill
        \begin{subfigure}[b]{0.48\textwidth}   
            \centering 
            \includegraphics[width=0.7\textwidth]{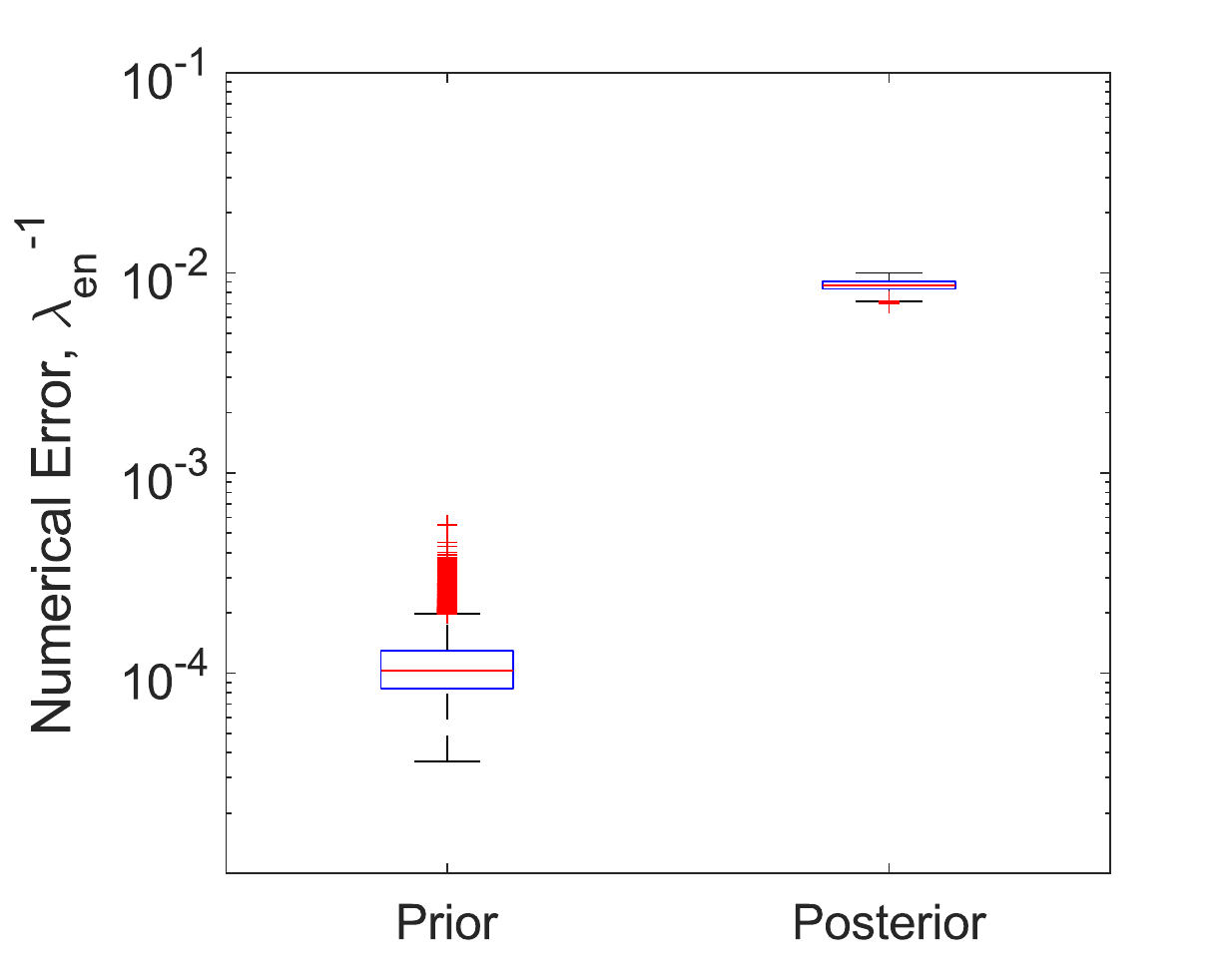}
            \caption{Numerical Error, $1/\lambda_{en}$}
            \label{fig:HPs_Numerical error}
        \end{subfigure}
        \caption{Toy problem: precision hyperparameters for KOH approach, period 1} 
        \label{fig:ToyKOHPrecisionHyperparameters}
    \end{figure}

One of the benefits of the KOH approach is that it gives posterior distributions not only for the calibration parameters but also for the model hyperparameters and hence the error terms (see description following Equation \ref{eqn:KOH}). Figure \ref{fig:ToyKOHPrecisionHyperparameters} shows a comparison of the prior and posterior distributions for the four precision hyperparameters for simulation of the first 20 data points. While the posterior distribution of the emulator precision hyperparameter is very similar to the prior, the model discrepancy or bias parameter has a median value lower than our prior assumption, whereas the random error and numerical error are higher.  This suggests that the calibration is not identifying a systematic error from the information provided, but instead is absorbing all differences between the model and the data into the random and numerical errors.  

The particle filter methodology has also been run with the synthetic test data.  In this approach 1000 particles are initiated comprising values for the two calibration parameters, $\theta$ and the lengthscale, $l$.  The $\theta$s are sampled from an initial uniform prior distribution over the possible range of values, whereas the values for $l$ are sampled from a lognormal distribution to ensure they are positive.  At the first timestep the particle filter compares the value of relative humidity that would be predicted by the simulation using the values from each particle against the true data. Here we do not run the simulation explicitly for each particle at each timestep but instead emulate the simulation output using a Gaussian process fitted to the simulation results over the range of calibration parameters run prior to the particle filtering.  The likelihood of each particle is then calculated based on the comparison, and weights are assigned to each particle according to the likelihood.  Particles are re-sampled from the prior distribution taking the particle weights into account and this new distribution forms the prior for the next step. The key difference between this approach and the KOH approach is that here each data point is considered sequentially whereas in the static KOH approach all data points are considered together.





\begin{figure}[t]
     \centering
     \begin{subfigure}[b]{0.3\textwidth}
        \centering
         \includegraphics[width=\textwidth]{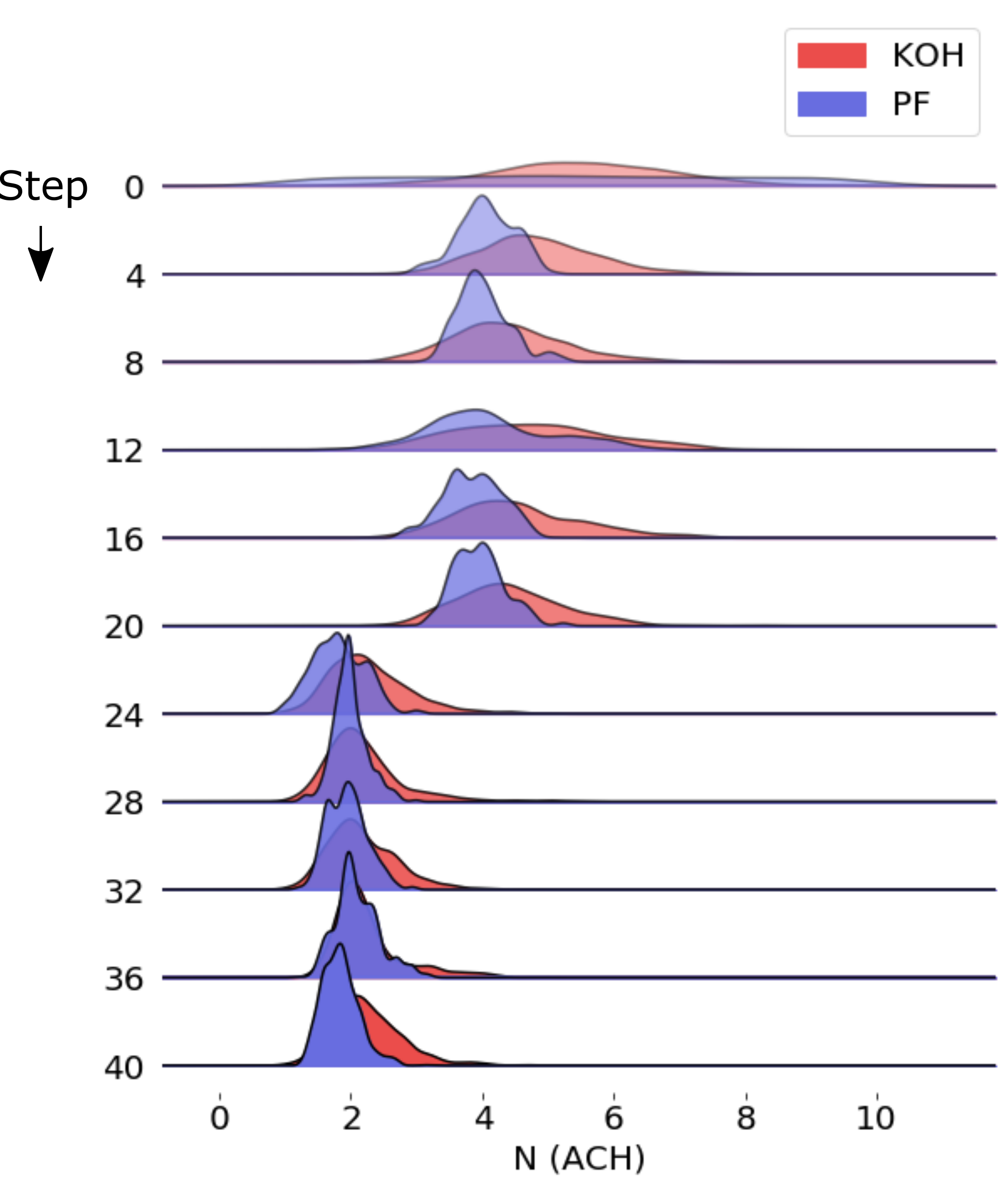}
         \caption{ Ventilation rate, N $(ACH)$}
         \label{fig:ToyPFtheta1}
     \end{subfigure}
     \hfill
     \begin{subfigure}[b]{0.3\textwidth}
         \centering
         \includegraphics[width=0.9\textwidth]{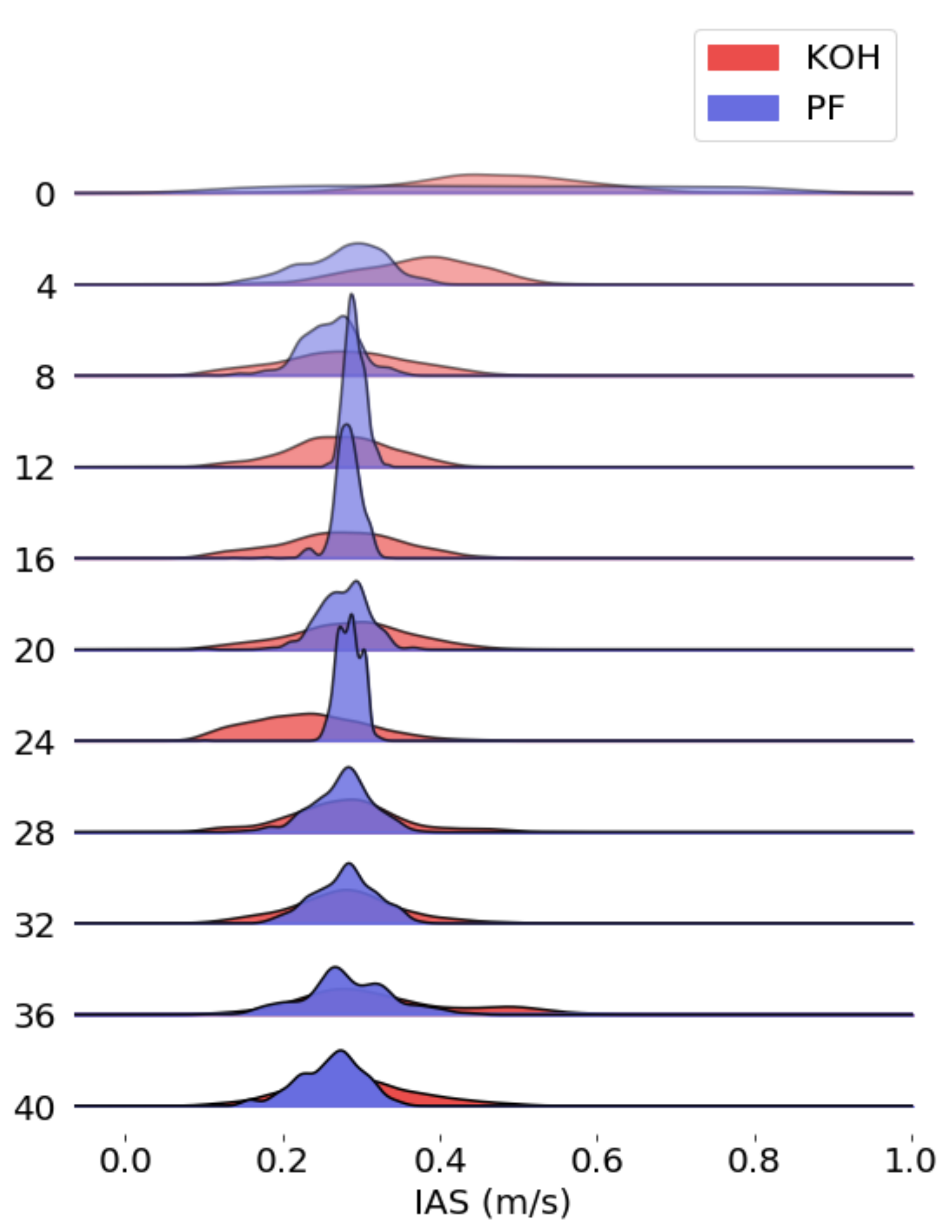}
         \caption{Internal air speed, IAS $(m/s)$}
         \label{fig:ToyPFtheta2}
     \end{subfigure}
     \hfill
     \begin{subfigure}[b]{0.3\textwidth}
         \centering
         \includegraphics[width=0.92\textwidth]{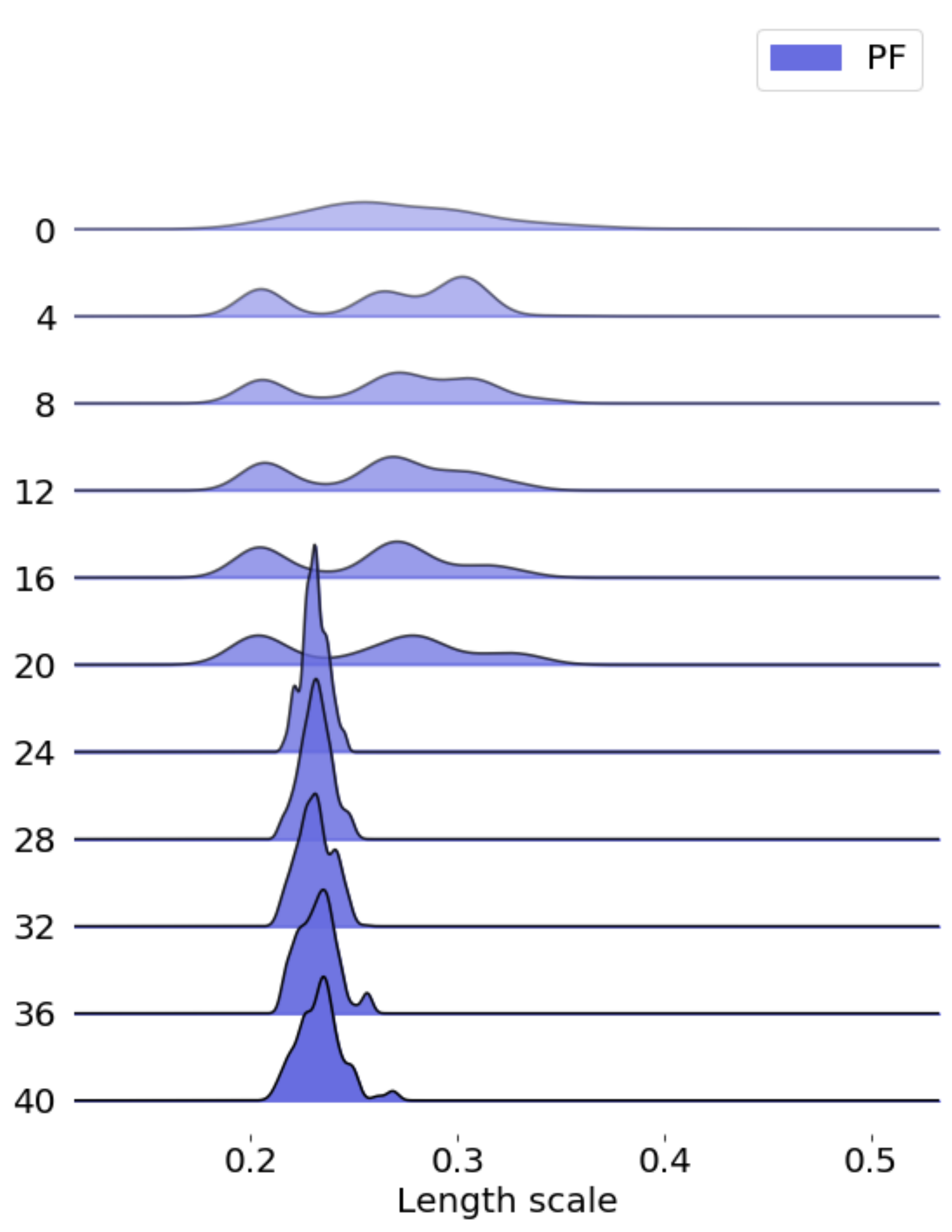}
         \caption{Length scale}
         \label{fig:ToyPFlength}
     \end{subfigure}
        \caption{Particle Filter approach: evolution of posteriors, showing a) $N (ACH)$, b) $IAS (m/s)$ and c) Length scale }
        \label{fig:ToyPFThetaEvolution}
\end{figure}

For this reason, in addition to the static Bayesian calibration, we have run the KOH approach sequentially over the whole time period. As we have already seen in Figure \ref{fig:ToyKOHRH}, using all the data gives posterior parameter estimates in line with the mean values over the whole time period but we wish to know whether we can use fewer data points and run the model sequentially and identify a sudden change in parameter value.  The sequential approach has been used with 2, 4 and 8 data points, as the methodology requires at least 2 data points in order to fit the emulator. In this approach, at each timestep the oldest data point is discarded and a new data point added - this means that for 2 data points there are 39 runs of the model, 4 data points there are 37 runs and 8 data points 33 runs.  In addition the mean of the posterior distribution from the previous timestep is used as the mean of the prior for the new timestep. Note we have not used the full posterior from the previous timestep as prior for the next as the standard deviation of the posterior becomes too narrow, making the prior too specific - instead we have maintained the standard deviation of the first prior. 

The evolution of the posterior probability distribution for each of the calibration parameters is illustrated in Figure \ref{fig:ToyPFThetaEvolution} for the particle filter model and the sequential KOH approach using 4 data points.  Each plot shows the evolution in the parameter value distribution, starting from the prior distribution at the top, progressing in steps of 4 data points to the end of the simulation at the bottom.  Note that all 40 data points were used in the simulation, but the output for the intermediate points is suppressed for brevity (as might be inferred from the figures, the suppressed values are very similar to those shown on either side).  The results are shown in blue for the PF model and in red for the sequential KOH approach.   Here, the change in ventilation rate at the mid-point of the simulation is clearly visible (Figure \ref{fig:ToyPFThetaEvolution}(a)), and although there is a small perturbation to the values of the internal air speed (Figure \ref{fig:ToyPFThetaEvolution}(b)), the distribution recovers quickly and the mean stays close to the true value throughout the simulation.  The similarity between the mean posterior values for $N$ and $IAS$ and the known true values ensures that the posterior predictions calculated using the mean parameter values are very close to the test data.  For this test case the root mean squared error (RMSE) values for the posterior predictions from the PF and sequential KOH models are $1.1\%$ and $0.9\%$ respectively.   

Figure \ref{fig:ToyPFThetaEvolution}(c) shows the evolution of the particle filter length scale parameter for this test case.  The distribution of values shows a wider spread at the start of the analysis, but this narrows as the analysis progresses and the mean value remains close to a value of 0.2 throughout. This is a true unknown parameter as it is a hyperparameter of the emulator, and approximately reflects the distance in the input space that corresponds to an impact on the output; the small value derived here reflects the high degree of variability from one data point to the next.

\begin{figure}[]
     \centering
     \begin{subfigure}[b]{0.45\textwidth}
        \centering
         \includegraphics[width=0.8\textwidth]{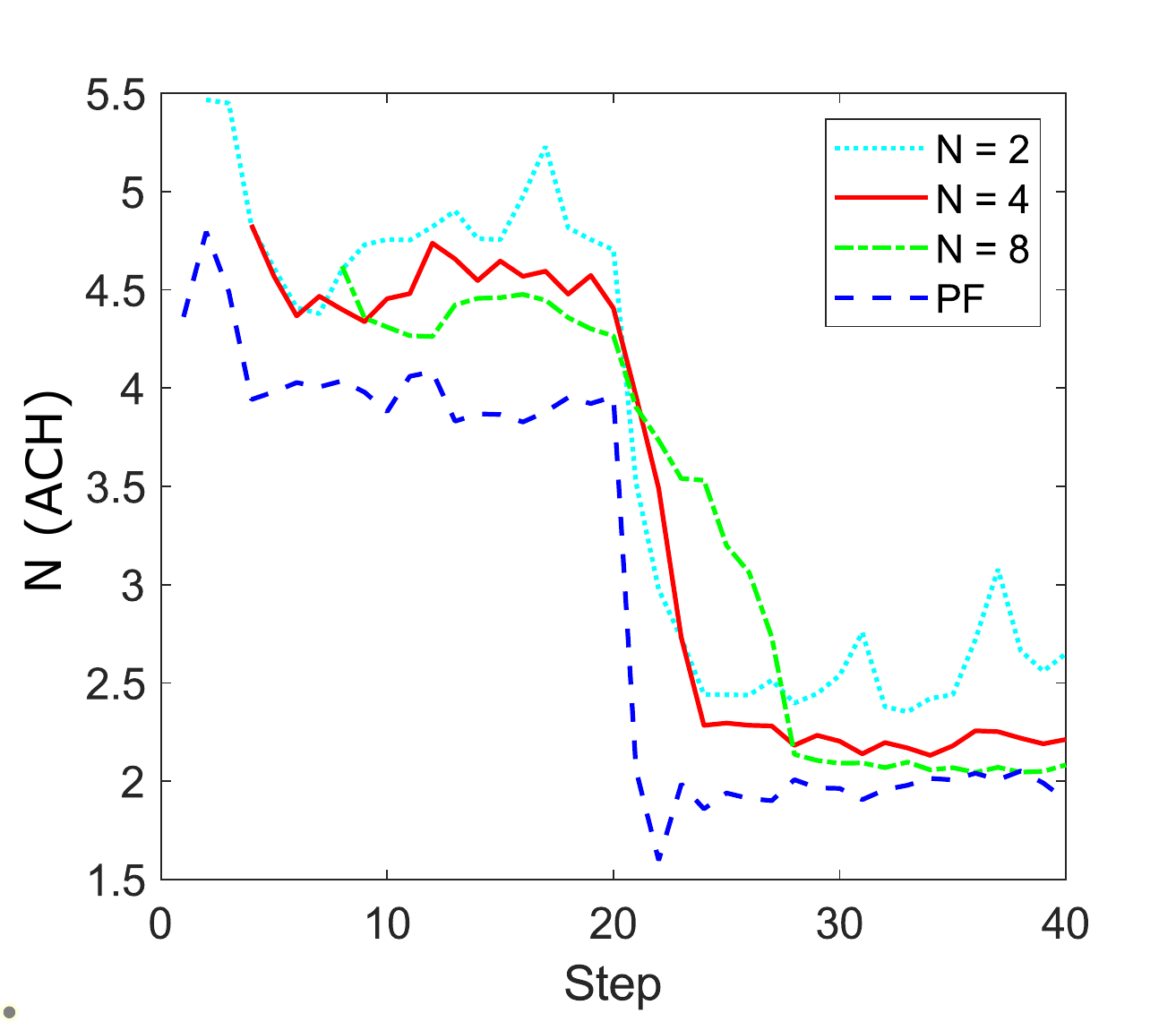}
         \caption{Ventilation rate, N $(ACH)$}
         \label{fig:ACHToySG}
     \end{subfigure}
     \hfill
     \begin{subfigure}[b]{0.45\textwidth}
         \centering
         \includegraphics[width=0.8\textwidth]{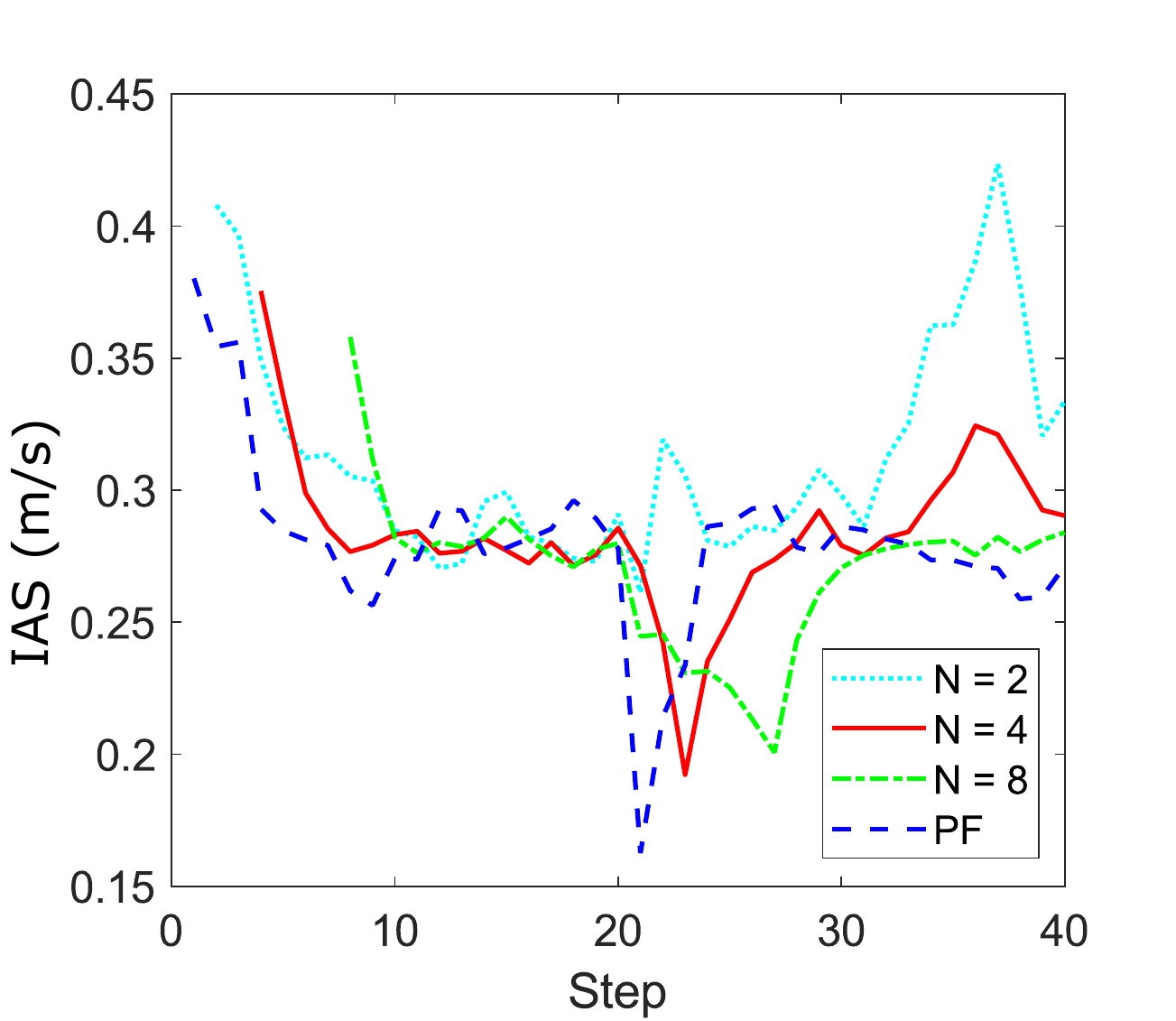}
         \caption{Internal air speed, IAS $(m/s)$}
         \label{fig:IASToySG}
     \end{subfigure}
        \caption{Sequential KOH: evolution of the mean posterior for a) ventilation rate, $N$ and b) internal air speed, $IAS$, showing the impact of increasing the number of data points included in each step from 2 to 8 }
        \label{fig:SequentialGASP}
\end{figure}

\begin{figure}[]%
    \centering
    \includegraphics[width=0.7\textwidth]{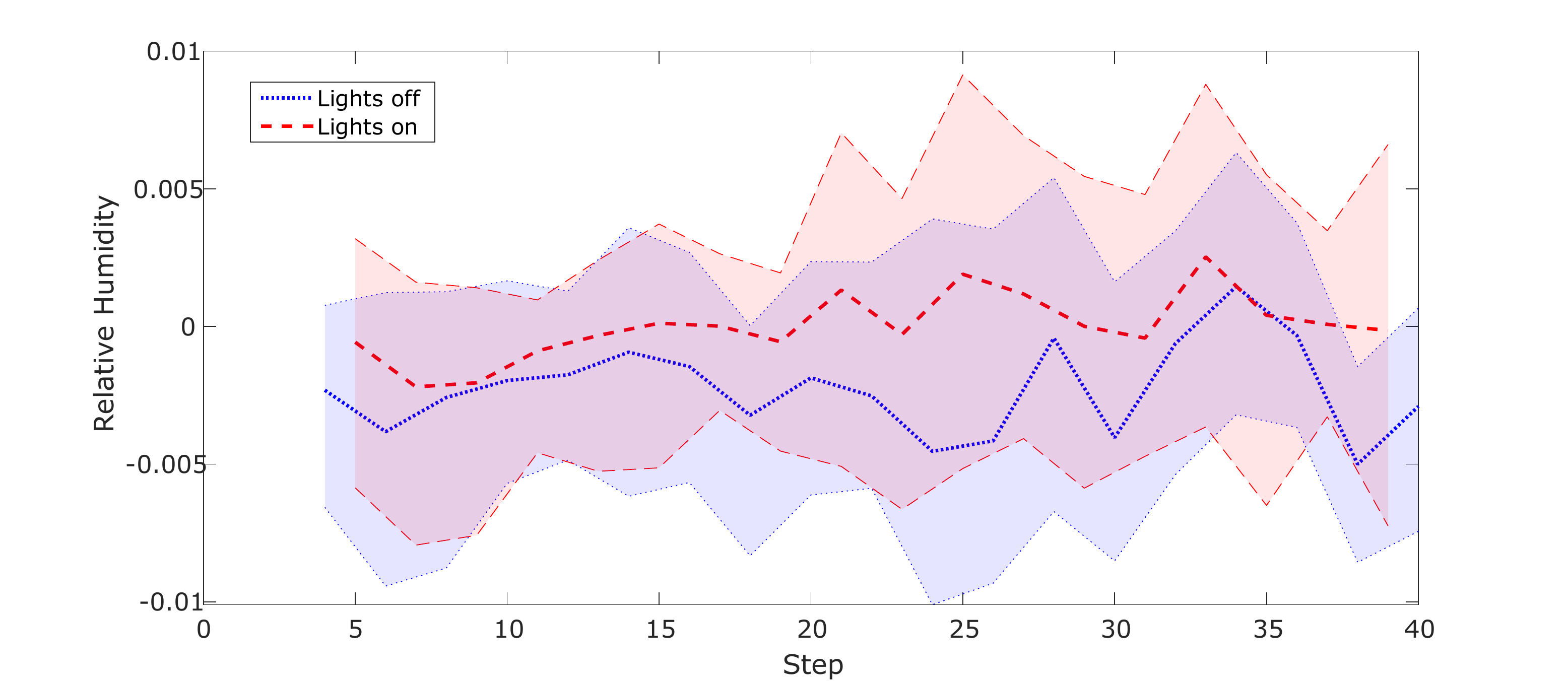}
    \caption{Sequential KOH: mean and 90\% confidence limits of the model bias function}
    \label{fig:KOHSequentialToyBias}
\end{figure}

 In figure \ref{fig:ToyPFThetaEvolution} it is clear that the results for the PF model exhibit a smaller variance than the sequential KOH approach, yet the mean values are in good agreement.  Figure \ref{fig:SequentialGASP} shows in more detail how the results of the sequential KOH approach compare against the particle filter model in terms of the mean posterior values for N and IAS.  Using 2 data points (N=2) the results show some fluctuations from the true value, particularly towards the end of the simulation for the internal air speed (Figure \ref{fig:IASToySG}), but pick up the timing of the change in ventilation rate well (Figure \ref{fig:ACHToySG}).  As the number of data points increases, this timing is harder to identify as the approach assumes a constant value over the data, so when for example 8 data points are used, the timing of the change in $N$ value is smoothed and appears not as a sudden change but as a smoother transition.  Using 4 data points appears to match the transition time well with a sharper transition than 8 data points, yet the mean of the posterior shows less fluctuation than 2 data points, hence this appears to be the best approach. Figure \ref{fig:KOHSequentialToyBias} shows the mean and $90\%$ confidence limits of the model bias function as a function of the step for the sequential KOH approach with $N=4$, separated according to whether the lights are switched on or off.  The model bias is small, less than $0.5\%$, hence will not make a significant difference to the results.  

\begin{figure}[t]
     \centering
     \begin{subfigure}[b]{0.48\textwidth}
        \centering
         \includegraphics[width=0.8\textwidth]{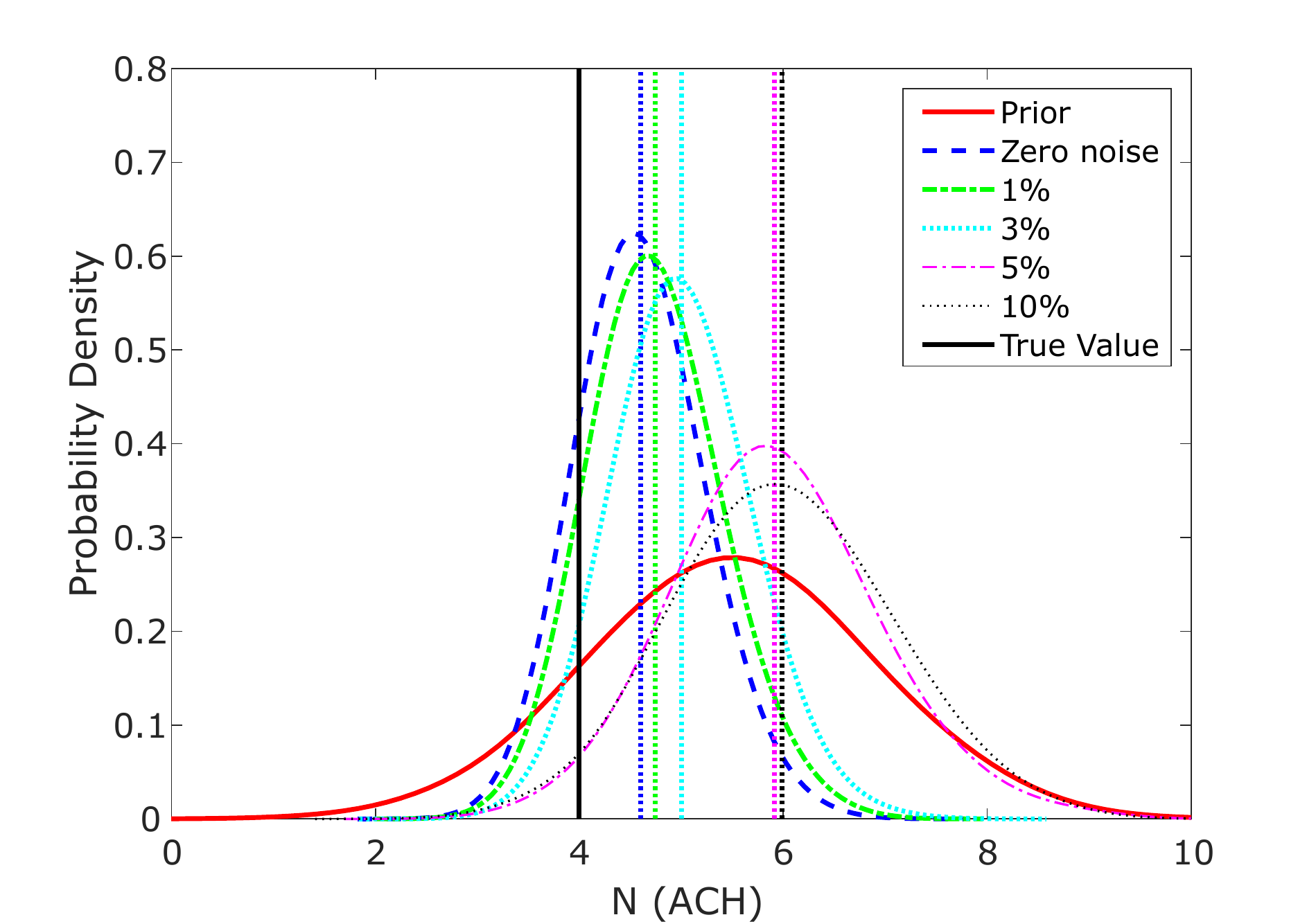}
         \caption{Ventilation rate, N $(ACH)$}
         \label{fig:ToyKOHRH_N_Noise}
     \end{subfigure}
     \hfill
     \begin{subfigure}[b]{0.48\textwidth}
         \centering
         \includegraphics[width=0.8\textwidth]{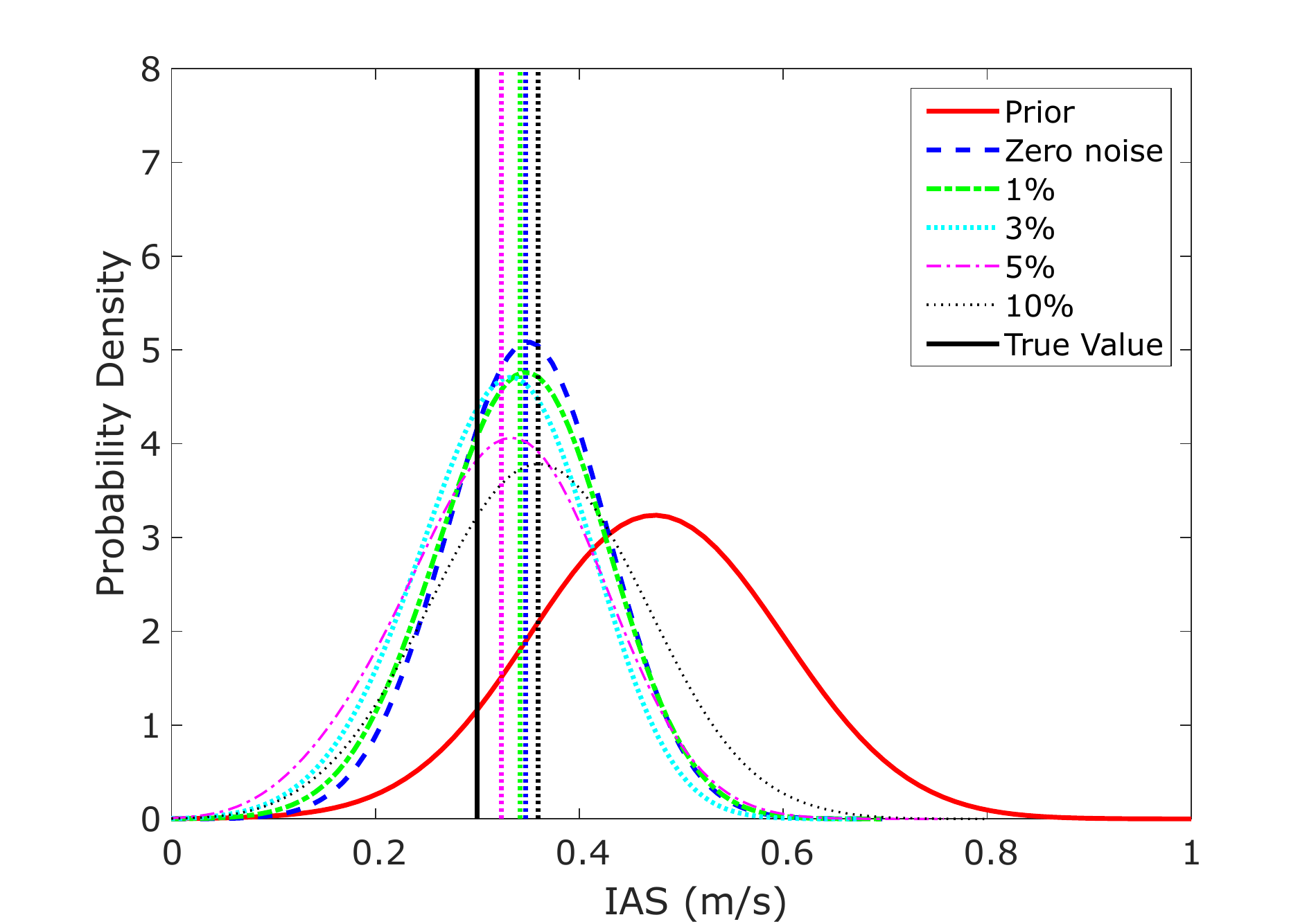}
         \caption{Internal air speed, IAS $(m/s)$}
         \label{fig:ToyKOHRH_IAS_Noise}
     \end{subfigure}
        \caption{Toy problem: results of the KOH calibration with period 1 data showing effect of increased noise in the data}
        \label{fig:ToyKOHRH_Noise}
\end{figure}

The synthetic test data are noise-free i.e. the data are generated directly from the simulation model.  But real data are typically noisy with errors arising from measurement - how do the approaches compare for noisy data?  To test this, the calibration exercise has been re-run with noise added to the data in the form of random values generated from a zero-mean normal distribution, with standard deviation $\sigma$, where for illustrative purposes $\sigma$ has been calculated such that $2\sigma$ is equal to relative humidity values of $0.01$, $0.03$, $0.05$ and $0.10$ i.e. $\pm 1\%$ to $10\%$, where $10\%$ is the difference in relative humidity observed for a change in ventilation rate from $4 ACH$ to $2 ACH$. Figure \ref{fig:ToyKOHRH_Noise} illustrates the comparison for the static KOH approach.  In this example the prior distributions for the two calibration parameters have been chosen as normal distributions as giving better information in the prior distribution should improve the ability of the methodology to identify the posterior.  However, as the noise level increases, the mean of the posterior moves increasingly further from the true value, more significantly for the ventilation rate, $N$ (Figure \ref{fig:ToyKOHRH_N_Noise}). 

\begin{figure}[t]
     \centering
     \begin{subfigure}[b]{0.3\textwidth}
        \centering
         \includegraphics[width=\textwidth]{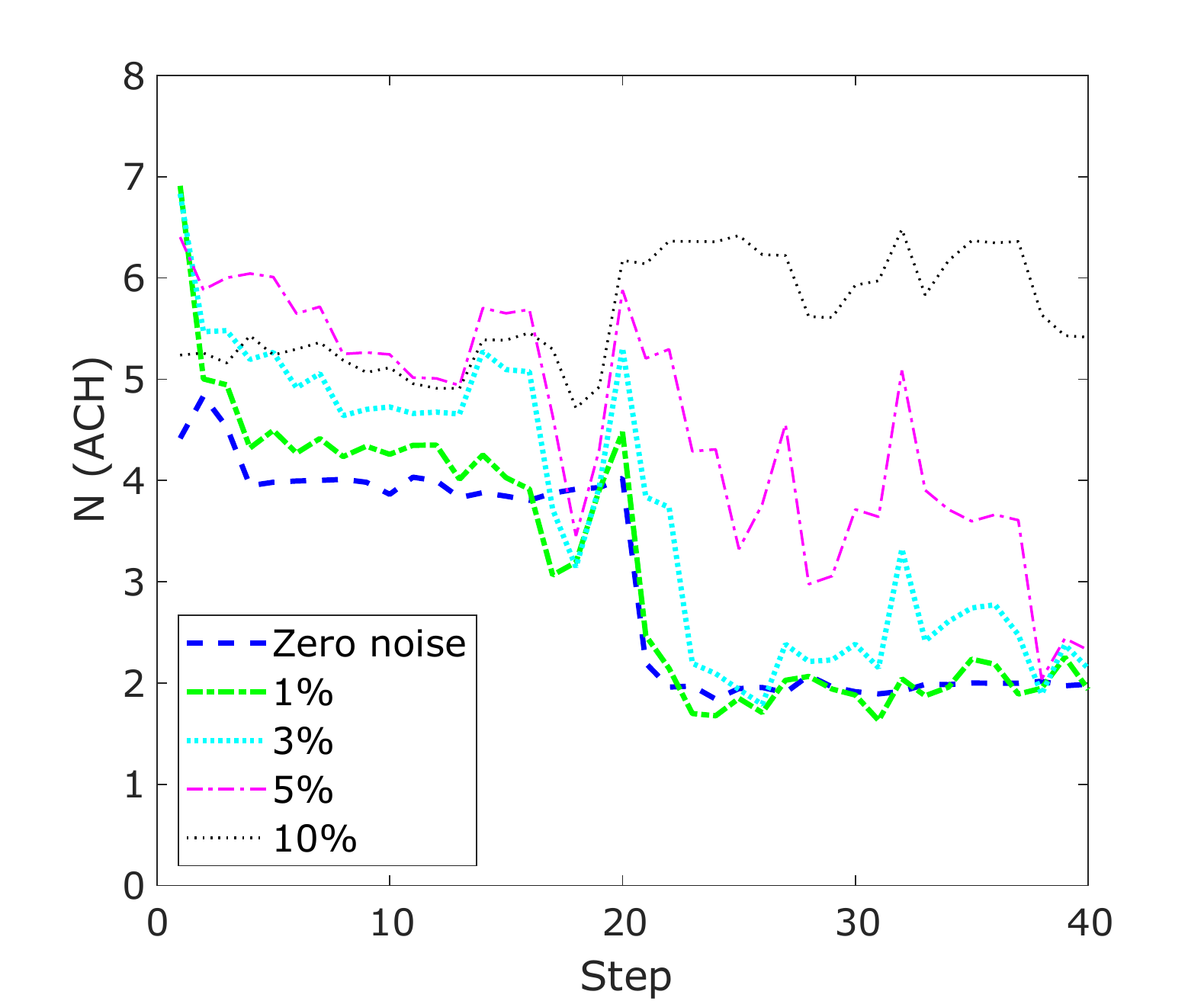}
         \caption{Ventilation rate, N $(ACH)$}
         \label{fig:ACHNoisePF}
     \end{subfigure}
     \hfill
     \begin{subfigure}[b]{0.3\textwidth}
         \centering
         \includegraphics[width=\textwidth]{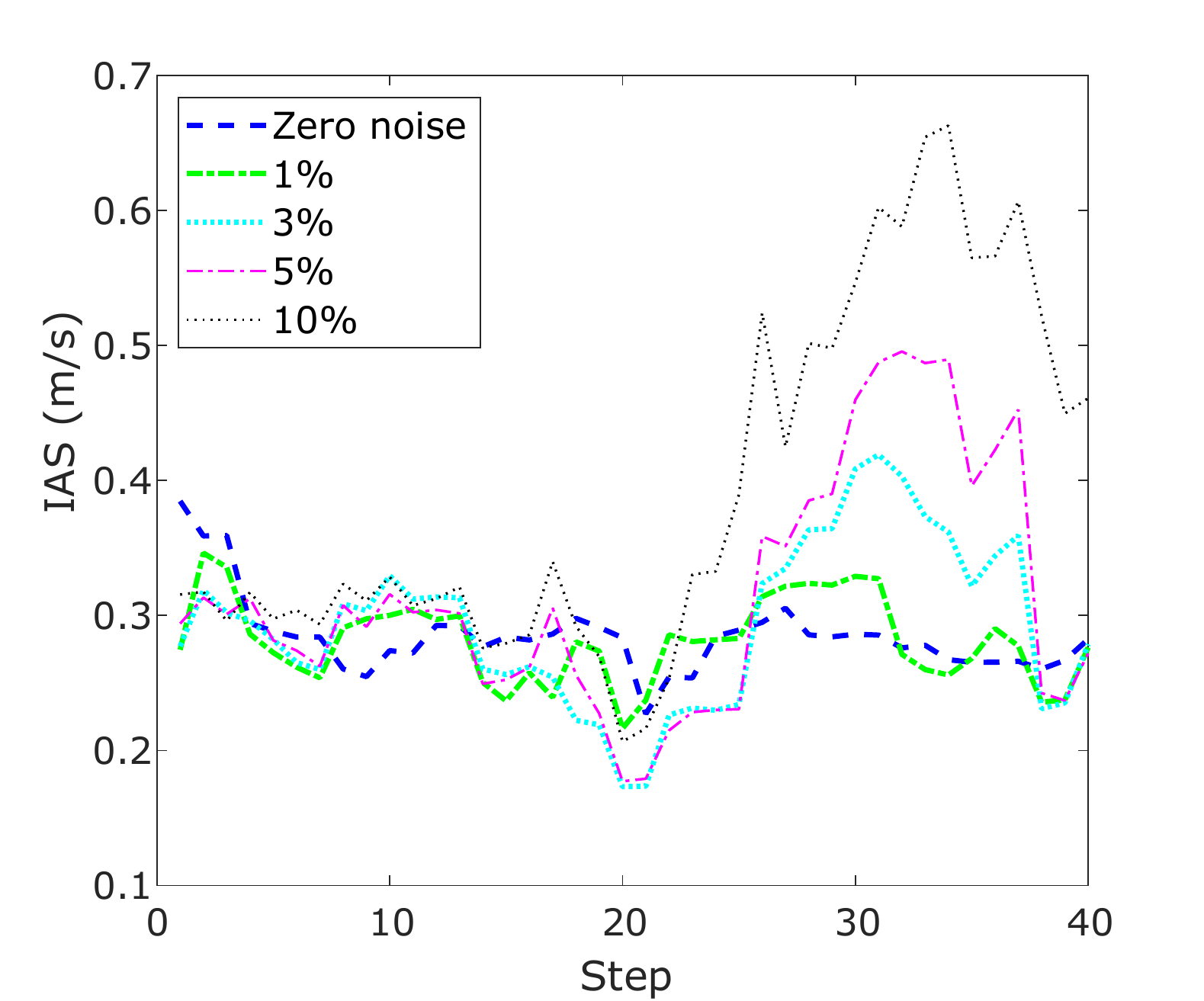}
         \caption{Internal air speed, IAS $(m/s)$}
         \label{fig:IASNoisePF}
     \end{subfigure}
     \hfill
     \begin{subfigure}[b]{0.3\textwidth}
         \centering
         \includegraphics[width=\textwidth]{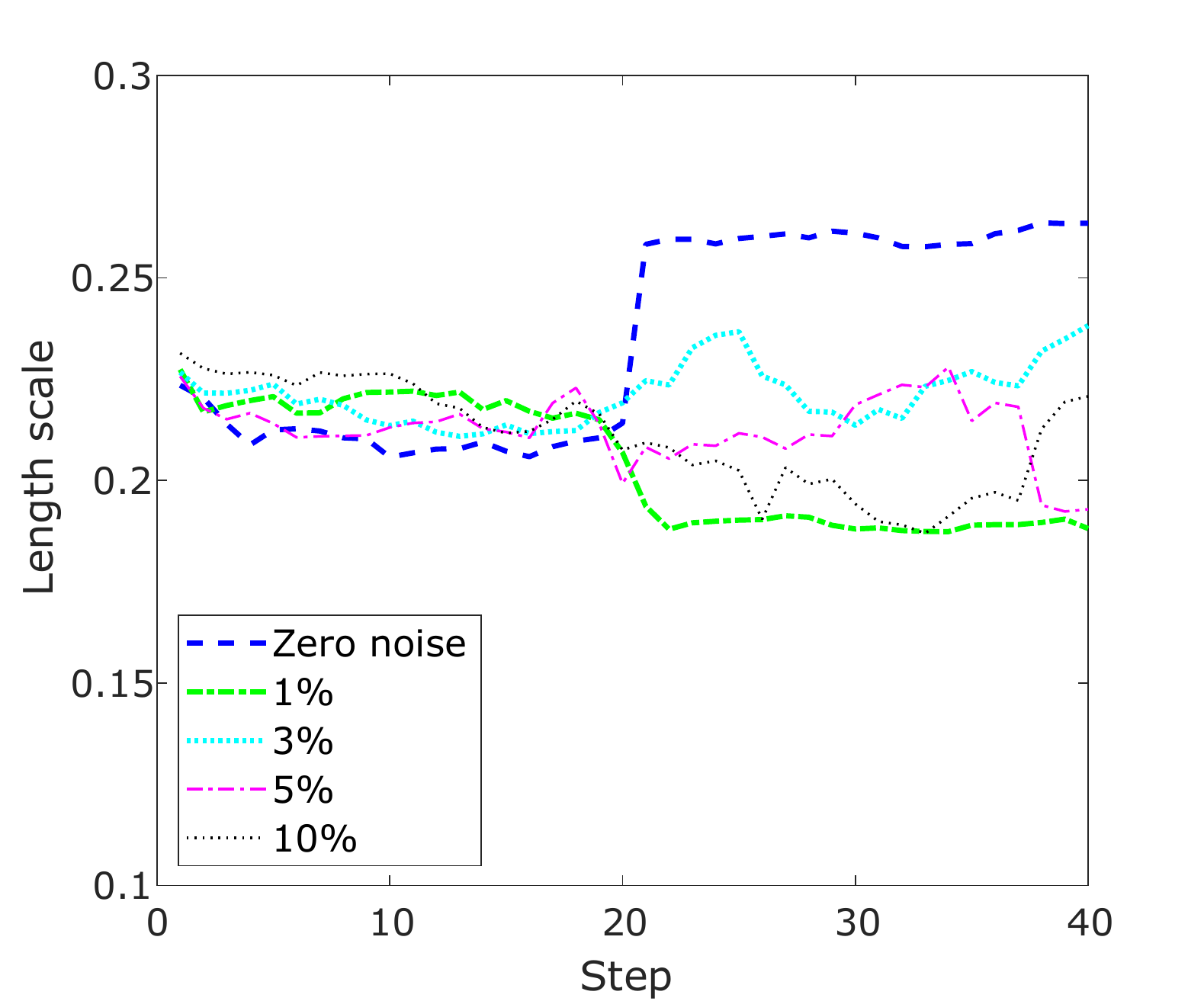}
         \caption{Length scale}
         \label{fig:LengthNoisePF}
     \end{subfigure}
        \caption{Toy problem: Mean of PF evolution demonstrating effect of noise in the data}
        \label{fig:NoisePF}
\end{figure}

To understand the impact of introducing noise in the data on the particle filter approach, consider Figure \ref{fig:NoisePF}, which plots the mean of the particle filter posterior distribution at each time step. Again the initial prior distribution for each $\theta$ was specified as a normal distribution. Looking first at the ventilation rate, Figure \ref{fig:ACHNoisePF}, the zero noise results, shown as a blue dashed line, clearly show the step change from a value of $N = 4 ACH$ to $N = 2 ACH$ at the mid-point of the analysis (step = 20).  The results for 1 and 3\% noise are not too dissimilar (green and cyan), and even at a noise level of 5\% the posterior finishes close to the true value after 40 steps, albeit without such a clear step-change.  At a noise level of 10\% however, the PF approach cannot identify the change in parameter value.  This is because the change in relative humidity due to a change in ventilation rate from 4 to 2 is close to 10\% and so the model cannot distinguish what is noise and what is a genuine parameter change. It is clear that if there is too much noise in the data then both sequential approaches fail to correctly infer the parameter values.  


\begin{figure}[t]
     \centering
     \begin{subfigure}[b]{0.3\textwidth}
        \centering
         \includegraphics[width=\textwidth]{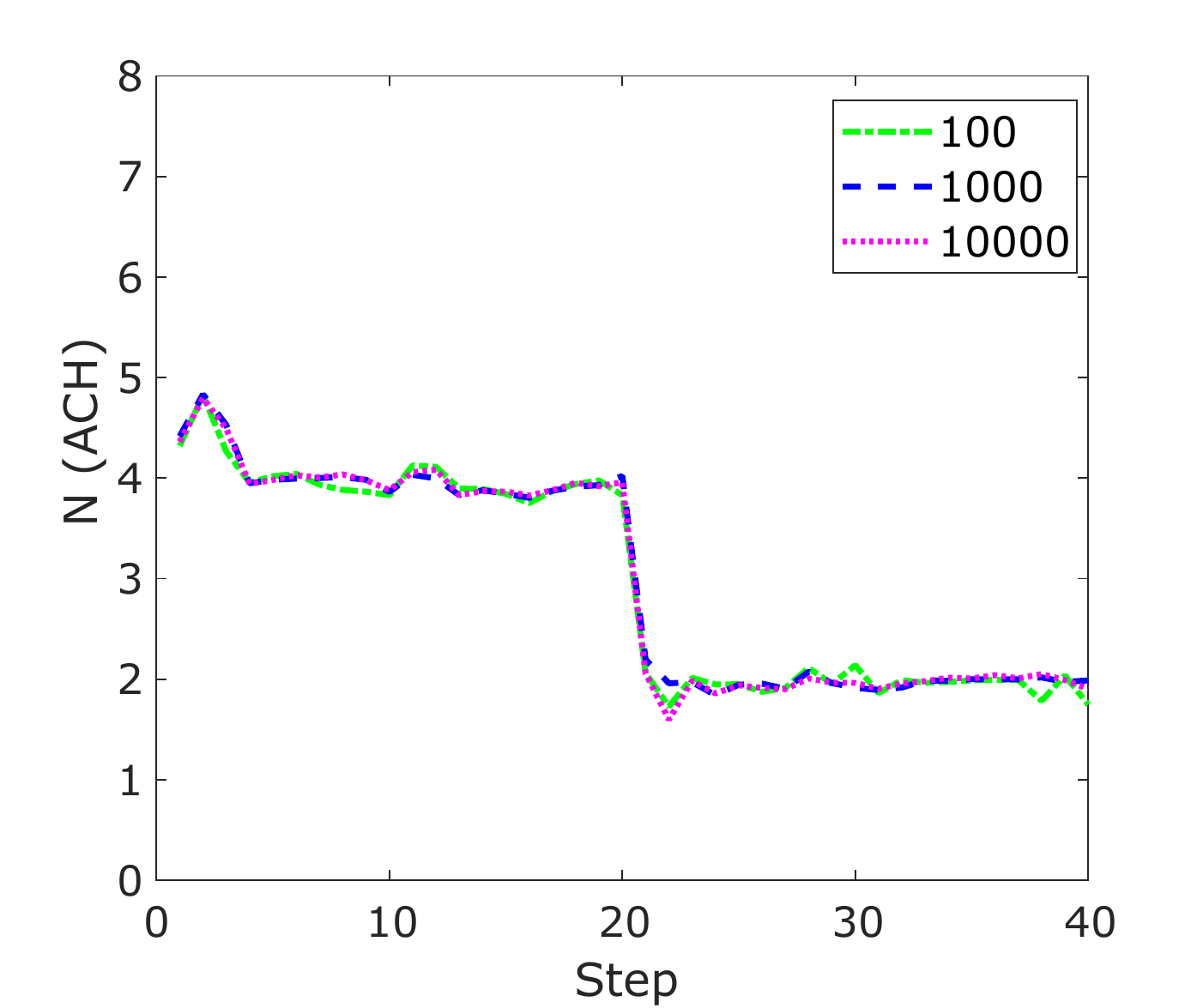}
         \caption{Ventilation rate, N (ACH)}
         \label{fig:ACHNumP}
     \end{subfigure}
     \hfill
     \begin{subfigure}[b]{0.3\textwidth}
         \centering
         \includegraphics[width=\textwidth]{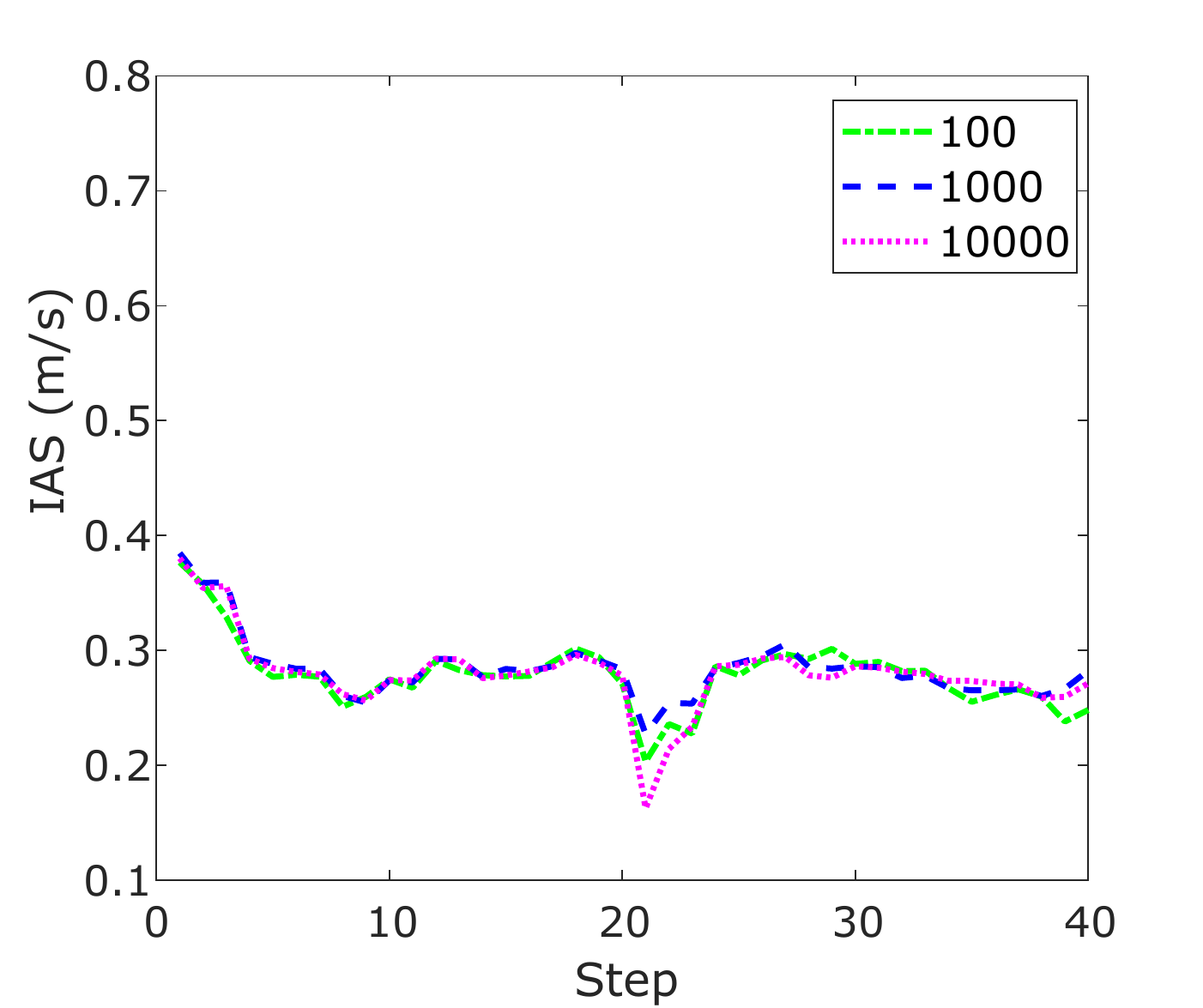}
         \caption{Internal air speed, IAS (m/s)}
         \label{fig:IASNumP}
     \end{subfigure}
     \hfill
     \begin{subfigure}[b]{0.3\textwidth}
         \centering
         \includegraphics[width=\textwidth]{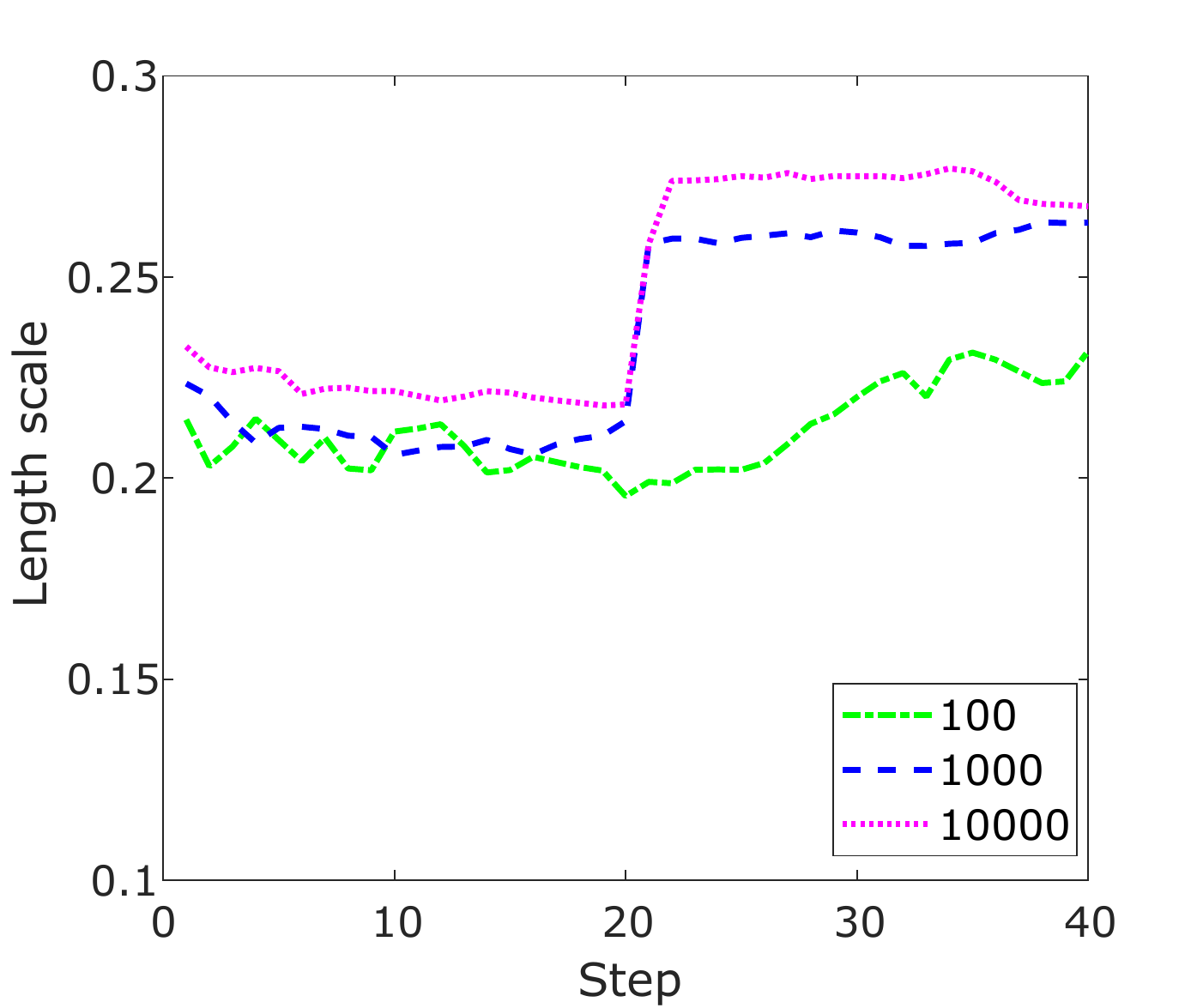}
         \caption{Length scale}
         \label{fig:LengthNumP}
     \end{subfigure}
        \caption{Mean of PF evolution demonstrating effect of the number of particles}
        \label{fig:NumP}
\end{figure}

Figure \ref{fig:IASNoisePF} shows the evolution of the posterior of IAS as a function of the step for the increasing noise levels.  The results are reasonably consistent up until time step 20, after which the mean IAS posterior increases, dependent on the noise level, before dropping back to the true value; for 10\% noise, the true value is not achieved by step 40.  Interestingly, although there are clear differences in the evolution of the mean length scale for the different noise levels, the absolute value varies most for zero noise, the mean value increasing from 0.21 to 0.26 (Figure \ref{fig:LengthNoisePF}). For higher noise levels the mean length scale value meanders close to 0.2.  This suggests that despite the noise the model is still detecting similar relationships between the data points.

The hyper-parameters of the particle filter approach have been set to be a similar magnitude to the KOH approach to ensure comparability.  One parameter that has no equivalence, however, is the number of particles.  This is important as it directly affects the run time of the model.  Figure \ref{fig:NumP} shows the impact of changing the number of particles on the output, with particle number equal to 100 (green), 1000 (blue) and 10000 (magenta). The results demonstrate that the number of particles has little impact on the outputs of this test case.


The more data points are used, however, the greater the time taken to run the analysis. Table \ref{tab:RunTimes} shows the run times for analysis of this toy problem using the different models. The models have been processed on a PC with an i7-6700 CPU processor and 32 MB RAM, and of course a more powerful PC would improve these times, but the relative difference between the run times for the models will still hold.  It is clear that the PF approach is substantially quicker than the static KOH approach as even with 10,000 particles the step time is only just over 3 minutes, giving a total time of just over 2 hours, whereas the KOH approach required a run time in excess of 6 hours for the same problem. The sequential KOH is substantially quicker, however, but using 4 data points the run time is still much slower than the PF approach with 1,000 particles.    

\begin{table}[t]
    \centering
    \begin{tabular}{lcc}
    \TCH{Model} & \TCH{Number of Data Points} & \TCH{Run Time (s)} \\
    \hline
    \TCH{KOH} & \TCH{} & \TCH{}  \\
    \TCH{} & \TCH{20} & \TCH{5,443}  \\
    \TCH{} & \TCH{40} & \TCH{22,572}  \\ 
    \hline
    \TCH{Particle Filter} & \TCH{} & \TCH{(step / total)}  \\
    \TCH{1000 particles} & \TCH{40} & \TCH{ 20 / 786}  \\
    \TCH{10,000 particles} & \TCH{40} & \TCH{ 197 / 7,915}  \\
    \hline
    \TCH{KOH-Sequential} & \TCH{} & \TCH{(step / total)}  \\
    \TCH{N=2} & \TCH{40} & \TCH{60 / 2,535}  \\
    \TCH{N=4} & \TCH{40} & \TCH{168 / 6,216}  \\
    \TCH{N=8} & \TCH{40} & \TCH{626 / 20,658}  \\
    \hline
    \end{tabular}
    \caption{Comparison of run times: the run times for the KOH approach are for the entire simulation, whereas for the PF and sequential KOH run times are given both for a single step and for the entire simulation. }
    \label{tab:RunTimes}
\end{table}


\subsection[Monitored]{Calibration with monitored data from the farm}
The toy problem has facilitated comparison between the three approaches, but the main aim of this study is to explore the applicability of using the particle filter approach for model calibration in the context of a digital twin.  To this end, real monitored data from August to October 2018 have been selected for further study.  This time period has been chosen as the data are complete and there were no known significant changes to operation except known changes to the ventilation settings.  This is important, as we would like to understand to what extent the two calibration approaches are capable of identifying the change in parameter. While we know what the changes are to the ventilation settings, we do not know exactly what that means in terms of changes to the parameter value (the ventilation rate).  We also do not know to what extent other operational conditions were in place that might have impacted on ventilation rate or air speed.

\begin{figure}[t]%
\FIG{\includegraphics[width=0.9\textwidth]{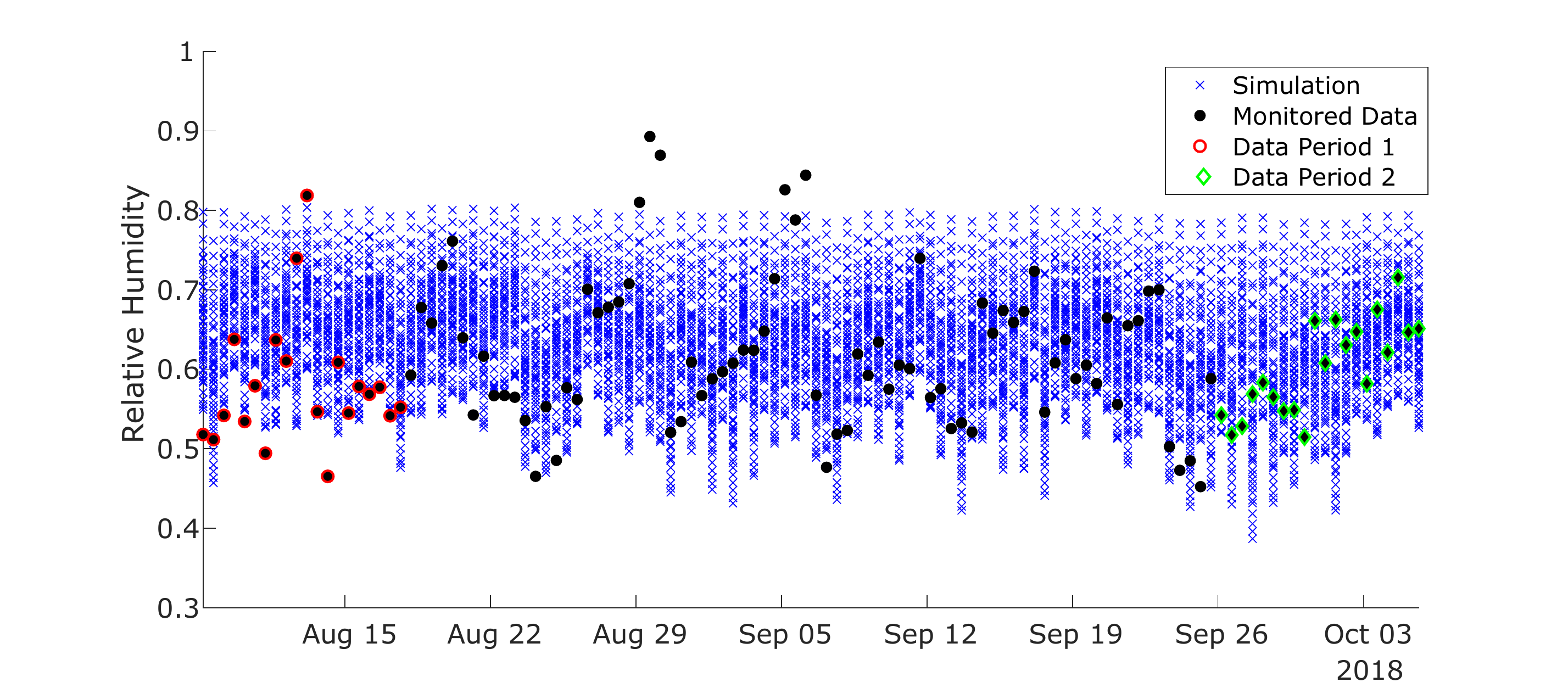}}
{\caption{Monitored data used for the calibration and the corresponding outputs from the calibration simulations over 3 months}
\label{fig:MonitoredData}}
\end{figure}

\begin{figure}[t]
     \centering
     \begin{subfigure}[b]{0.49\textwidth}
        \centering
         \includegraphics[width=\textwidth]{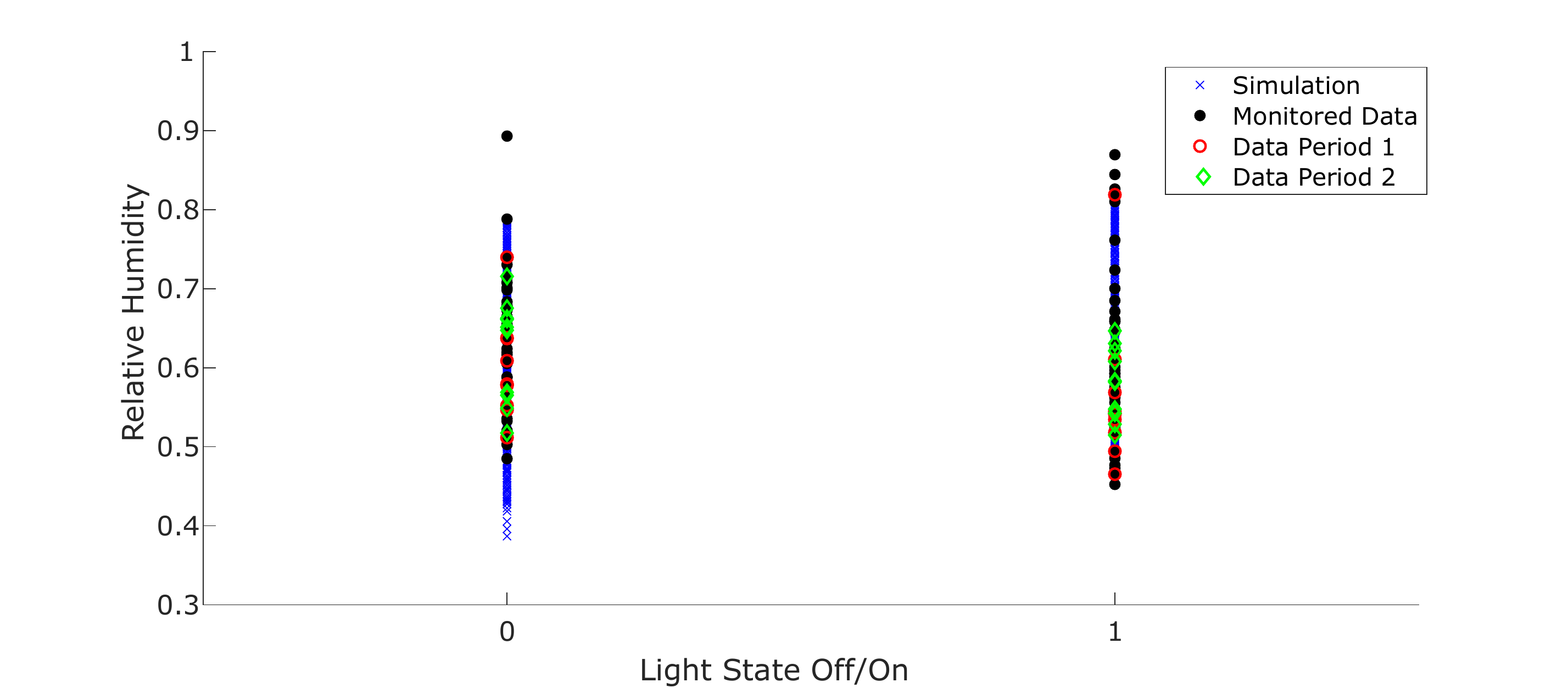}
         \caption{Light State Off=0, On=1}
         \label{fig:MonitoredLightState}
     \end{subfigure}
     \hfill
     \begin{subfigure}[b]{0.49\textwidth}
         \centering
         \includegraphics[width=\textwidth]{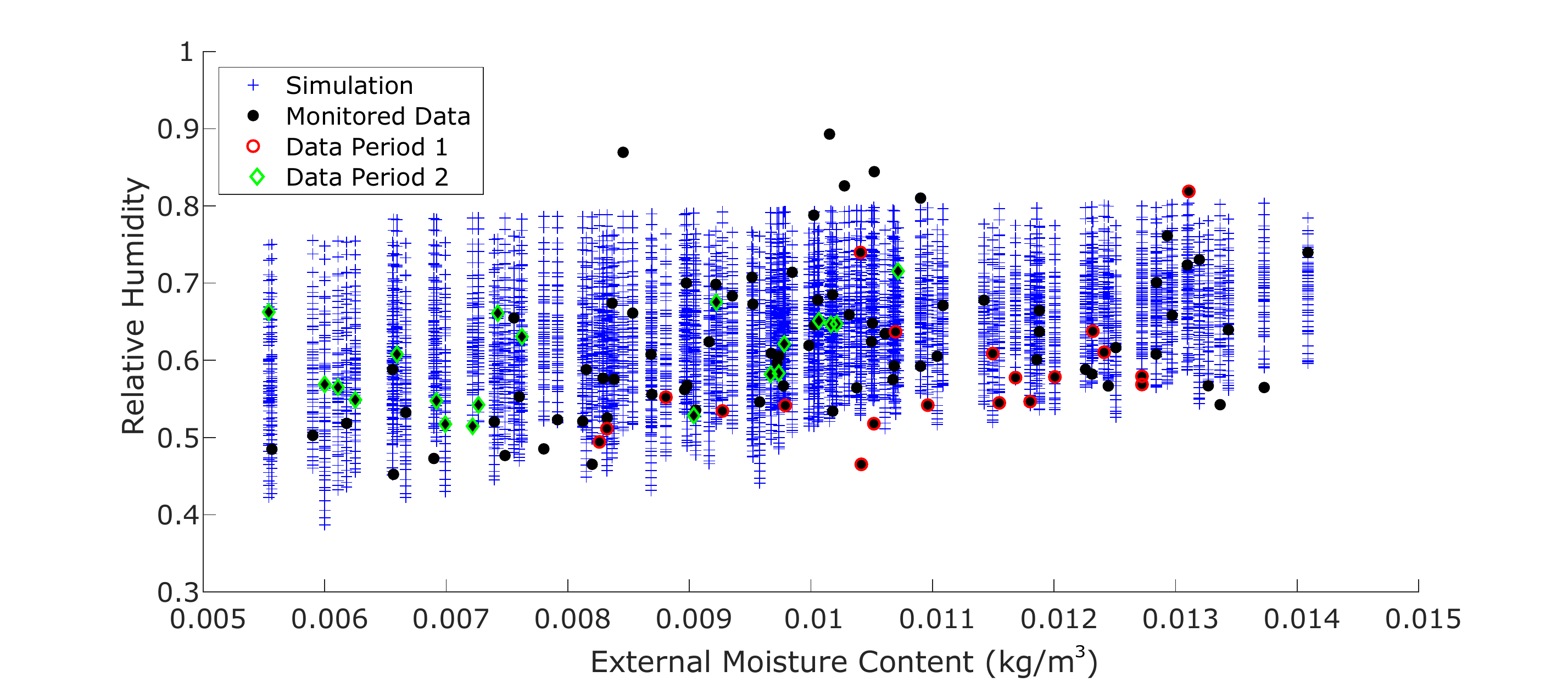}
         \caption{External Moisture Content, $C_w (kg/m^3)$}
         \label{fig:MonitoredExtCw}
     \end{subfigure}
        \caption{Monitored data and corresponding simulation outputs for KOH approach, showing the scenarios a) light state, and b) external moisture content}
        \label{fig:MonitoredDataKOH}
\end{figure}

Implementation of the models with the monitored data has followed the same approach as described for the toy problem above.  The monitored data and the corresponding simulation outputs across the input parameter ranges are illustrated in Figures \ref{fig:MonitoredData} and \ref{fig:MonitoredDataKOH}. The dots in the first figure illustrates how the data evolve in time, whereas in the second figure they show how the data vary as a function of the light state and external air moisture content, the two scenarios used in the KOH approach.  As for the toy problem, data points have been extracted at 4pm and 4am, corresponding to the lights being off and on.  Each black dot in Figure \ref{fig:MonitoredData} is a data point, and we have selected two separate periods of 20 data points highlighted in red and green for analysis using the standard BC approach. Period 1 (red) is chosen as it lies before the change in ventilation settings, and Period 2 (green) after the change.  It is not feasible to use all 118 data points for the static KOH approach as the run-time would be too long.  Also shown in the figure are the outputs from the simulation runs executed for the calibration (blue crosses).  Unlike the toy problem, here there are several observations which lie outside the range covered by the simulation outputs, specifically there are points of low and high relative humidity that are not predicted by the model. This is likely due to simplifications in the model that do not adequately represent the sources of humidity, for example the irrigation of the plant trays is represented by a single mean saturation level that does not account for daily fluctuations in moisture level.


\begin{figure}[t]
     \centering
     \begin{subfigure}[b]{0.49\textwidth}
        \centering
         \includegraphics[width=0.8\textwidth]{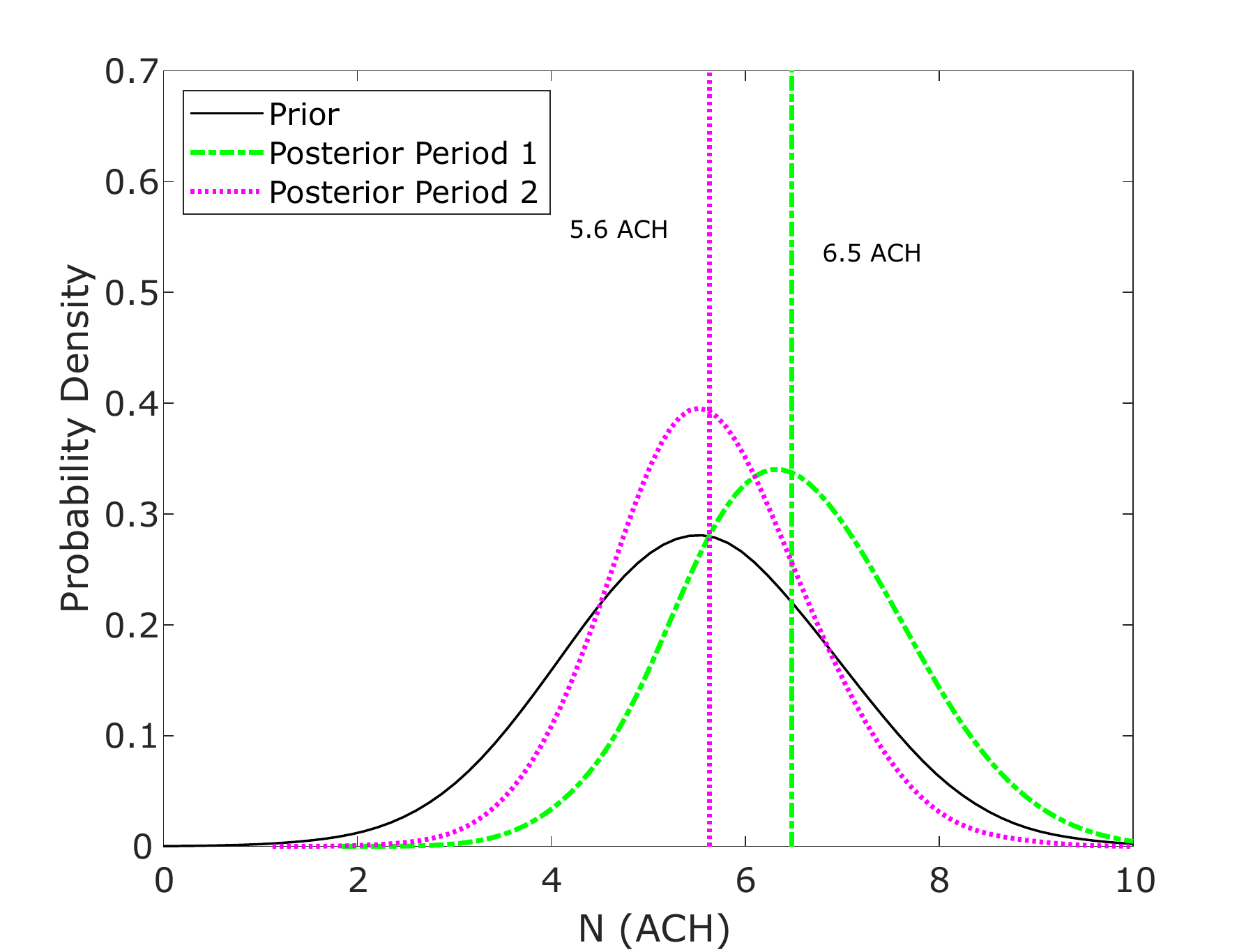}
         \caption{Ventilation rate, N $(ACH)$}
         \label{fig:PosteriorKOH_ACH}
     \end{subfigure}
     \hfill
     \begin{subfigure}[b]{0.49\textwidth}
         \centering
         \includegraphics[width=0.8\textwidth]{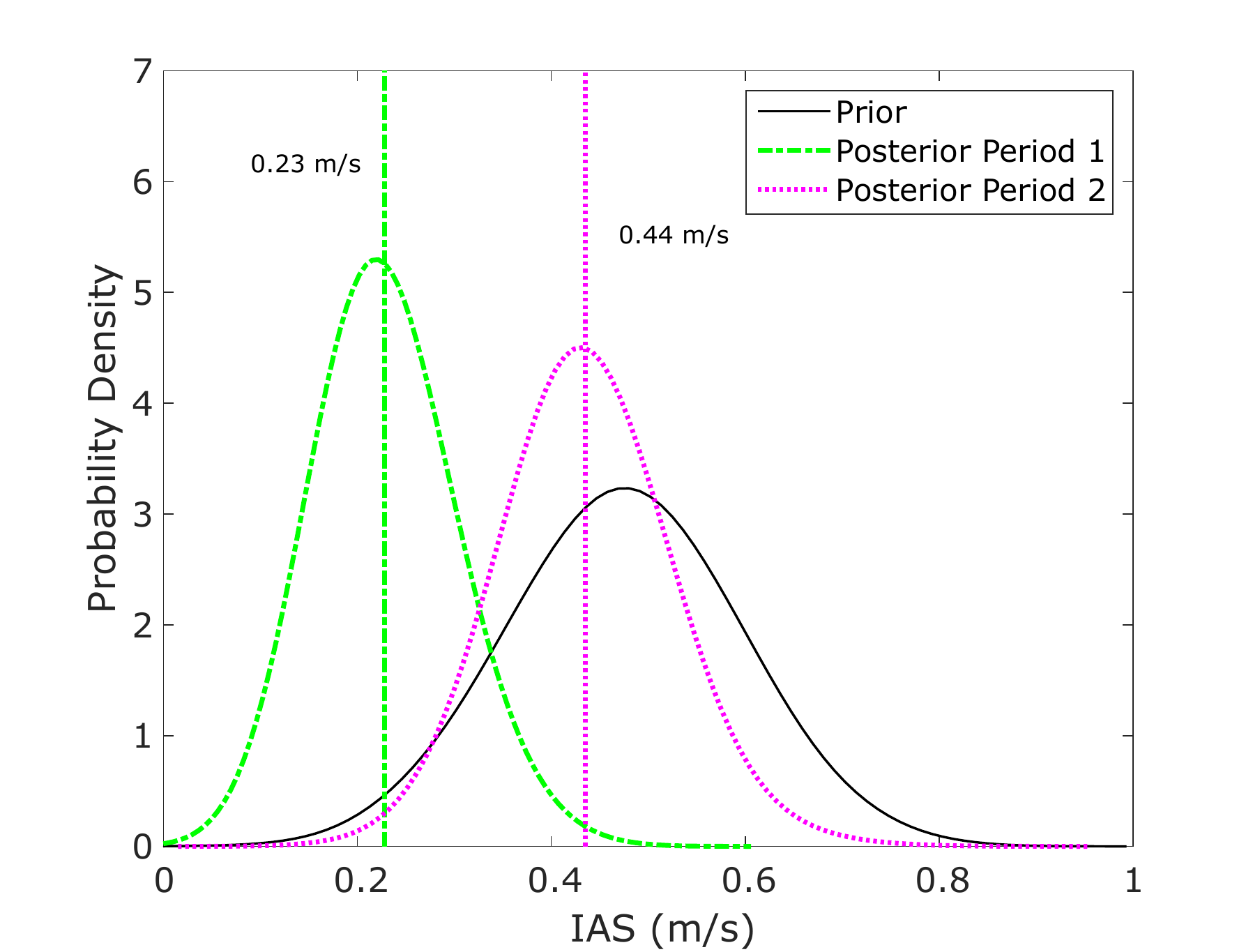}
         \caption{Internal air speed, IAS $(m/s)$}
         \label{fig:PosteriorKOH_IAS}
     \end{subfigure}
        \caption{Prior and posterior distribution of ventilation rate, $N$ and internal air speed, $IAS$ using the KOH approach with monitored data from the farm}
        \label{fig:MonitoredKOHPriorPosterior}
\end{figure}

\begin{figure}[t]
     \centering
     \begin{subfigure}[b]{0.49\textwidth}
        \centering
         \includegraphics[width=\textwidth]{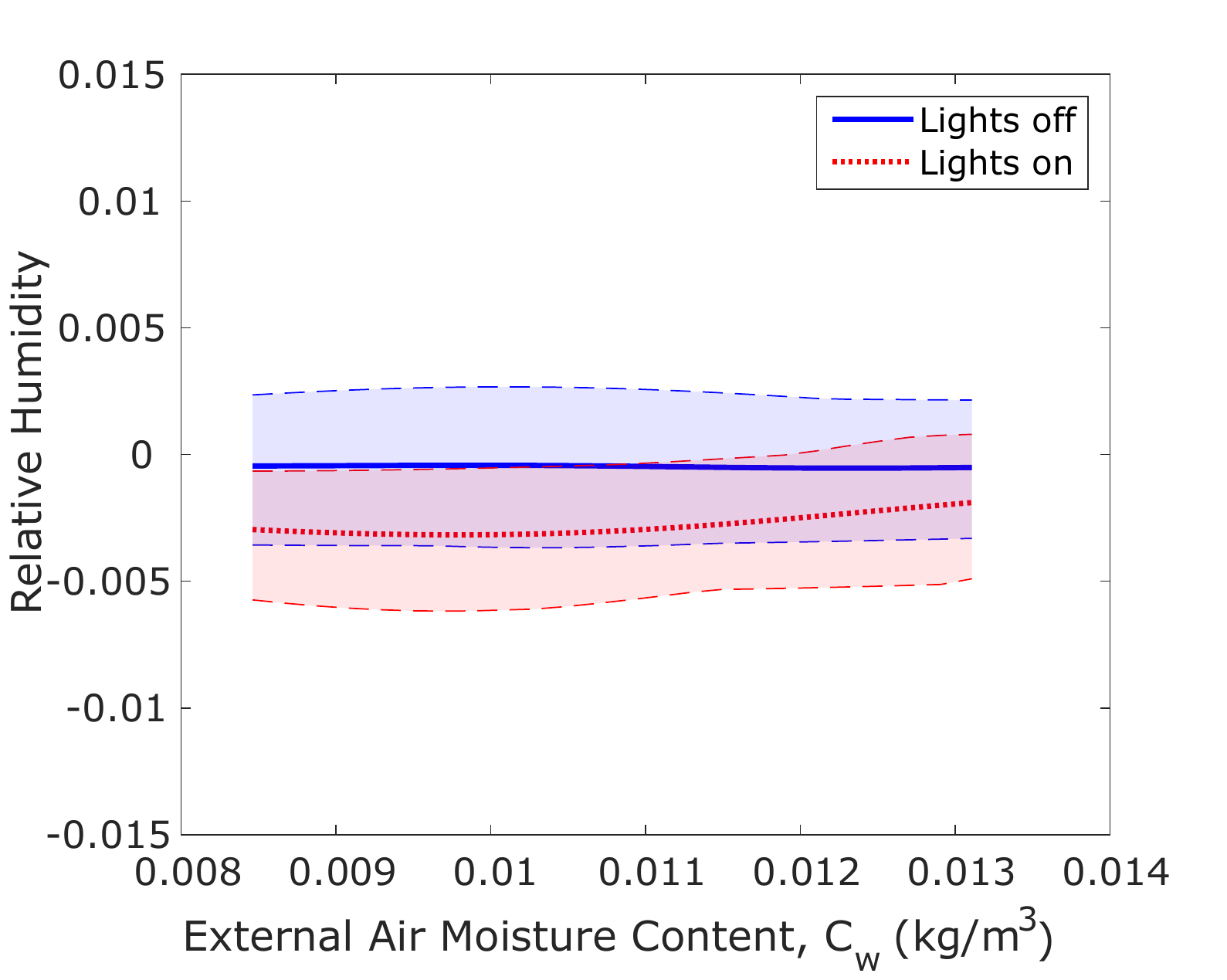}
         \caption{Model bias, period 1}
         \label{fig:BiasAug}
     \end{subfigure}
     \hfill
     \begin{subfigure}[b]{0.49\textwidth}
         \centering
         \includegraphics[width=\textwidth]{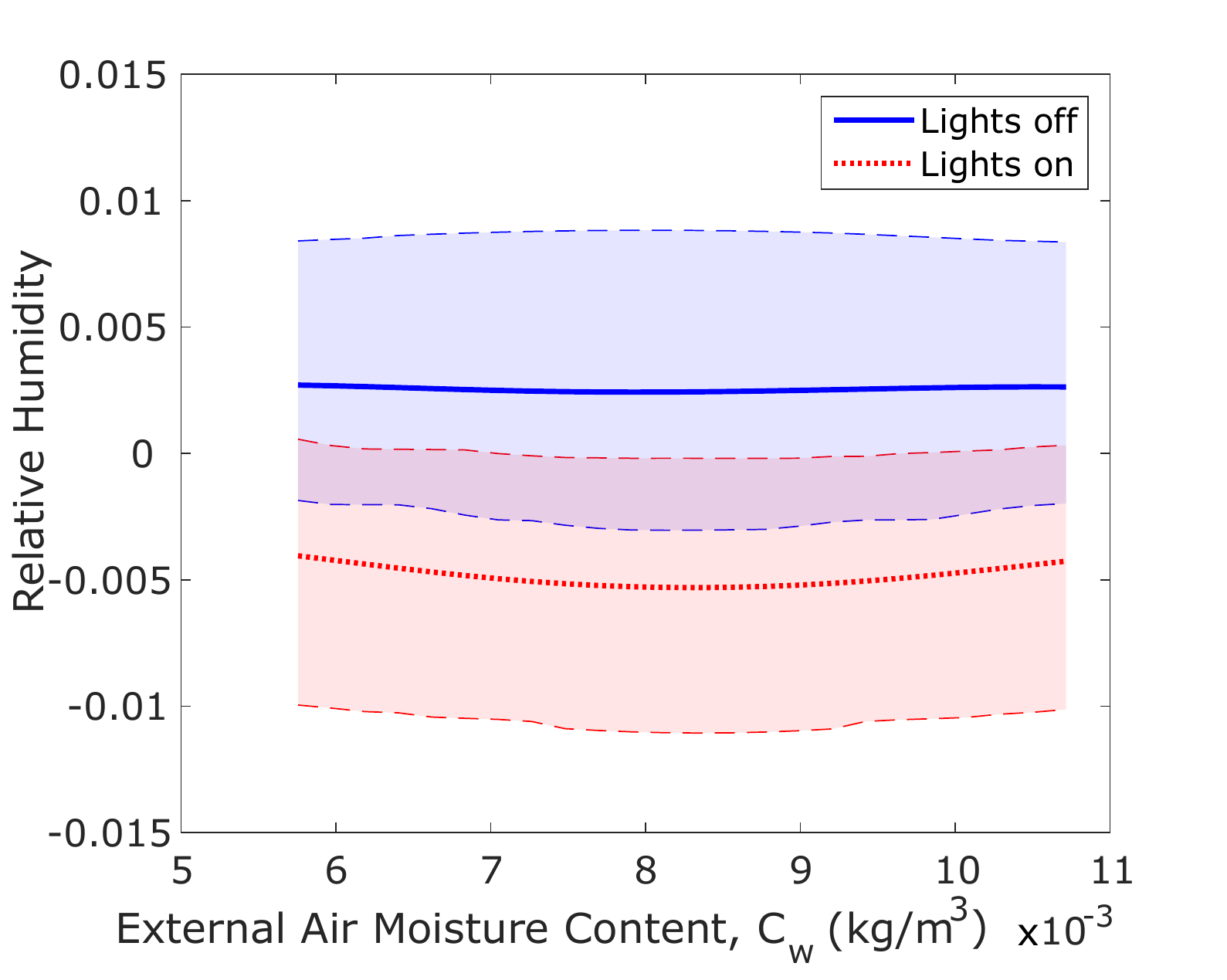}
         \caption{Model bias, period 2}
         \label{fig:BiasSep}
     \end{subfigure}
        \caption{Model bias function}
        \label{fig:ModelBiasKOHS}
\end{figure}

As a first step, the KOH approach has been used to estimate the model input parameters for periods 1 and 2. Figure \ref{fig:MonitoredKOHPriorPosterior} shows the posterior distributions for both parameters for both time periods, with the first time period shown in green (-.) and the second in magenta (..).  Here a normal prior distribution has been used (shown in black) to give a more informative prior as we expect the observations to have some degree of measurement error. We have selected prior mean and standard deviation values to position the prior at the centre of the possible range of parameter values, but with sufficient variance to generate samples across the range, as we anticipate the parameter values being towards the centre of the possible range of values.  The posterior parameter distributions suggest that the ventilation rate drops from a mean value of $6.5 ACH$ in August to a mean value of $5.6 ACH$ in October, whereas the mean internal air speed increases from 0.23 to 0.44 $m/s$. These results are plausible given the increase in relative humidity observed towards the end of period 2 where both a reduction in ventilation rate and an increase in internal air speed tend to result in an increase in relative humidity. 

It is particularly useful when using monitored data to explore the model bias function as it can give some idea as to where the physics-based model might be improved.  Figure \ref{fig:ModelBiasKOHS} shows the mean and 90\% confidence limits of the model bias for periods 1 and 2 plotted separately for the two light states - off and on - as a function of external moisture content on the x axis. As for the toy problem the magnitude of the bias is small, suggesting that differences between the model and the data are primarily accounted for by measurement error which has a mean posterior value of 0.77.  However there is a clear difference between the bias function for lights on and lights off. When the lights are on, the absolute magnitude of the bias function is greater implying that the agreement between the model and the data is better when the lights are off, particularly in period 1. The fact that the bias function is negative when the lights are on suggests that the model is over-predicting the relative humidity when compared with the data.  Relative humidity is a combined effect of temperature and air moisture content, so a value that is too high can be due to either temperatures that are too low or too much moisture in the air.  This information gives us valuable insight into how to improve the physics-based model. 

As for the test case, we run the particle filter approach for the entire period from August to October, a total of 118 data points.


\begin{figure}
    \centering
     \begin{subfigure}[b]{\textwidth}
        \centering
         \includegraphics[width=0.8\textwidth]{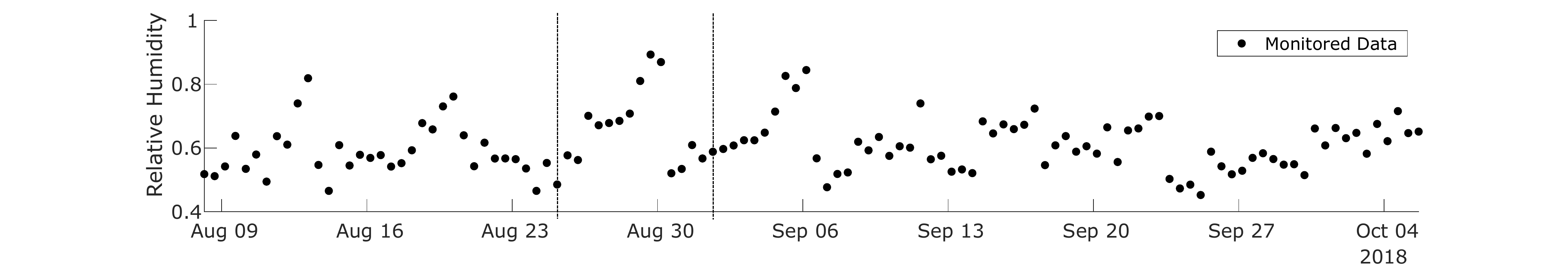}
         \caption{Monitored relative humidity, RH}
         \label{fig:RHMon}
     \end{subfigure}
     \vfill
     \centering
     \begin{subfigure}[b]{\textwidth}
        \centering
         \includegraphics[width=0.8\textwidth]{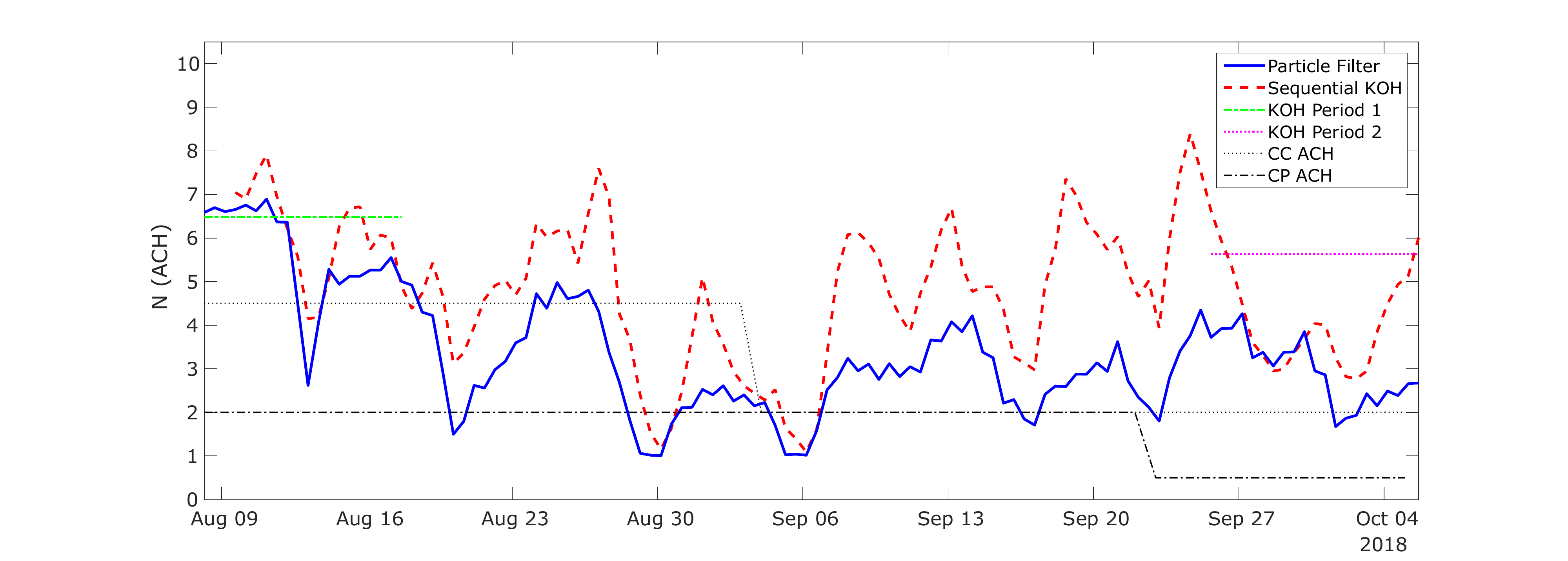}
         \caption{Evolution of mean ventilation rate, N $(ACH)$}
         \label{fig:ACHMon}
     \end{subfigure}
     \vfill
     \begin{subfigure}[b]{\textwidth}
         \centering
         \includegraphics[width=0.8\textwidth]{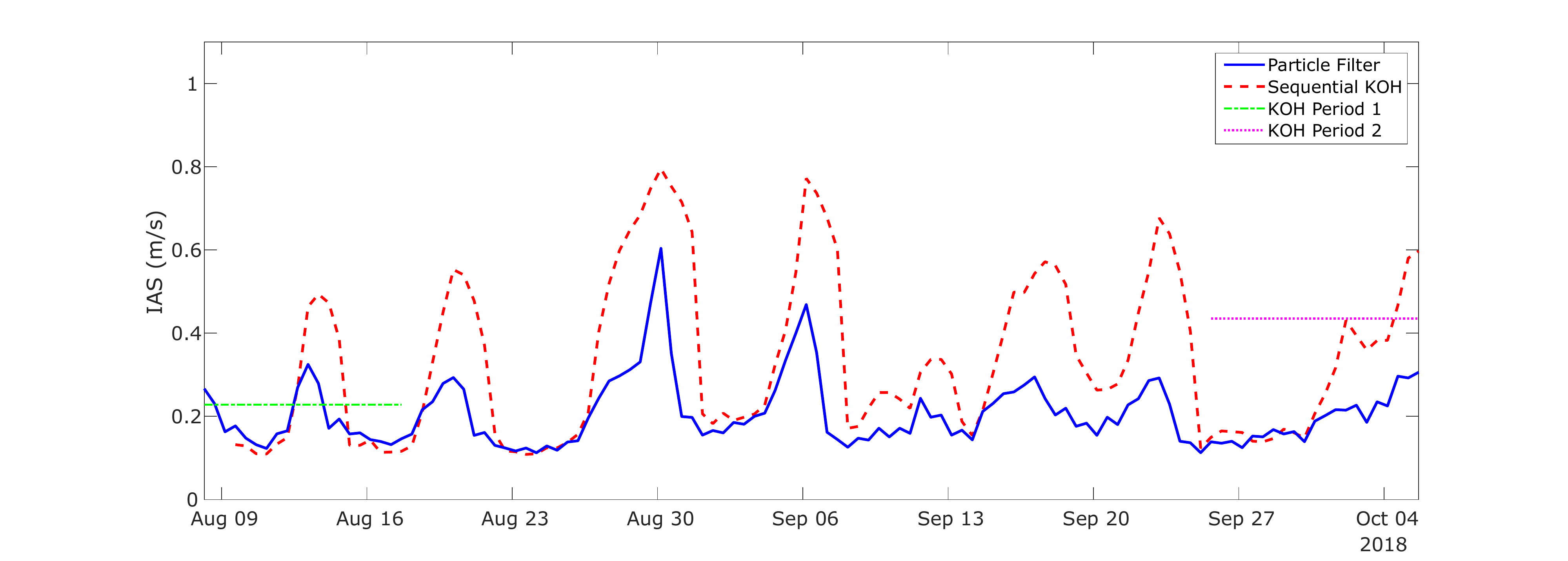}
         \caption{Evolution of mean internal air speed, IAS $(m/s)$}
         \label{fig:IASMon}
     \end{subfigure}
     \vfill

     \centering
     \begin{subfigure}[b]{0.49\textwidth}
        \centering
         \includegraphics[width=0.7\textwidth]{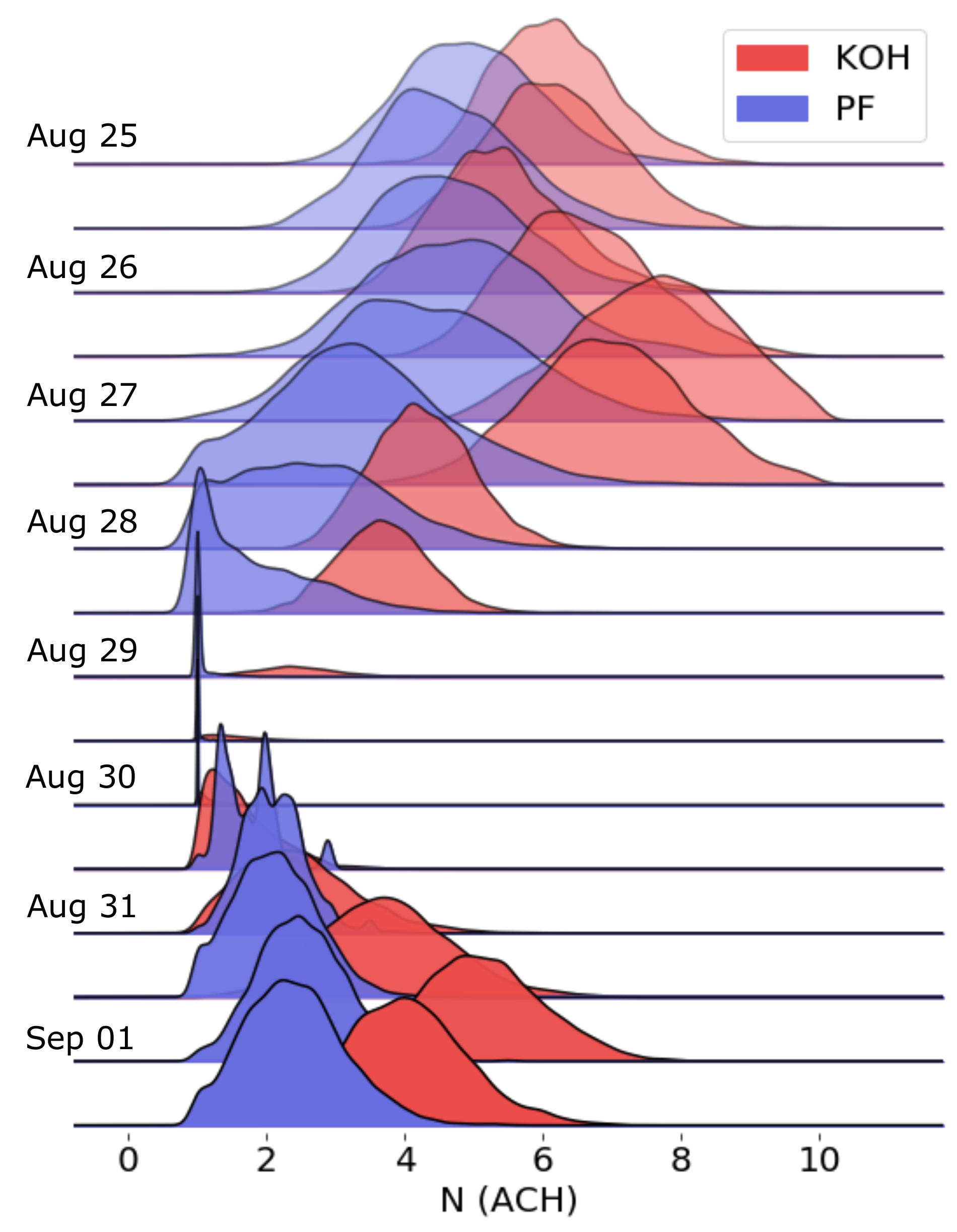}
         \caption{Evolution of posterior distribution over 30th August, ventilation rate, N($ACH$)}
         \label{fig:ACH_35_50}
     \end{subfigure}
     \hfill
     \begin{subfigure}[b]{0.49\textwidth}
         \centering
         \includegraphics[width=0.7\textwidth]{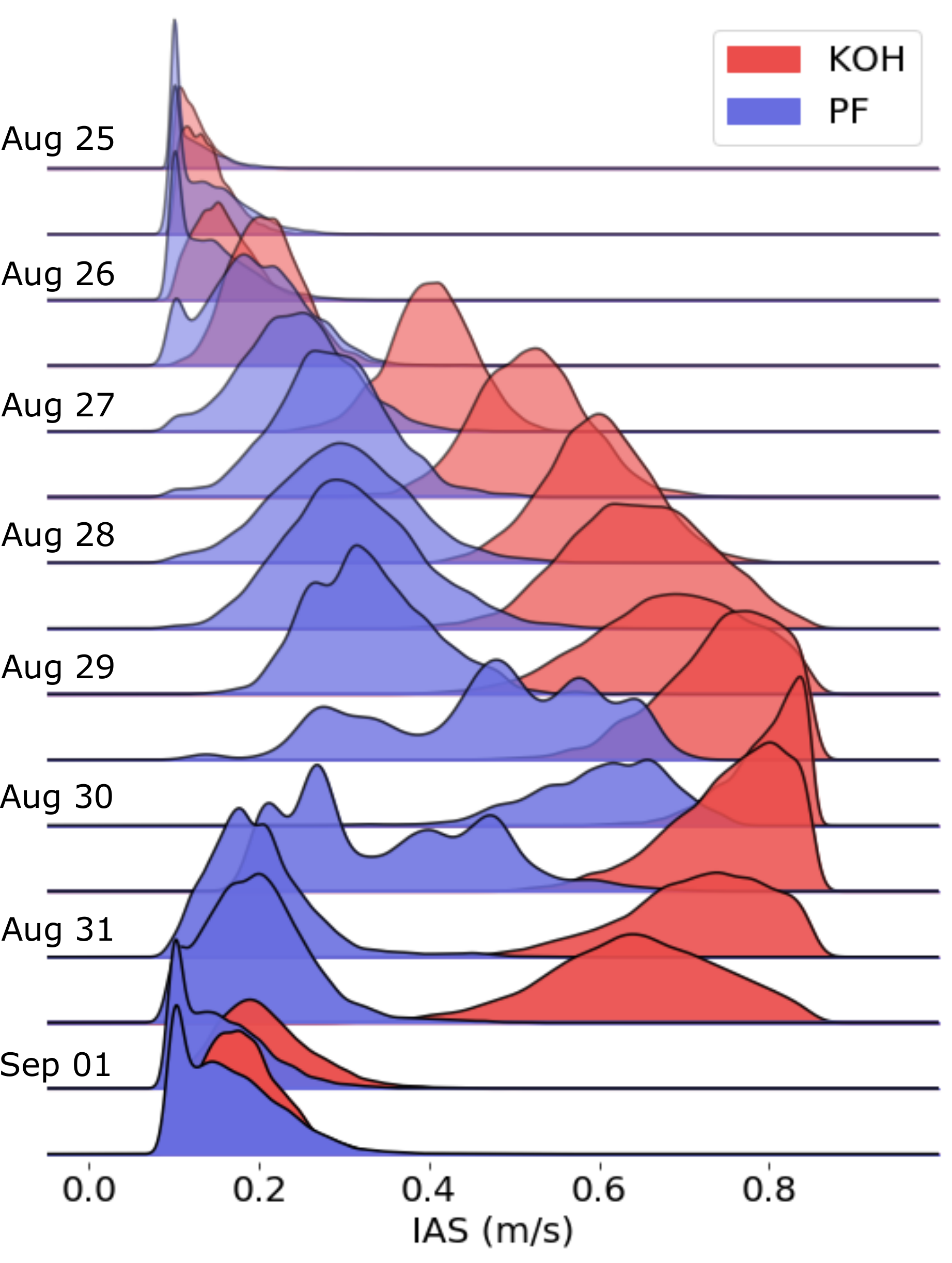}
         \caption{Evolution of posterior distribution over 30th August, internal air speed, IAS ($m/s$)}
         \label{fig:IAS_35_50}
     \end{subfigure}
        \caption{Calibration input (a) and outputs (b-e) for monitored RH data }
        \label{fig:PFMon}
\end{figure}

The outputs of the calibration process in terms of the evolution of the posterior distributions for the ventilation rate and internal air speed are indicated in Figure \ref{fig:PFMon}, the PF length scale remained fairly constant at a value of just under 0.2.  Figure \ref{fig:PFMon} shows first the monitored data used for the calibration in Figure \ref{fig:RHMon}, and the mean of the posterior parameter values are shown in Figures \ref{fig:ACHMon} and \ref{fig:IASMon} as a solid blue line.  We have also run a sequential Bayesian calibration updating after 4 data points similar to the toy problem case, shown in the figures as the dashed red line. The mean of the two static Bayesian calibrations are indicated on the figures as green (-.) and magenta (..) lines over the periods of interest.  Also indicated on Figure \ref{fig:ACHMon} are the settings of the two ventilation systems over this period.  The CP ventilation setting is constant until September 22nd, at which point it is reduced from $2$ to $0.5 ACH$, whereas the CC setting is higher, at $4.5 ACH$, during the initial period, reduced in early September to a value of $2 ACH$.


Considering Figures \ref{fig:ACHMon} and \ref{fig:IASMon}, the first point is that the results for the two sequential calibrations are consistent.  The sequential KOH results are more extreme, particularly for the internal air speed (Figure \ref{fig:IASMon}) where the value peaks on several occasions, reaching a high of $0.8m/s$ on August 30th. The PF approach has peaks at similar times, but they do not reach the same magnitude, only approaching $0.6m/s$ on August 30th. To match observed high peaks in monitored relative humidity, the model requires a low ventilation rate and a high internal air speed - the first prevents moisture from leaving the system, and the second increases evaporation and transpiration rates from the growing system. The more extreme peaks arise from the sequential KOH approach due to a combination of factors that can be appreciated with a consideration of Figures \ref{fig:ACH_35_50} and \ref{fig:IAS_35_50} which show the evolution of the posterior distributions for each of the two calibration parameters over the highest peak in the monitored data on the 30th August. The first factor is that the sequential KOH approach uses four monitored data points at each time step, whereas the PF approach considers each data point individually.  This means that for the sequential KOH approach the influence of each data point extends over a longer time period and makes the approach more likely to attain higher parameter values. But it also means the sequential KOH approach is slower to respond to a sudden change in value than the PF approach; note for example in the PF approach the internal air speed drops on 30th-31st August responding to the sudden drop in monitored relative humidity while in the KOH approach this drop does not happen until the 1st September when the influence of the high monitored RH on 29th-30th August is no longer felt.  The second factor giving rise to the more extreme peaks is the fixing of the posterior variance in the sequential KOH approach which as described previously was necessary to avoid unacceptable narrowing of the posterior distribution. Reducing the fixed variance value for the sequential KOH value yields lower peak mean parameter values as the smaller variance gives less credibility to the higher values. By comparison the PF approach has no such constraint so when a sudden change in the monitored data occurs the variance of the posterior distribution can increase to encompass a wider range of possible parameter values.  This can lead to a lower mean value relative to the sequential KOH approach, as for example illustrated for the internal air speed for the second data point on the 29th August (Figure \ref{fig:IAS_35_50}). 

The second point observable from Figure \ref{fig:PFMon} is that the sequential KOH results are very different from the static KOH results, particularly for period 2 where the static values of $5.6 ACH$ and $0.43 m/s$ for the ventilation rate and internal air speed respectively are substantially higher than the sequential values which are closer to $3 ACH$ and $0.2 m/s$ for most of period 2.

Thirdly, neither of the sequential models strongly reflect the known changes to the settings of the ventilation system.  The total ventilation rate is the combination of both controlled and uncontrolled ventilation.  That neither model identifies changes corresponding to the changes to the controlled ventilation may suggest that the non-controlled ventilation arising from lift shafts, door infiltration etc. is dominant.  
Equally it may reflect the fact that the data are quite noisy, so the true values are masked by the noise in the data.

Even if the models are not able to identify 'true' values for these parameters, are we able to infer parameter values that enable the model to accurately simulate the farm environment? To assess this, we have run the model using parameter values for $N$ and $IAS$ equal to the outputs of the sequential models, and also equal to the static values for period 1.  

\begin{figure}[t]
     \centering
     \begin{subfigure}[b]{0.9\textwidth}
        \centering
         \includegraphics[width=\textwidth]{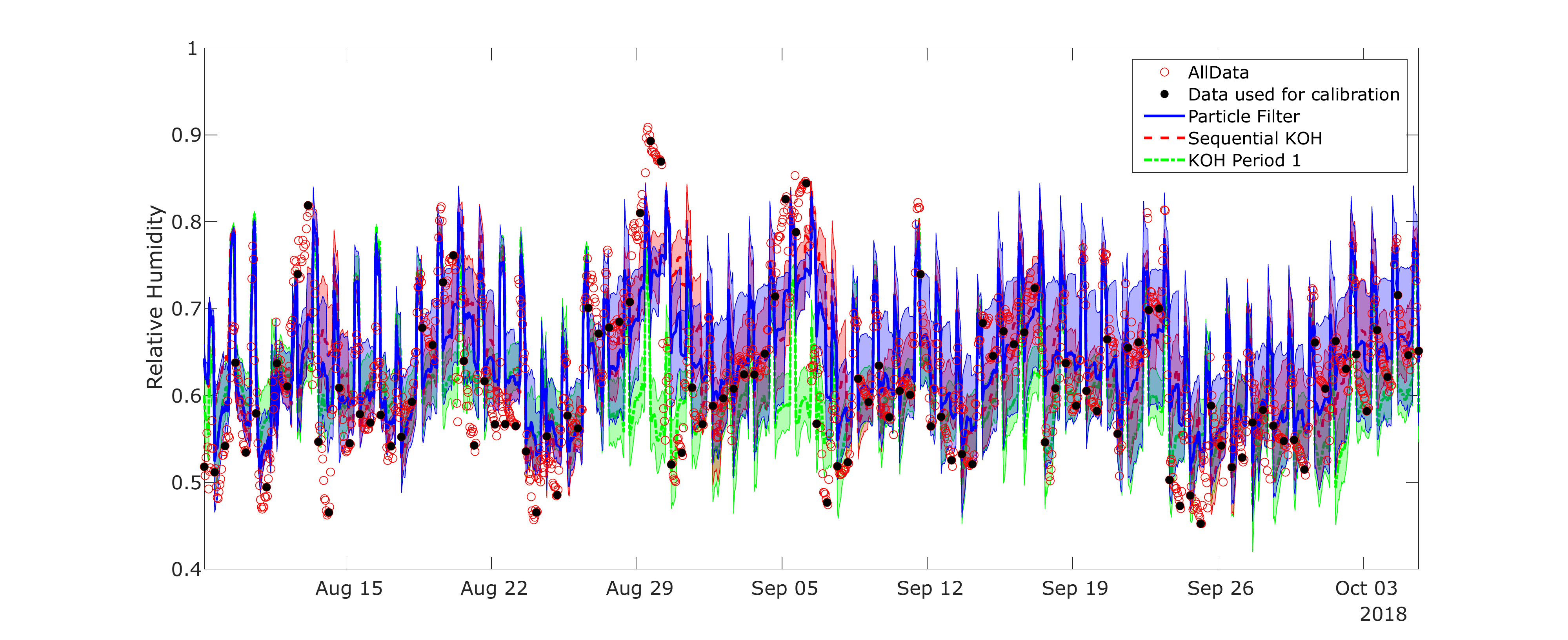}
         \caption{Entire period}
         \label{fig:RHAll}
     \end{subfigure}
     \vfill
     \begin{subfigure}[b]{0.9\textwidth}
         \centering
         \includegraphics[width=\textwidth]{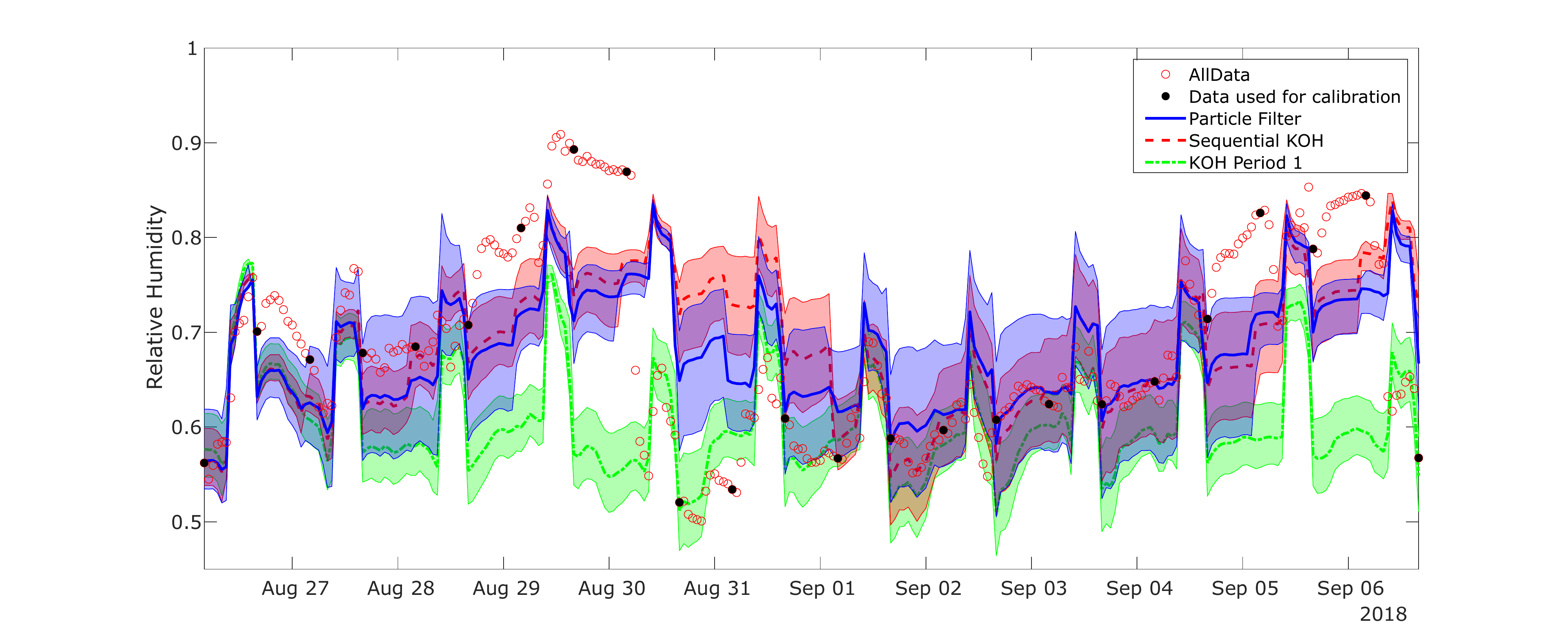}
         \caption{Central period}
         \label{fig:RHCentral}
     \end{subfigure}
        \caption{Relative Humidity results for the PF, sequential KOH and static KOH approachs showing a) the entire simulation, and b) a central period}
        \label{fig:RHResults}
\end{figure}

\begin{figure}[h]
    \centering
    \includegraphics[width=0.8\textwidth]{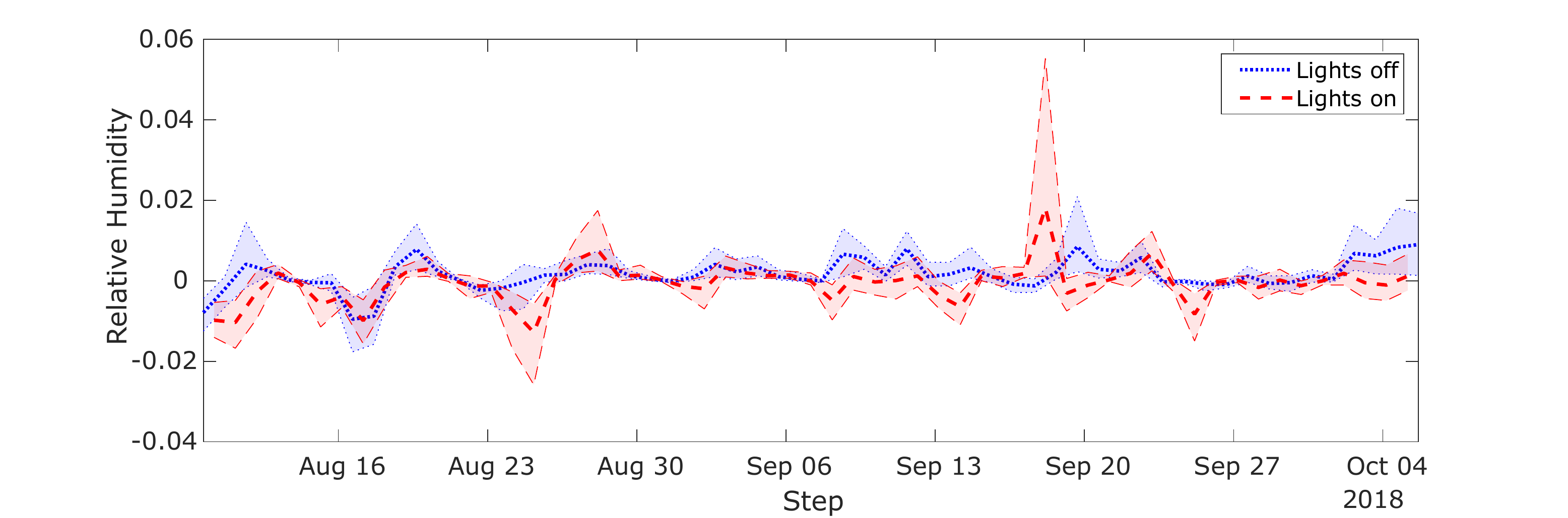}
    \caption{Sequential KOH: mean and 90\% confidence limits of the model bias function}
    \label{fig:KOHSequentialMonitoredBias}
\end{figure}

The mean and 90\% confidence limits of the relative humidity values simulated using the three different approaches are shown in Figure \ref{fig:RHResults}. Figure \ref{fig:RHAll} shows the simulation results across the entire time period compared against the monitored data, with the complete data set shown as red circles and the data used for the calibration picked out as black dots. Specifics are difficult to pick out, but the overall trend for the sequential models to give a much closer approximation to the data than the static model is clear, as the PF approach results in blue and the sequential KOH approach results in red show a much greater degree of variation than the static KOH results in green, in line with the data. This is particularly clear over the central section of the simulation around August 30th, and this section of the plot is enlarged for clarity in Figure \ref{fig:RHCentral}.  Here we can clearly see that using the parameters derived for period 1 using the static KOH approach gives a poor approximation to the data in the days before August 30th, whereas the two sequential approaches give an increase in relative humidity over this period in line with the data.  That they do not get even closer is due to the fact that these data lie outside the range of the simulation runs illustrated in Figure \ref{fig:MonitoredData}.  Another point of interest from Figure \ref{fig:RHAll} is that there is a gradual increase in variance of the PF results as time progresses i.e. the 90\% confidence limit shown at the beginning of August is much narrower than that shown in October.  This feature of the particle filter that the variance of the particle weights increases over time means that for longer term simulations some form of correction will need to be implemented. This is not an issue for the sequential KOH as we have not allowed the variance of the parameter values to change as the simulation progresses.

The KOH approach yields a model bias function for periods 1 and 2 which suggests that the model may overestimate the relative humidity when the lights are switched on (Figure \ref{fig:ModelBiasKOHS}).  Figure \ref{fig:KOHSequentialMonitoredBias} shows the mean and 90\% confidence limit of the bias function for the sequential KOH approach plotted as a function of time.  The bias function when the lights are on tends to be opposite in sign to that when the lights are off, and again the sign of the bias suggests the model tends to overestimate the relative humidity when the lights are on and underestimates when the lights are off.  The absolute magnitude is small however, only peaking over 0.01 when the lights are off on August 19th, and at 0.02 when the lights are on in the morning of September 18th. The mean estimated measurement error is greater at 0.25 and peaks where the monitored RH values are high on the 30th August, suggesting that the differences between the model and the data are not systematic.  


\section[Discussion]{Discussion}
In the context of a digital twin, calibrating the simulation element of the twin with current data is essential for ensuring that the simulation is representing reality as closely as possible. Manual calibration is time consuming and not suitable when parameter values are changing.  This study explores the suitability of a particle filter approach for calibration and compares it against static Bayesian calibration following the \cite{KOH_2001} approach, and also uses the KOH approach sequentially as a further comparison.

The static KOH approach assumes that the calibration parameters do not change over the timespan of the data and so is only able to give good estimates if there is confidence that the parameter values are unchanging. We have seen using a toy problem that if the parameter values do change the KOH approach can only give an average of the true value (Figure \ref{fig:ToyKOHRH}). If the parameters are unchanging however, it is still the 'gold standard' for calibration against which we compare other approaches.  It is particularly useful as the KOH approach is formulated in such a way as to give estimates not only of the calibration parameters but also of the discrepancy between the model and the data (model bias), and the measurement and numerical errors. We have seen in the toy problem (Figure \ref{fig:ToyKOHPrecisionHyperparameters}) that the posterior distributions of the error terms have low median values as expected when the data are derived directly from the model, but for monitored data the approach gives a more useful insight, particularly into the model bias (Figure \ref{fig:ModelBiasKOHS}).  The figure illustrates that while the bias only shows a small variation with the external air moisture content, there is a more significant difference according to whether the lights are switched on or off.  With the lights on - a direct source of heat - the bias is negative, suggesting that the model is predicting relative humidities that are higher than observed in the data.  Relative humidity increases as temperature decreases for the same air moisture content, so if it is too high that suggests either the predicted temperature is too low, or the predicted air moisture content is too high. This gives direct insight into how to improve the physics-based model going forwards.

The sequential approaches offer the potential to track parameter values, which is what we desire for the digital twin. We have applied the different approaches to both a toy problem, with data derived from the simulation with known parameters and a controlled change in the value of ventilation rate, and to real monitored data. The difference is very clear; whereas with the toy problem the sequential methods were able to track a clearly defined change in ventilation rate accurately (Figure \ref{fig:ToyPFThetaEvolution}), the monitored data gives a much more variable output (Figure \ref{fig:PFMon}). We have also seen that in the monitored data there are data points that fall outside the range of the simulation runs.  The very high relative humidity values occurring around the 29th August and September 5th (Figure \ref{fig:MonitoredData}) seem to be not following the diurnal pattern typically seen on other days so could be due to something happening within the farm which we are not aware of - a change to the operation of the dehumidifiers for example.  Equally, there are values which are lower than the simulation runs, particularly in early August.  More work is needed to explore the model inadequacies, particularly for extreme values of relative humidity.

Since we know the settings of the ventilation system, we might expect a clearer link between the ventilation rate parameter values and the system settings, but this is not observed. One possible reason for this is that the data are subject to a high degree of measurement error - we saw in the toy problem how noisy data can prohibit the ability of the sequential approaches to track the parameter change. But we should consider what we are tracking: we are tracking the parameter values that give the best agreement of the model with the data.  The relationship between the tracked parameters and reality depends on the extent to which the model represents reality.  For example, here the ventilation rate assumed in the model must encompass all controlled and uncontrolled components of the ventilation and is not simply equal to the setting on the dial. 

This highlights the balance to be considered in the complexity of the simulation model.  Here we use a simple physics-based model calibrated using monitored data to give the best agreement between the model and the data.  But there is limited insight into why the ventilation rate and internal air speeds appear to fluctuate so dramatically - this could be a real effect, or as is more likely it could be that other events are happening that impact on the relative humidity but are not incorporated in the model.  In this instance the calibration approaches are seeking to match the observed data with a simulation based on an incomplete description of events and will push the parameter values to give as close a match as possible. The calibration parameters then become a proxy for all events affecting their values i.e. as outlined above the ventilation rate has to encompass both controlled and uncontrolled ventilation, and the internal air speed will be affected by factors such as blockages in the tunnel which affect air flow.  A more complex model could include a more detailed representation of the system processes explicitly, but a more complex model would take longer to run and longer to calibrate. This balance must be assessed in the design and development of a digital twin.  

In this study we have compared the PF approach against a sequential version of the KOH approach, and it is clear that they give similar results.  The main differences between them are time and information - on the one hand, the PF is substantially quicker depending on the approach details - for the PF, a higher number of particles causes the run time to increase linearly, whereas in the KOH approach the run time increases approximately as the square of the number of data points for the same number of MCMC iterations. On the other hand the sequential KOH approach estimates the model bias and the measurement and numerical error terms in addition to the calibration parameters.  The model bias extracted for the sequential KOH study of the monitored data (Figure \ref{fig:KOHSequentialMonitoredBias}) reinforces the trend observed for the toy problem i.e. that the model tends to overpredict relative humidity when the lights are on and underpredict when the lights are off.  This study was based purely on two data points each day, one from when the lights are switched on and one from when they are off; a more targeted choice of data or scenario could give more insight into the model discrepancy and the error terms (\cite{Menberg_2019}). Further work will assess the sensitivity of the approaches to the choice of data with reference to maximising the impact on the accuracy of the digital twin.  

The particle filter approach works well for the purposes of providing quick estimates of parameter probability distribution in real time and responds quickly to sudden changes in the monitored data.  The sequential KOH approach also works well, but in this study has been shown to be more computationally intensive with longer run-times, give more extreme results and react more slowly to a sudden change in the monitored data than the PF approach.  This latter point is a result of the need to include more than one data point per time step for the sequential KOH approach.  As discussed in reference to Figure \ref{fig:PFMon} the influence of each data point extends over multiple time steps and response to a sudden change in the monitored data only occurs once the influence of the value pre-change is no longer felt.  

In the implementation of the sequential KOH approach we have prescribed the variance of the prior for all time steps i.e. we have not allowed changes in variance to propagate through time.  This was a necessary assumption to ensure the prior remained sufficiently diverse.  By comparison, in implementing the PF approach we have allowed the variance of the prior to propagate. 
As discussed with reference to Figure \ref{fig:PFMon}, the artificially prescribed variance for the sequential KOH approach contributes to the more extreme peaks in mean IAS observed in Figure \ref{fig:IASMon}.  At the locations of the peaks the monitored RH values are greater than the model would predict with the parameter value ranges provided and as a consequence the models extract the highest values possible within the parameter bounds that have been set.  The magnitude of the peak parameter value is dependent on the likelihood of that value given the data and the calibration run outputs.  In the sequential KOH approach, we have observed that the magnitude of the peaks is dependent on the assumed prior variance - if a smaller variance is assumed the peaks are significantly lower and more in line with the PF results. This is because a higher variance gives more credibility to a wider range of values, affording higher values a greater likelihood. 

Both approaches make use of an emulator fitted to the simulation output in order to map the inputs of the model to the output of interest, and for both approaches Gaussian Process models have been used.  This requires the physics-based model to be run a considerable number of times with parameter values covering the ranges of interest prior to running the calibration. Within the digital twin framework this approach will require modification such that the model predictions are generated as the calibration progresses, rather than in advance. The PF approach may be more suited to this framework than the KOH approach as the PF considers each data point separately rather than mapping the entire parameter space. 

\section[Conclusions]{Conclusions}
This study makes an essential contribution to the development of digital twins of the built environment. Combination of data and a model - in this case the physics-based model - is at the heart of what is meant by a digital twin. To maximise the benefits it is essential to ensure that the data are used to continuously update the model.  In this study the continuous calibration of the model has been explored using a particle filter approach.  This has been compared against Bayesian calibration using the framework put forward by \citet{KOH_2001}, and against a sequential form of the KOH approach.  

The development of the digital twin for an underground farm provides an exemplar for the built environment more broadly.  It is necessary to maintain favourable environmental conditions in the farm and to that end an extensive programme of monitoring has been undertaken and is still ongoing. A simple physics-based model of the farm has also been developed which can aid prediction of environmental conditions and provide useful information for future planned farm expansion.  

The PF approach has been shown to give a good estimate of the mean and variance of the model input parameters ventilation rate and internal air speed which were selected owing to their impact on the relative humidity in the farm.  When using the calibrated time varying parameters in the model a much closer agreement with the monitored data is observed than when using parameters inferred from the static BC approach.  It is clear, however, that the physics-based model is unable to predict the full range of relative humidities observed in reality, in particular there are data points that fall both above and below the predicted range of relative humidity.  This model inadequacy cannot be explored using the PF approach alone - the KOH approach is required in order to quantify the model bias function and error terms. 

In terms of suitability for incorporation in a digital twin, the PF approach is more suitable than the sequential KOH approach used here as it is quicker and is more suited to combination with execution of the physics-based model at each data point.  It also quantifies the variance in the posterior parameter probability distributions, although the inevitable increase in variance over time will need to be addressed. A combined approach, using the PF approach in real time and the KOH approach for periodic assessment of the model bias function would yield the best result for ensuring continuity of model accuracy.  Further work will address the practical implementation of the particle filter within the digital twin framework. 

\begin{Backmatter}

\paragraph{Acknowledgments}
We are grateful for the continued support of Growing Underground without whose input this study would not have been possible.

\paragraph{Funding statement}
This research was supported by AI for Science and Government (ASG), UKRI's Strategic Priorities Fund awarded to the Alan Turing Institute, UK (EP/T001569/1) and the Lloyd's Register Foundation programme on Data-centric Engineering.

\paragraph{Competing interests}
None

\paragraph{Data availability statement}
The data used were made available by Growing Underground for the purpose of this study.

\paragraph{Ethical standards}
The research meets all ethical guidelines, including adherence to the legal requirements of the UK.

\paragraph{Author contributions}
Conceptualization: M.G.; R.C.; A.G. Methodology: R.W.; A.G; Data visualisation: R.W. Writing original draft: R.W; R.C; Data Curation: M.J-S. All authors approved the final submitted draft.

\paragraph{Supplementary material}
No supplementary information has been provided.

\bibliographystyle{apalike}
\bibliography{References}

\end{Backmatter}

\end{document}